\documentclass{egpubl}
 
 \ConferenceSubmission   
\usepackage[T1]{fontenc}
\usepackage{dfadobe}  

\usepackage{cite}  
\BibtexOrBiblatex
\electronicVersion
\PrintedOrElectronic
\ifpdf \usepackage[pdftex]{graphicx} \pdfcompresslevel=9
\else \usepackage[dvips]{graphicx} \fi

\usepackage{egweblnk} 

\usepackage{framed, multirow}
\usepackage[utf8]{inputenc}
\usepackage{subcaption}
\graphicspath{{images/}}
\usepackage{amsmath,bm}
\usepackage{amssymb}
\usepackage{hyperref}
\usepackage{comment}
\usepackage[title]{appendix}
\usepackage[normalem]{ulem}
\usepackage{algorithm}
\usepackage{algorithmic}
\usepackage{xcolor}

\newcommand{\bf}[1]{\mathbf{#1}}

\newcommand{\n}{\mathbf{n}}

\title[FAKIR : An algorithm for revealing the anatomy and pose of statues from raw point sets]%
      {FAKIR : An algorithm for revealing the anatomy and pose of statues from raw point sets}

\author[T. Fu \& R. Chaine \& J. Digne]
{\parbox{\textwidth}{\centering Tong Fu$^{1}$, Rapha\"elle Chaine$^{1}$ and Julie Digne$^{1}$
        }
        \\
{\parbox{\textwidth}{\centering $^1$Université de Lyon, UCBL, CNRS, France}}
}

\begin{document}

\teaser{
 \includegraphics[width=\linewidth]{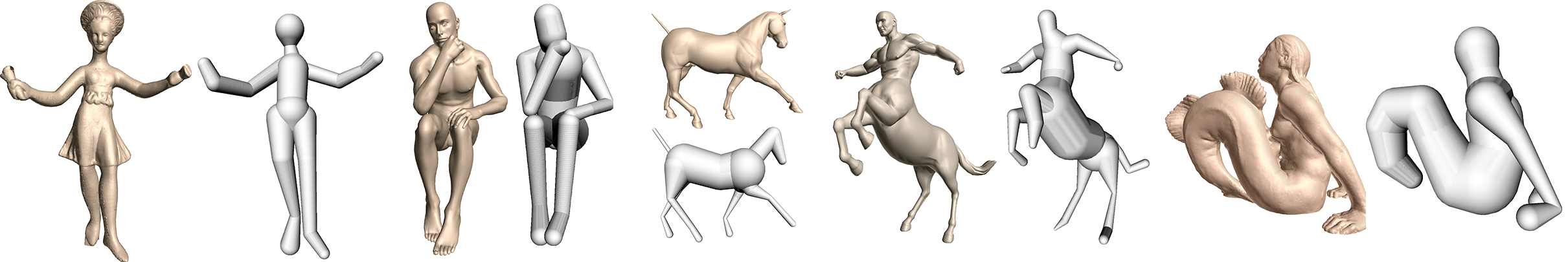}
 \centering
  \caption{Our method is able to detect the morphology and pose of articulated shapes, given an elementary anatomical model, from a single static scan. Our approach is equally effective for human, animal shapes and even imaginary creatures.}
\label{fig:teaser}
}

\maketitle
\begin{abstract}
3D acquisition of archaeological artefacts has become an essential part of cultural heritage research for preservation or restoration purpose. Statues, in particular, have been at the center of many projects. In this paper, we introduce a way to improve the understanding of acquired statues representing real or imaginary creatures by registering a simple and pliable articulated model to the raw point set data. Our approach performs a Forward And bacKward Iterative Registration (FAKIR) which proceeds joint by joint, needing only a few iterations to converge. We are thus able to detect the pose and elementary anatomy of sculptures, with possibly non realistic body proportions. By adapting our simple skeleton, our method can work on animals and imaginary creatures.


\end{abstract}  


\section{Introduction}

With the progress of 3d scanning techniques, it is now common to create digital replicas of artworks, which will remain forever intact, while their real-world counterparts will slowly decay due to time damage or human activity.
As part of the automatic processing of scanned statues, it is often necessary to identify the pose and anatomy of the model. 
Indeed, registering a model to a statue is useful for many applications. For example, one can bring statues to a common pose to better compare their style. It can also serve to combine statue parts to restore broken statue virtually or to animate a statue.

While pose recognition has been efficiently addressed for human models, in particular using Machine Learning algorithms, these  methods can only work if the model fits the training dataset of shape models. A challenge of artistic human statues compared to real human scans lies in the difference in aesthetic perception. Indeed many sculptors favored the perceived beauty of their work over the realism of human proportions \cite{143371,newell_1934}. As we will see, this has drastic consequences for example-based machine learning algorithms, which fail at adapting to these statues. As for imaginary creatures, no such training database exist. Hence there is a need for simple anatomical models, that can be fitted without requiring an extensive training on various creatures, animals or humans with nonrealistic body proportions.

In this work, we focus on human statues with no or few garments, animals, and imaginary creatures. Furthermore, we consider that the digitized statues are provided as point sets.
We propose a method for calibrating and registering a simple articulated model to a point set, which we named Forward And bacKward Iterative Registration (FAKIR). FAKIR works directly on the point cloud, avoiding thus the tedious meshing step. 
FAKIR iterates between assigning each point to its best corresponding model part and optimizing the anatomical model pose and proportions accordingly, and converges in only a few iterations.

To summarize, our contributions are the following:
\begin{itemize}
    \item A simple articulated model efficiently representing a statue pose and anatomy.
    \item An efficient calibration and registration process based on inverse kinematics principles. 
\end{itemize}

\section{Related work}

\paragraph*{Anatomical Model.} 
Designing anatomical models for human shapes has raised a lot of interest in Computer Graphics. 
The most common representation consists in a more or less detailed graph of bones such as the ones used in the MakeHuman framework~\cite{makehuman}, while some methods go beyond this kind of elementary skeleton representation and model every single muscle to increase realism~\cite{Lee12}. Due to our nonrealistic context, we will focus here on basic skeletons, which are pliable and efficient enough for our purpose.
A skeleton-based model consists of two components: a skeletal structure~\cite{855884} and a representation for the volume surrounding it. This representation can be either mesh surfaces or volumetric primitives. Recent surface-based models \cite{Anguelov05scape:shape,Dyna:SIGGRAPH:2015,SMPL:2015,Hasler09cgf,Zuffi:CVPR:2015} are learned from numerous scans of real people.
Among those, SMPL ~\cite{SMPL:2015} is a human anatomical mesh model in which a set of parameters control non-rigid deformations resulting from a statistical study on a large number of humans and positions. However, to our knowledge, there is no approach that allows us to position this model from a static point cloud without positioning the model close to the data or without using 2D views and deep learning.
It works well for the capture of human motion and shape in a video~\cite{Zhang17} or images~\cite{Bogo:ECCV:2016, MuVS:3DV:2017}. 
But this model has a poor performance when the data is unrealistic which is common for archaeological statues. Another possible shape representation is based on volumetric primitives, e.g. using medial axis transform (MAT)~\cite{Blum:1967:ATF,medialaxis}, metaballs~\cite{metaballs}, B-meshes\cite{b-mesh} or other primitives~\cite{Gavrila19963DMT,loose_limbed,viper}.
Among these models, the sphere-mesh model~\cite{Thiery13}, a variant of convolution surfaces~\cite{Bloomenthal:1991:CS:122718.122757}, has been introduced for representing mesh models by packing spheres into it and encoding their structure. Conceptually, the sphere-mesh model can be seen as a piecewise linear simplification of the computational geometry skeleton~\cite{Tagliasacchi16}. Although sphere-meshes were developed to extract the shape structure from an input mesh, they can be used to represent an anatomical model by imposing constraints on them. As such, it has been used successfully for representing hand skeletons~\cite{Tkach16,Remelli17}. This model is light and pliable and we will also rely on it.

\paragraph*{Skeleton registration} 
Registering a model to a shape is an important task which has received much research interest. The goal can be to animate a shape by skeleton \emph{rigging} and \emph{skinning}, or to detect human poses.
Skeleton rigging can be performed manually~\cite{855884, b-mesh}, but a few methods have investigated automatic processes. In particular, the Pinocchio  algorithm~\cite{Baran07} adapts a skeleton to a static mesh by defining an objective function and maximizing it. It works by packing spheres into the mesh and by considering their centers, gathered in a graph, as the admissible joint positions. This pre-computation makes the skeleton pose estimation tractable. On points sets, a $\ell^1$-medial skeleton could be used \cite{Huang2013} alternatively to sphere-packing, but it is not suitable for noisy or incomplete data, as shown by our experiments.
If the input data is dynamic, it is possible to infer, or track, a skeleton from it. Most tracking approaches~\cite{Sigal:IJCV:10b,gall2009,Theobalt2004} focus on the capture of the positions of the joints, and deduce pose parameters (angles) and intrinsic parameters (\emph{e.g.} bone lengths) from it.
Many of such tracking methods~\cite{gaussiansbodymodel,motioncapture} start with a calibrated skeleton but the calibration itself can be performed from a depth video and a set of known admissible poses~\cite{Tkach16,Remelli17,onlinehand}. 
Such methods require a dynamic scene and cannot apply to the static shape rigging problem. 
It is also possible to rely on a database of people scans to learn the pose and deformation of human bodies using the SMPL model \cite{10.1007/978-3-642-33783-3_18,home3dbody,alldieck2018video}. Recently, CNN-based detectors, such as DeepCut\cite{Deepcut} and OpenPose\cite{cao2018openpose} were used for 2D joints detection in images or videos. Bogo et al.~\cite{Bogo:ECCV:2016} estimate the 3d human pose and proportions from a single image by fitting a SMPL model to DeepCut estimated joint positions. 
Using multi-view images over time \cite{MuVS:3DV:2017} improves the pose accuracy, however the human proportions remain approximate. Human tracking can also be done without needing a model \cite{Bogo:ICCV:2015}. Similarly it is possible to register two models using manifold-harmonics based non rigid registration \cite{puzzles}, but this would not help for skeleton-based registration. Learning approaches can also work from a single image \cite{lazova2019360degree,alldieck2019learning}.

Registering an SMPL skeleton directly to a point cloud has been addressed using deep learning directly on point sets augmented with feature detection~\cite{Jiang19}, but this method only targets human shapes, which it learns from a database, while our method can work on nonrealistic anatomies and on various animals, as will be demonstrated in our experiments. Deforming a point cloud to match a template mesh has also been tackled using auto-encoders~\cite{LBS}, but this requires a full template mesh for each model. Our required skeleton model is much lighter.
Finally, recently, the FARM\cite{FARM} method builds on the functional map framework to register a parametric model (such as the SMPL one) to a mesh or a point set in a fully automatic way. This approach reaches state of the art results while being the closest in goal to ours, and we will compare to it.

Finally, some methods \cite{Hasler09smi,Zhang17,Pons-Moll:Siggraph2017,yu2019simulcap} aim at finding a person's pose despite its sometimes loose clothing, but this is outside the scope of our paper.

Our registration algorithm makes extensive use of kinematic chains, processing them alternatively forward and backwards. In spirit, this is related to inverse kinematics, and in particular the Fabrik \cite{Aristidou:2011:FABRIK} and CCD \cite{86079} algorithms. Indeed, both methods define kinematic chains and compute their transformation from an input pose to a target pose by updating pose parameters one after the other alternatively forward and backward along each chain. However the similarity ends here, since our goal is to estimate not only the pose but also the proportions of the model limbs using data-attachment constraints in a static framework.

\section{Anatomy and Pose estimation}

  \subsection{Human model}
   
In our context of artistic statues, it is necessary to devise a human model with few constraints, allowing to fit a sculpture which does not follow the human proportion beauty canons. The existing fully detailed human templates modeling every single limb and muscle in a very realistic way are too constrained for our purpose.
 
\begin{figure}[ht]
    \centering
     \begin{subfigure}[ht]{0.22\textwidth}
        \centering
        \includegraphics[trim = 0mm 50mm 0mm 60mm, clip, width=\textwidth]{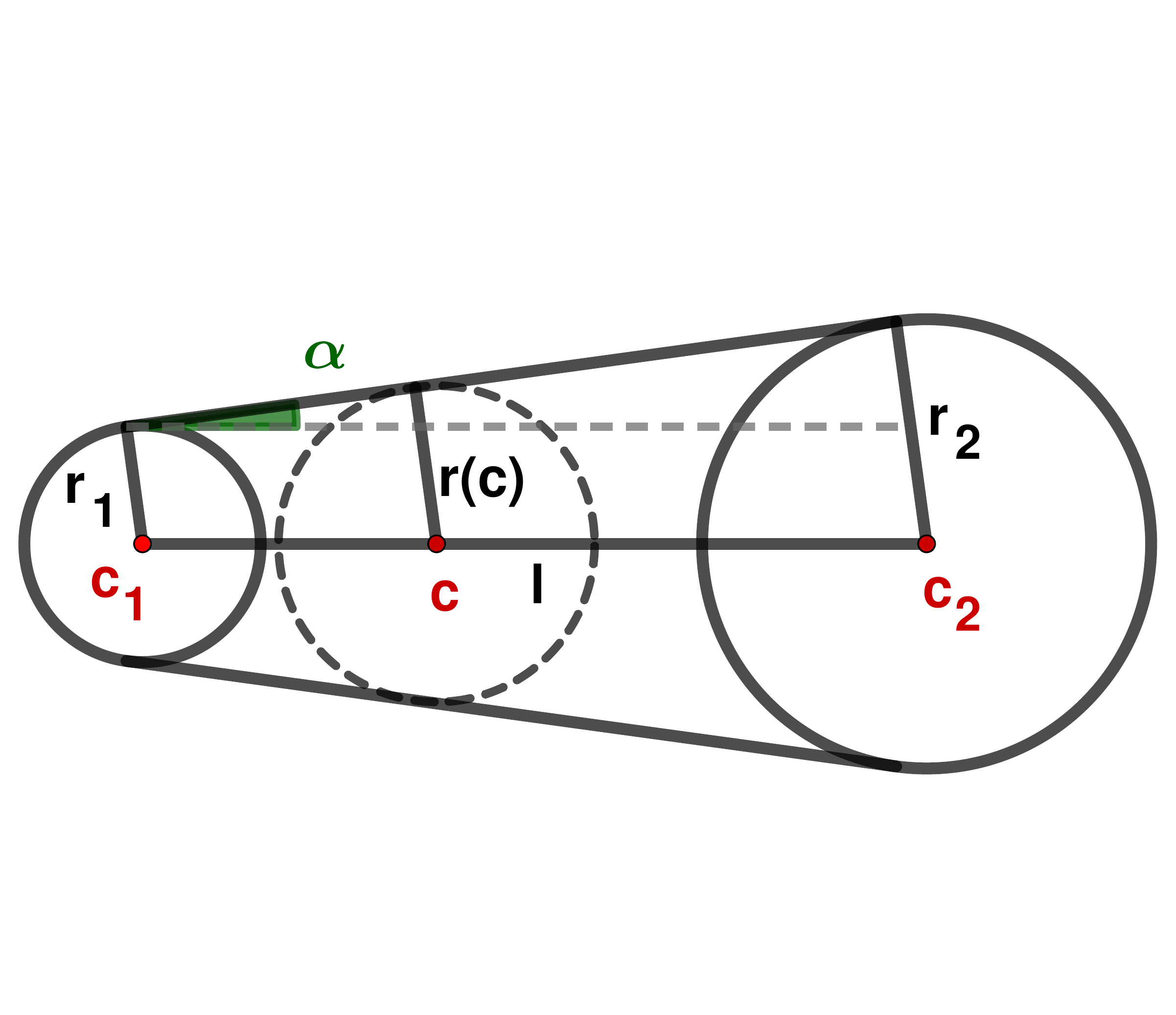}
        \caption{Cross-section of a bone}
        \label{fig:spmesh2d}
    \end{subfigure}
    ~
    \begin{subfigure}[ht]{0.22\textwidth}
        \centering
        \includegraphics[width=\textwidth]{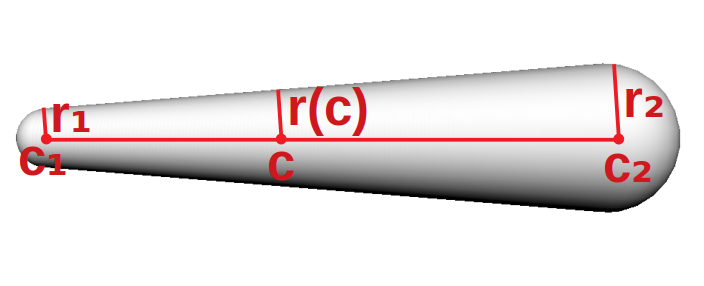}
        \caption{Sphere-mesh of a 3D bone.}
        \label{fig:sphere}
    \end{subfigure}
    \caption{The sphere-mesh of a bone is the union of the spheres centered on segment $[c_1c_2]$, with radius varying linearly between the two extremities of the segment.}
    \label{one_bone}
\end{figure}

We introduce an anatomical model inspired by the sphere-mesh model~\cite{Thiery13}, already successfully used for hand tracking~\cite{Tkach16, Remelli17}, using only one-dimensional elements.
In this model, each bone is represented by a sphere-mesh $B$ corresponding to the envelope of the union of a set of spheres centered on a segment and with a linearly varying radius (Figure \ref{fig:sphere}).
Each bone is defined by two end sphere centers
$c_1$ and $c_2$ with associated radii $r_1$ and $r_2$ respectively. The segment $[c_1c_2]$ is the medial axis of the bone. For each point $c \in[c_1c_2]$,  the radius of the sphere centered at $c$ is $r(c) = (1-\tau) r_1 + \tau r_2$, with $\tau= \frac{\|c_1c\|}{\|c_1c_2\|}$.

The sphere-mesh model is controlled by the length $l= \|c_1c_2\|$ and the pair of sphere radii  $\mathbf{r}=\{r_1,r_2\}$. Consequently, we denote the sphere-mesh model for one bone as $B(l,\mathbf{r})$. We also denote by $\alpha$ the angle of the conic part of the bone, as illustrated on Figure \ref{fig:spmesh2d}.
Importantly enough, the bones we are defining do not correspond to anatomical bones, but more to limbs (\emph{i.e.} it includes a coarse description of the flesh volume around the anatomical bone). By analogy to inverse kinematics, we keep the word \emph{bone} instead of \emph{limb}.

\begin{figure}[ht]
 \centering
     \begin{subfigure}[t]{0.45\linewidth}
        \includegraphics[trim = 0mm 5mm 5mm 0mm, clip, width=0.45\textwidth]{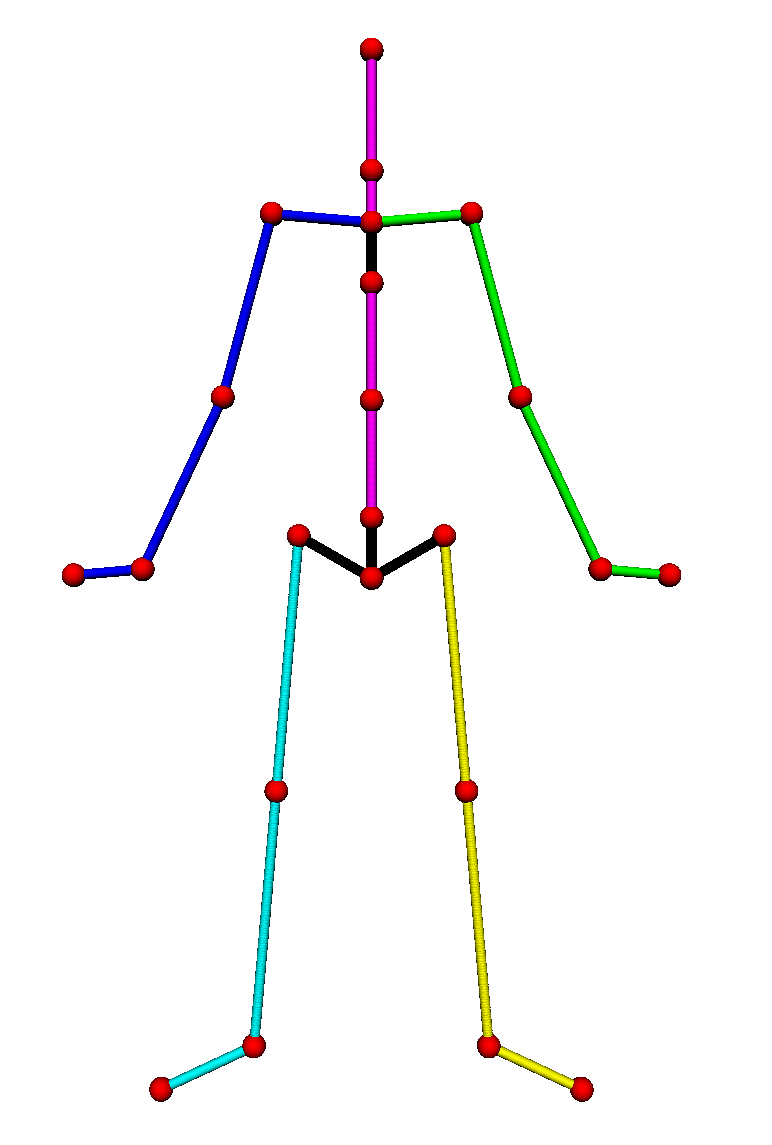}
        \includegraphics[trim = 0mm 5mm 5mm 0mm, clip, width=0.45\textwidth]{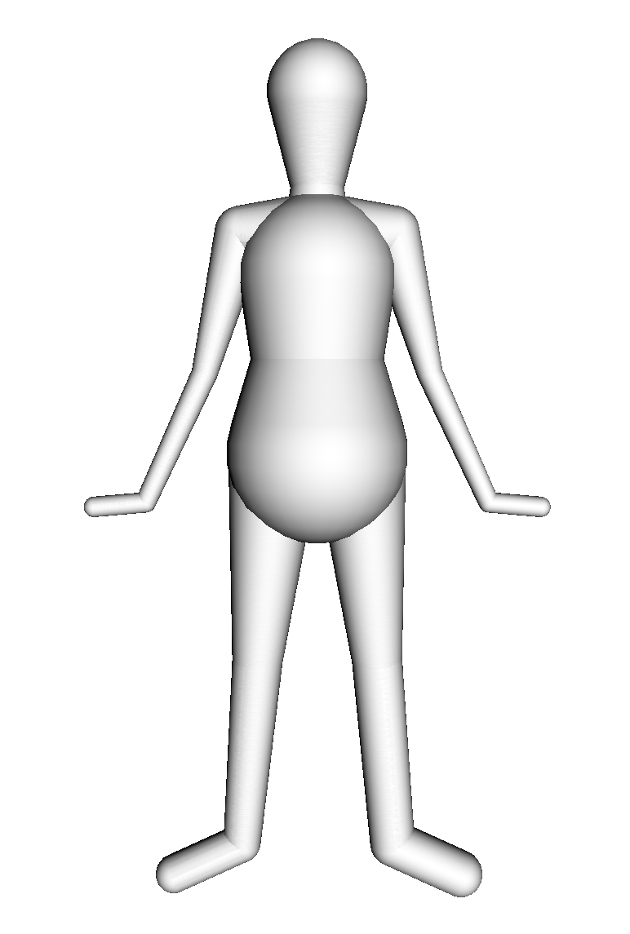}
        \caption{Human model}
        \label{fig:hmodel}
    \end{subfigure}
     \begin{subfigure}[t]{0.45\linewidth}
        \includegraphics[trim = 0mm 5mm 5mm 0mm, clip, width=0.45\textwidth]{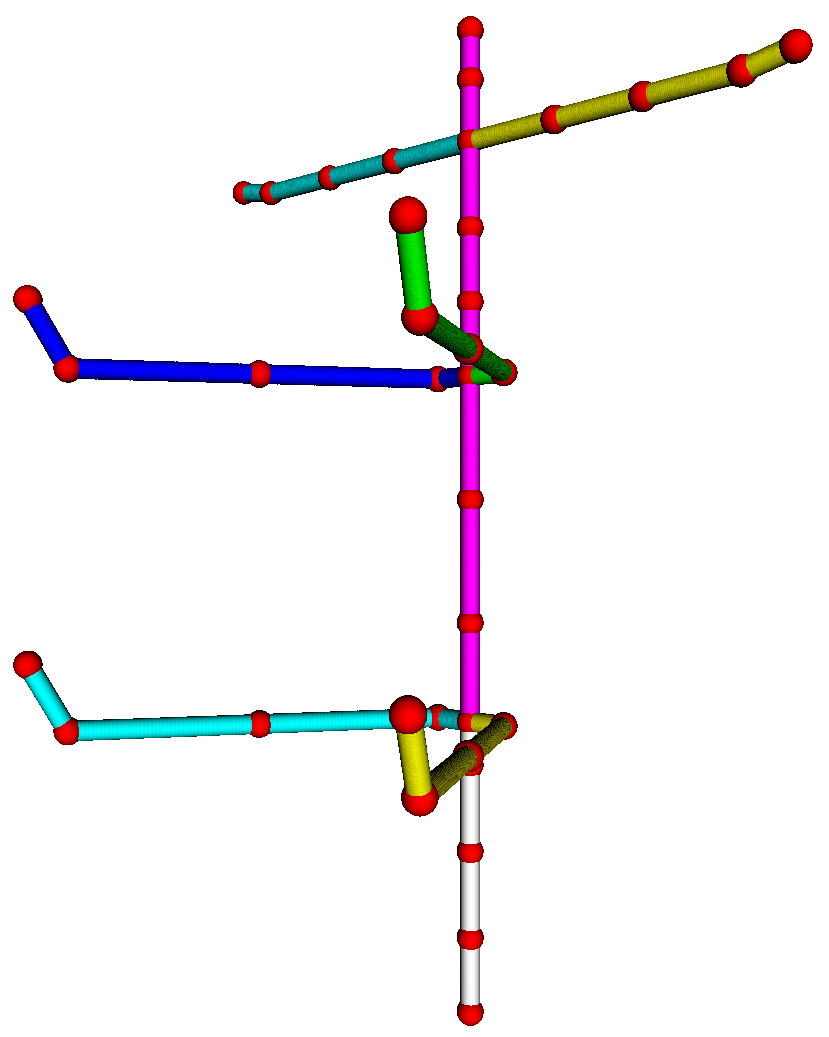}
        \includegraphics[trim = 0mm 5mm 5mm 0mm, clip, width=0.45\textwidth]{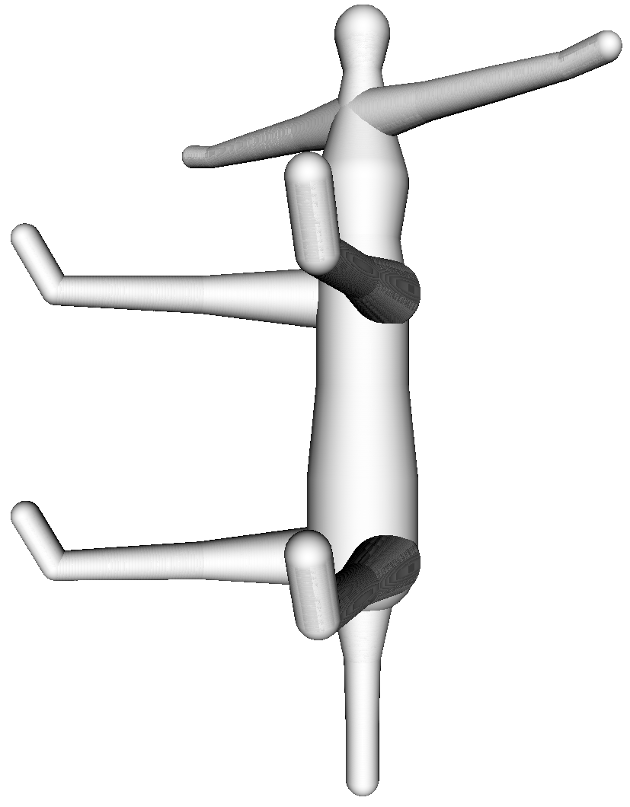}
        \caption{Centaur model}
        \label{fig:centaur_skeleton}
    \end{subfigure}
    \caption{Skeleton and sphere-mesh models for humans and centaurs. For the human model: the bones are organized into 5 chains shown in different colors. $4$ additional bones are drawn in black: the pelvis which is a constrained triplet of bones, and the connection bone between the spine and the neck. Our model can also be adapted to various creatures, even imaginary ones such as a centaur.}
    \label{fig:human_model}
\end{figure}

With this type of bone element, we construct a simple human body template with a very coarse respect of human proportions as an initial body shape (Figure \ref{fig:hmodel}) but we can construct a template for any other animal or imaginary creature (Figure \ref{fig:centaur_skeleton}) as well.
Our human body template contains 22 bones $\{B_k\}_{k=1..22}$.
Three of those correspond to the pelvis and have no relative motion: their length is fixed up to a common scale parameter that will be determined during the registration, along with the orientation of the triplet. Additionally, a special bone is used to connect the spine bone to the neck, and its length and orientation directly depend on the adjacent spine bone. The other bones have no constraint on their relative proportions.
The bones are organized into 5 chains, depicted in different colors in Figure \ref{fig:human_model}. 
These chains are independent with the only constraint that some extremities must remain anchored to the spine.
The orientation of the chains is used to define the predecessor and successor of each bone. The ordering will be reversed to process the chain forward and backward several times during the registration process.
Each bone is thus fully defined by its intrinsic parameters (length and two radii) and by its extrinsic parameter (rotation with respect to its predecessor). 
Because of the simplicity of the sphere-mesh bone model, the distance from a point to the model can be easily computed. In contrast, using a mesh model would make these computations much more demanding.

\subsection{Distance between the model and a point set} \label{computation}

To capture the anatomy and the pose of a statue, we need a distance function to measure how the sphere-mesh model fits a point set $P$, even if the points are far from the bones they should be attached to. The sphere-mesh model calibration and registration strives to reduce the distance between the sampled points and their corresponding bones on the sphere-mesh. The problem is that the bone to which a point should be assigned is unknown, especially if the model has not been calibrated beforehand and if it is far from the data. Therefore, the target bone is usually replaced by the closest bone.
We assume that the coordinates of the points are provided with a coarse approximation of the oriented normal. This speeds up the registration when the model is not close to the data, but it remains possible to implement our registration algorithm with a simple Euclidean distance from the points to the model.

\paragraph*{Distance from a point to one bone.}
We start by defining the normal-constrained projection of a sampled point $p$ on a single bone $B$ by using the oriented normal vector $\n_p$ to disambiguate the choice between several orthogonal projection possibilities. 
Given a point $p$ in the ambient space with oriented normal $\n_p$, 
we consider the lines passing through $p$ and orthogonally intersecting the sphere-mesh surface (possibly crossing its interior) at some points. 

Whenever it is possible, we select the projection $\tilde{p}$ whose normal $\n_{\tilde{p}}$ has positive scalar product with $\n_p$. Considering this normal-constrained orthogonal projection allows for a faster convergence and better results (see the supplementary material for more details).
Since each point $p$ has a normal-constrained projection on all the bones, we refer to its normal-constrained projection on bone $B_k$ as $\tilde{p}_k$. If no subscript is provided, $\tilde{p}$ refers to the normal-constrained projection of $p$ on the closest bone. 

\paragraph*{Distance from a point set to a sphere-mesh chain.}
Given a point set $P$ and a sphere-mesh chain of $K$ bones, we first need to approximate the subset of current points that project on each bone. In the following, we define the point set $P_k$ as the subset of points $p\in P$ which are closest to bone $B_k$ using the distance $d_k=\|p-\tilde p_k\|, k=1\cdots K$. 
Once the assignment is computed, the one-bone distance function $E_k$ is defined as the sum of squared distances from points of $P_k$ to bone $B_k$:
\begin{equation}
E_k(P_k,B_k(l_k,\boldsymbol{r}_k),\boldsymbol{\theta}_k) = \underset{p\in P_k}{\sum}\|p - \tilde{p}_k\|^2
\end{equation}
Importantly enough, the subset $P_k$ and the one-bone energy $E_k$ depend on the position of the initial extremity of the chain of bones involving $B_k$, as well as the parameters of the other
bones in the chain.

The sum of one-bone distance functions measures the fitness of the model and serves as an objective function that we aim to minimize in order to capture the anatomy and pose of the sphere-mesh that best corresponds to our point set. 
\begin{equation}
E = \sum_{k=1}^K \sum_{p\in P_k}\|p - \tilde{p}_k\|^2.
\end{equation}

In the next sections, we will also be interested in the distance restricted to two adjacent bones $B_k$ and $B_{k+1}$, which we call two-bones energy:
\begin{equation}
E_{k, k+1} = \sum_{p\in P_k}\|p - \tilde{p}_k\|^2 + \sum_{p\in P_{k+1}}\|p - \tilde{p}_{k+1}\|^2.
\end{equation}

\subsection{FAKIR : Forward And bacKward Iterative Registration}

To register our anatomical model, we propose a kinematic approach taking into account the way the model is articulated. Contrarily to many methods that work from videos or multiple views~\cite{Tkach16,Remelli17}, our method requires only one joint center to be close to its optimal position, the rest of the skeleton pose being arbitrary.
Inspired by the FABRIK~\cite{Aristidou:2011:FABRIK} and CCD~\cite{86079} algorithms, our registration algorithm successively loops forward and backward through the chains of bones so as to rotate and scale them to match the data, refining the parameters while temporarily fixing the extremities of some bones. Hence our algorithm is named Forward And bacKward Iterative Registration (FAKIR). 
An originality of our method is that bones are mainly considered by consecutive pairs, which allows for a more robust estimation of the pose and skeleton parameters along a chain. 
The optimization of the parameters related to bone $B_k$ requires that there are relevant points attracting that bone into the data attachment term, which justifies a special order for the optimizations (see the supplementary for details).

\begin{figure*}[ht]
        \begin{subfigure}[ht]{0.135\textwidth}
            \includegraphics[width=\textwidth]{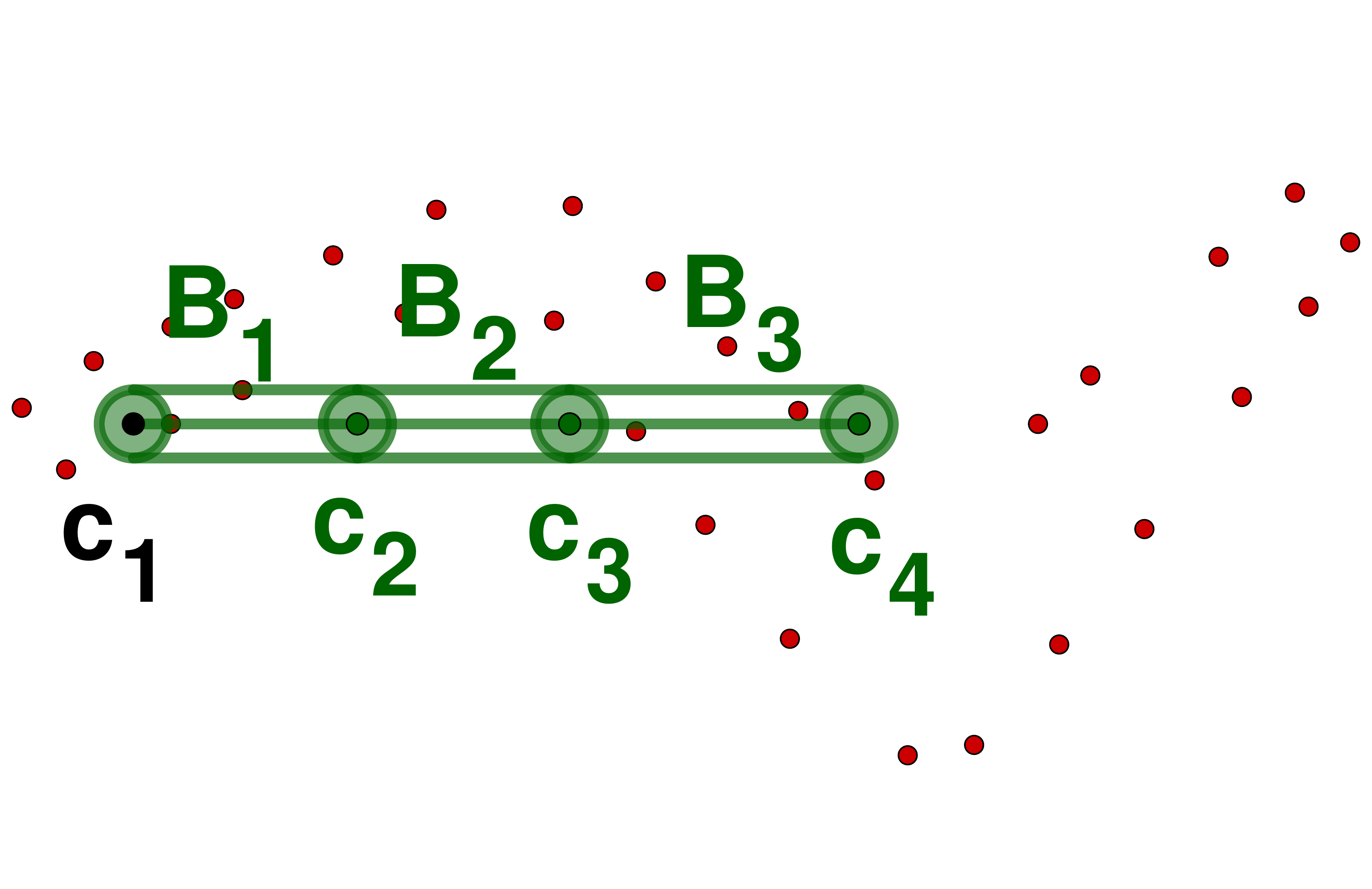}
            \caption{}
        \end{subfigure}
        \begin{subfigure}[ht]{0.135\textwidth}
            \includegraphics[width=\textwidth]{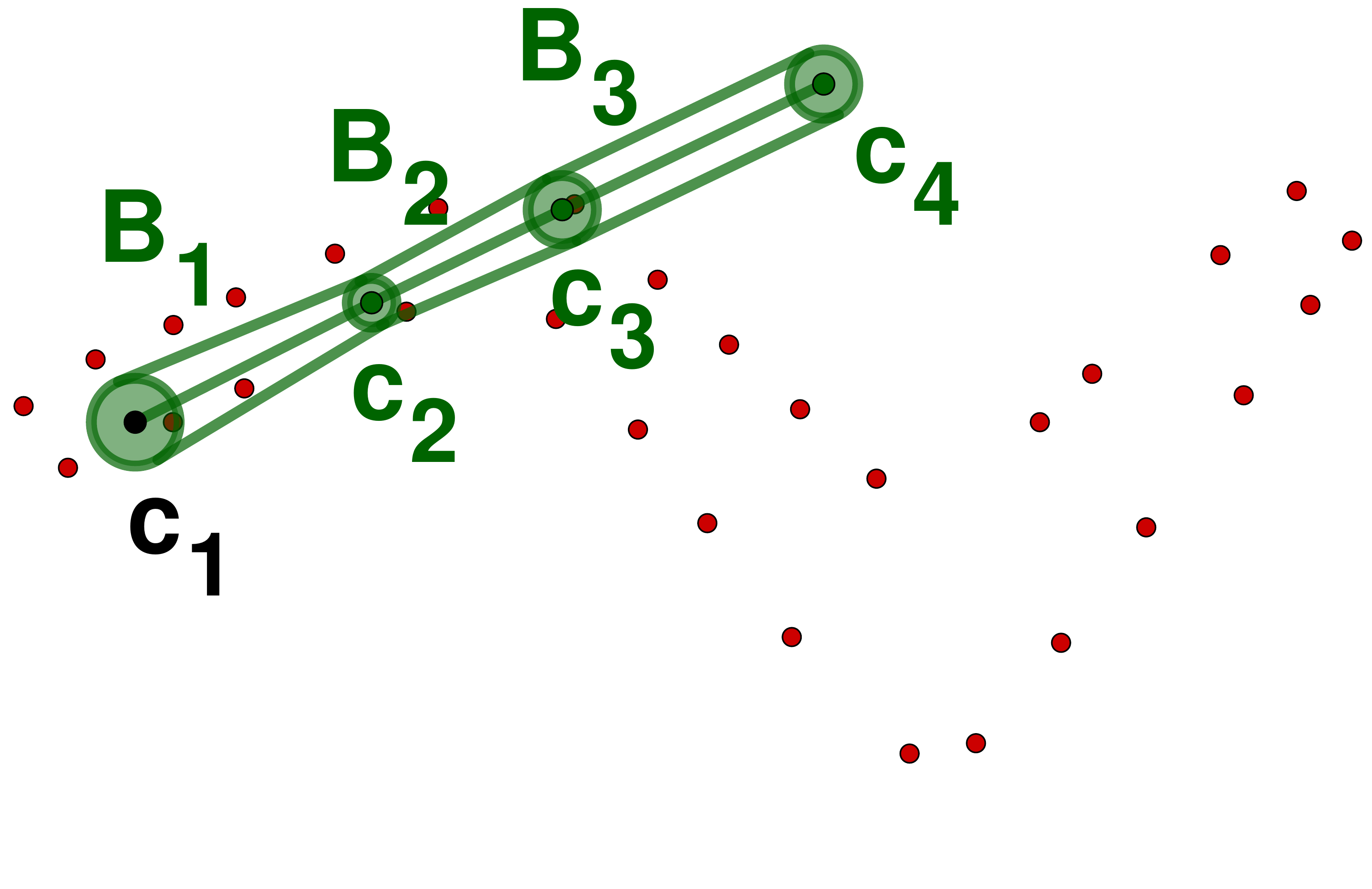}
            \caption{}
        \end{subfigure}
        \begin{subfigure}[ht]{0.135\textwidth}
            \includegraphics[width=\textwidth]{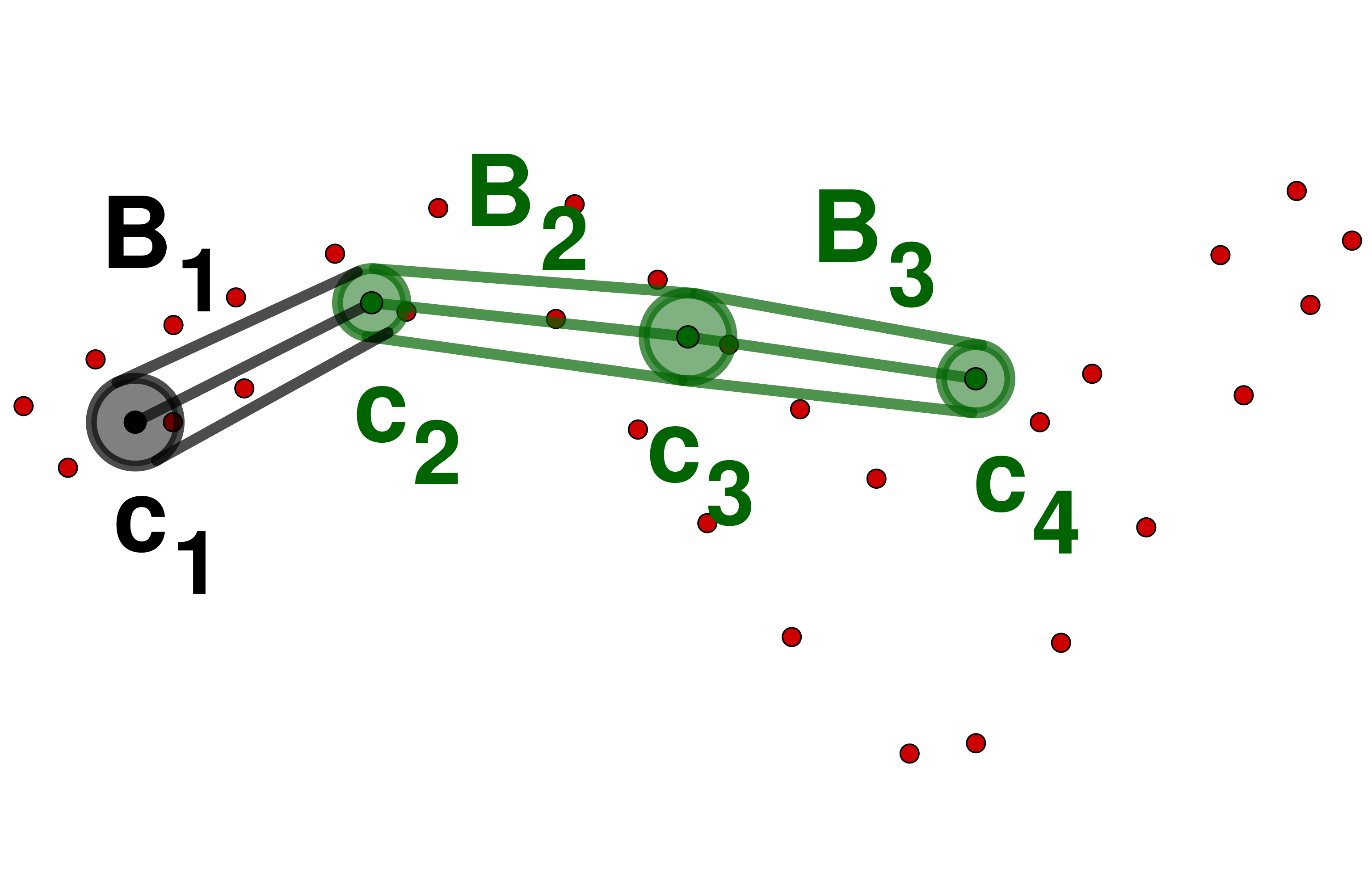}
            \caption{}
        \end{subfigure}
        \begin{subfigure}[ht]{0.135\textwidth}
            \includegraphics[width=\textwidth]{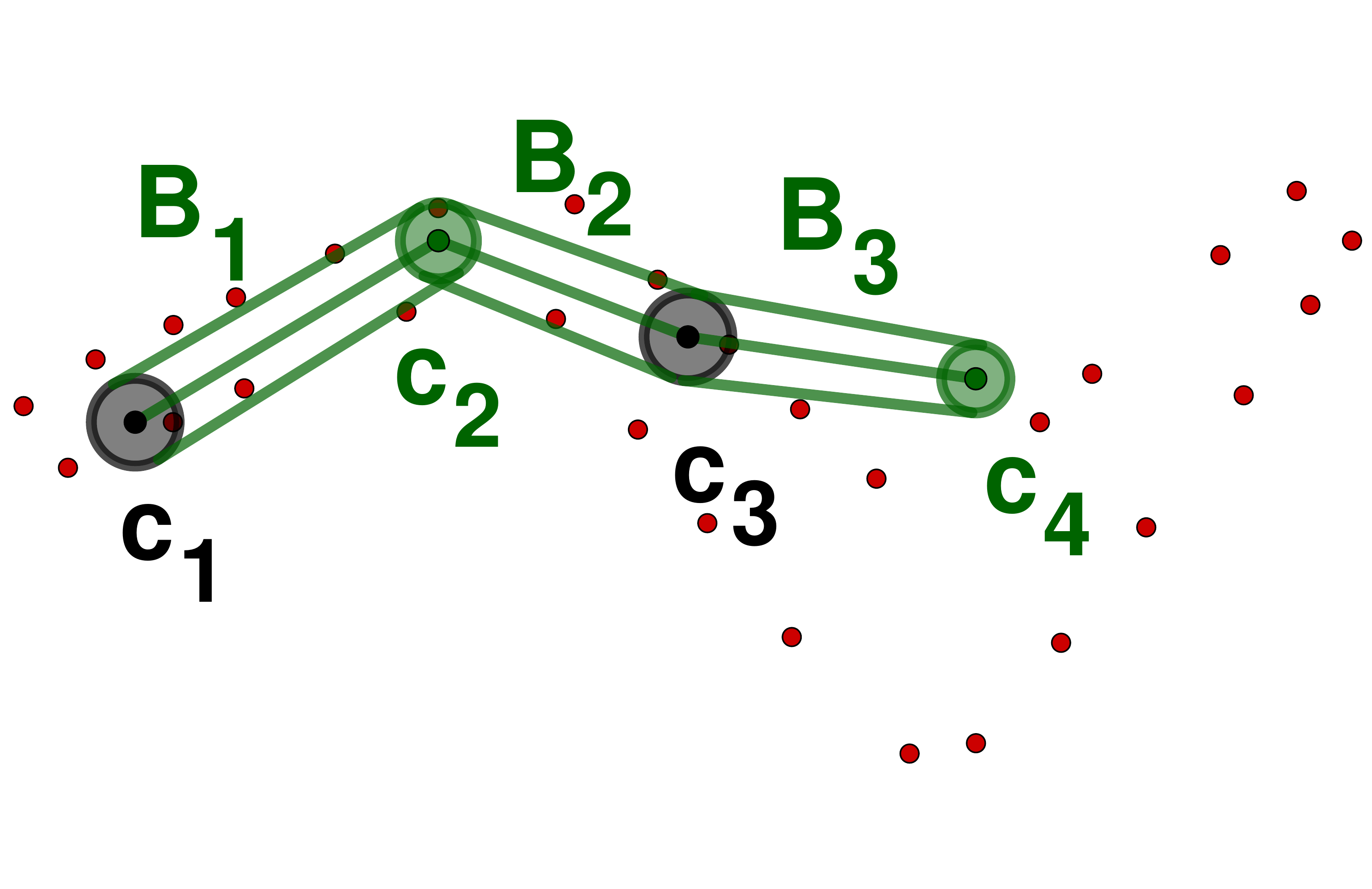}
            \caption{}
        \end{subfigure}
        \begin{subfigure}[ht]{0.135\textwidth}
            \includegraphics[width=\textwidth]{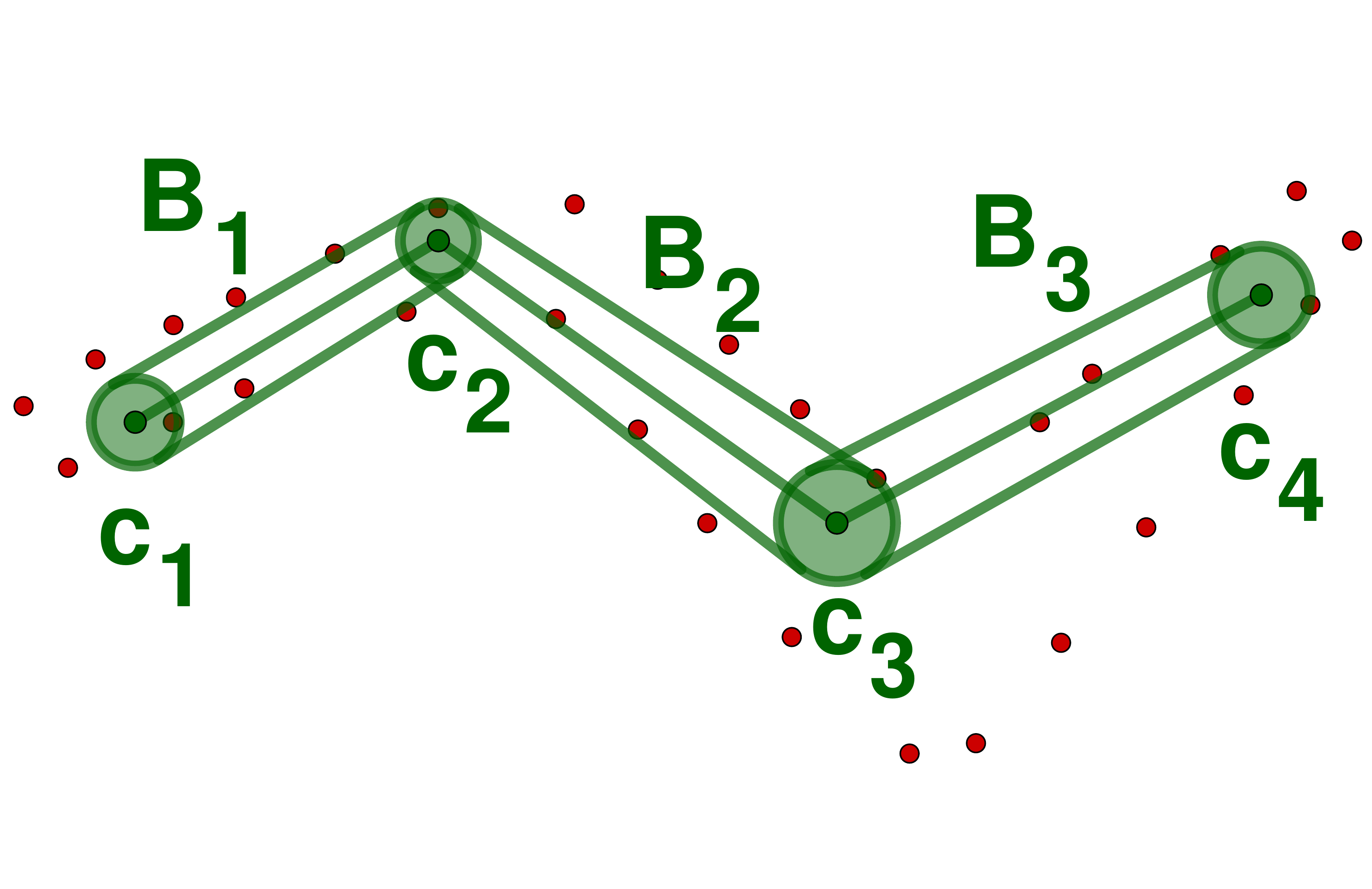}
            \caption{}
        \end{subfigure}
        \begin{subfigure}[ht]{0.135\textwidth}
            \includegraphics[width=\textwidth]{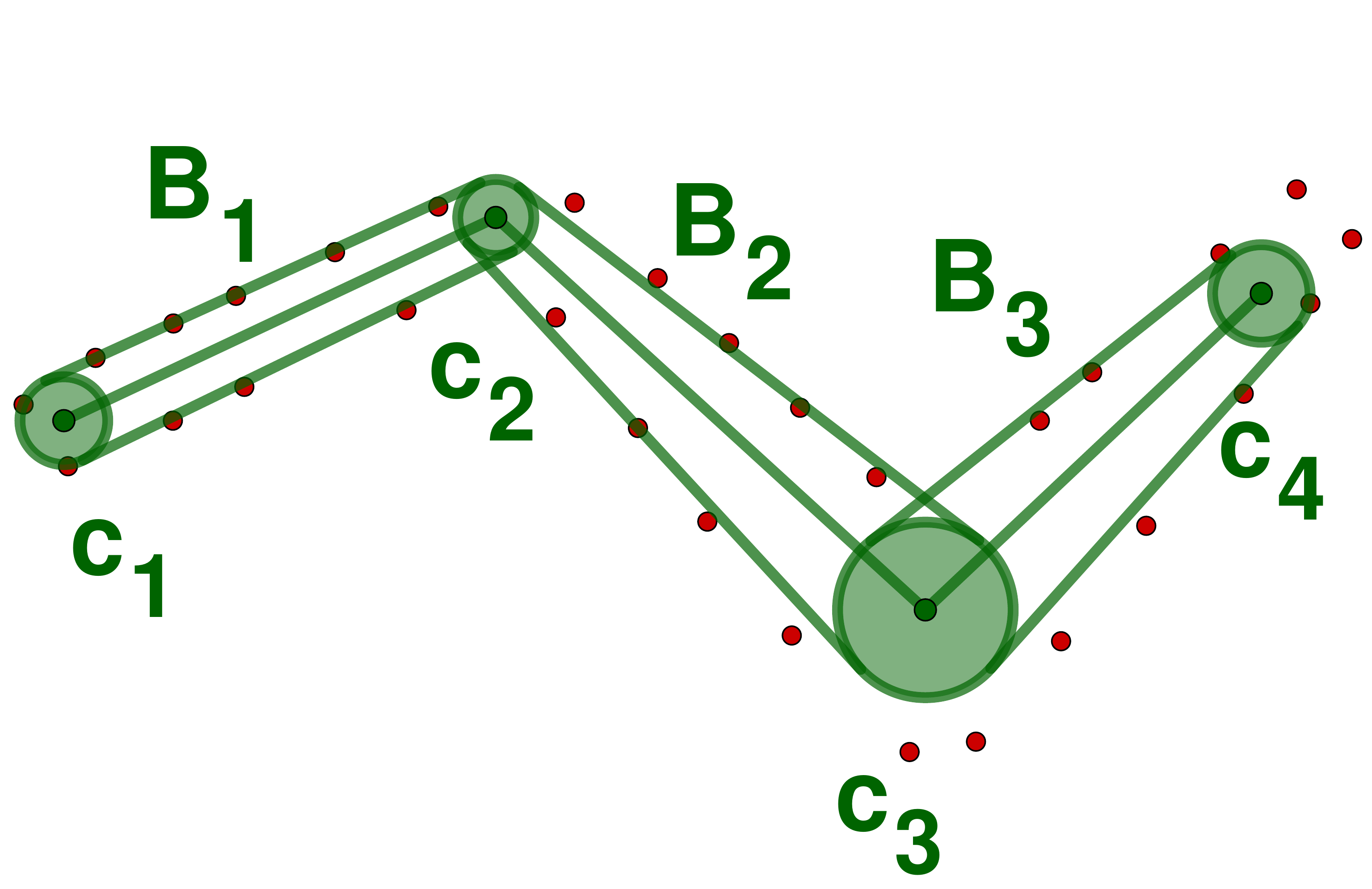}
            \caption{}
        \end{subfigure}
        \begin{subfigure}[ht]{0.135\textwidth}
            \includegraphics[width=\textwidth]{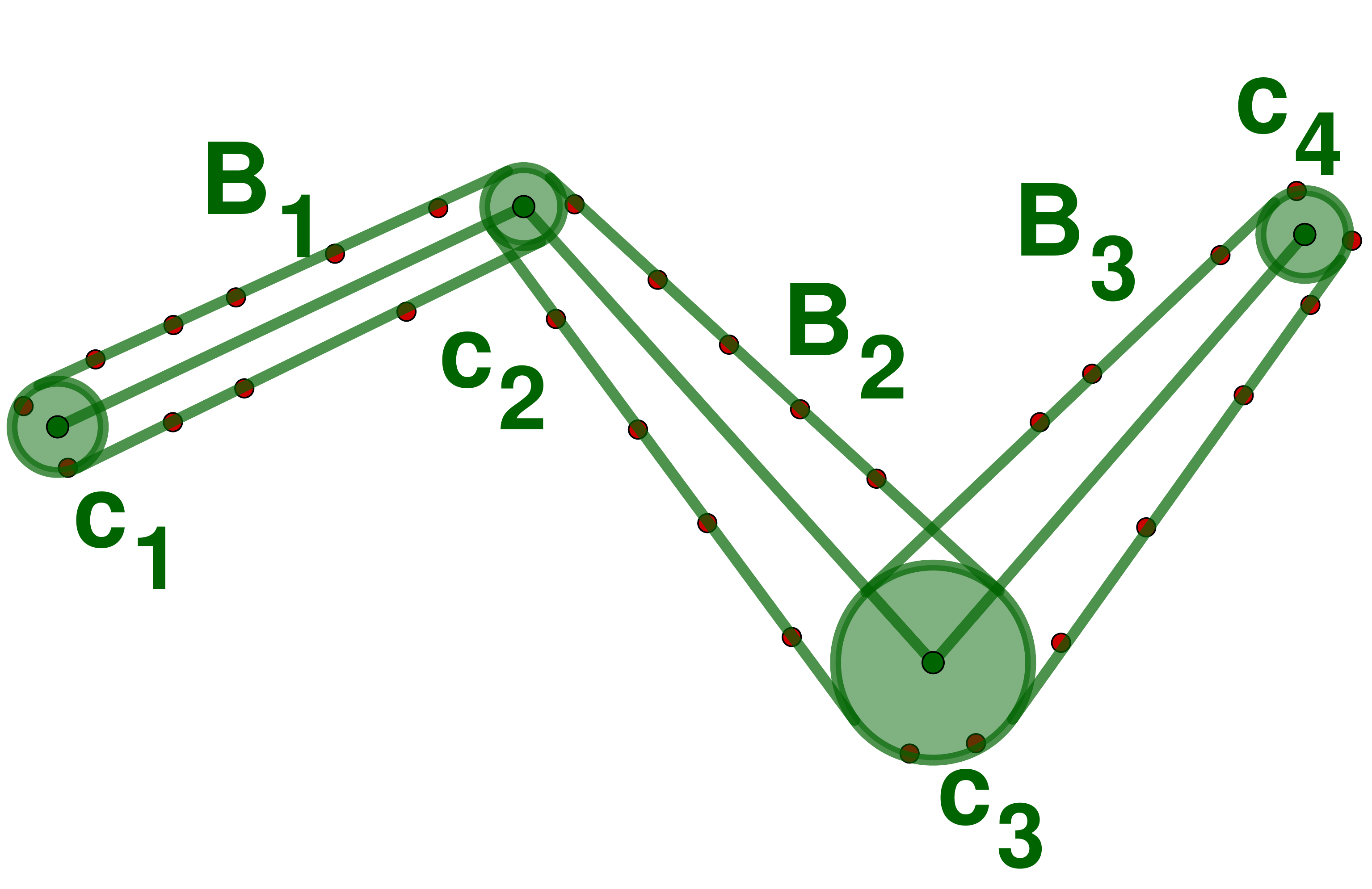}
            \caption{}
        \end{subfigure}
    \caption{
    Overview of the forward and backward iterative registration for a 3-bone chain. From an initial position (a), the chain extremity $c_1$ is fixed and the first bone $B_1$ is rotated and scaled to roughly calibrate  its dimensions and pose through the optimization of the one-bone energy (b); the bone $B_1$ is fixed and the parameters of the second bone $B_2$ are roughly calibrated in turn (c); joint $c_2$ which is common to the first two bones is scaled and its position is optimized, by using the two-bones energy, the other joints being fixed (d); The position and length of the third bone $B_3$ are then coarsely calibrated through one-bone optimization and the process continues by alternating single bone optimization and two-bones optimization, until the last bone of the chain (e). After this first coarse calibration forward pass finishes, a backward pass using only two-bone optimizations is performed (f) permitting to refine the pose and skeleton parameters and solve for the chain extremity position. With few forward and backward pass involving two-bone optimization only, the model is registered (g).}
    \label{fig:process}
\end{figure*}

\paragraph*{Registration process for a chain of bones.}  
If the parameters of bone $B_k$ have not yet been initialized, and that it is close to a subset of points to which it should ideally be associated, the estimation of the one-bone energy is meaningful and the minimization of this energy can be used to initialize the position and radii of that bone with respect to the data.
Our algorithm gradually rotates and scales the current bone $B_k$ with respect to its predecessor, updating $P_k$ after each step, so that $P_k$ gradually contains more relevant points. However, if $P_k$ is empty, the bone is first rotated around the three axis until some input points are projected onto it to bootstrap the optimization.
Once the position of $B_{k}$ has been approximately found, the algorithm turns to the coarse estimation of the position of $B_{k+1}$. 
All these computations are driven by the minimization of the one-bone energy. However, the one-bone energy alone might be inefficient to approximate the full length of a bone accurately.
To alleviate that, in an intertwined manner, a finer local registration is performed each time two consecutive bones $B_{k}$ and $B_{k+1}$ have been processed, by minimizing the two-bone energy. This process optimizes the common joint position and radius while keeping the two other joints fixed. It is the most important component in our algorithm.
Once a chain of $K$ bones has been positioned and scaled over its entire length, we repeat the process forward and backward in the chain in order to further refine the joints positions and radii between pairs of consecutive bones, using only two-bone energies optimizations. Extremity bones are optimized based on the one-bone energy after each forward or backward pass. 
The full process is summarized in Algorithm \ref{algo:full_algo} and illustrated with a chain of three bones in Figure \ref{fig:process}.

Notice that if two limbs are aligned, the joint position can not be guessed from the data and it may cause several limbs to be included in a single primitive of our sphere-mesh. To avoid this, very loose constraints on the differences in proportions between consecutive bones can be set.

\begin{algorithm}
\footnotesize
\caption{Forward and backward iterative registration}
\begin{algorithmic}[1]

\REQUIRE A point set $P$ and a sphere-mesh chain of $K$ bones with one chain extremity close to its optimal position
 \ENSURE  The registered sphere-mesh chain.

  \STATE \textbf{Initialization:}
     \STATE Fix the center of the first extremity of the chain. Rotate the first bone and adjust its radii and length by minimizing the one-bone energy;  
     \FOR {$k:=1$ \TO $K-1$}
         \STATE Consider the pair of bones $B_k,B_{k+1}$ :
         \STATE Fix the position of the joint common to $B_k$ and $B_{k+1}$;
         \STATE Alternate between the optimization of $B_{k+1}$'s rotation w.r.t $B_k$, optimization of $B_{k+1}$'s intrinsic parameters and update of $P_{k+1}$;
         \STATE Fix the positions of the 2 joints that $B_k$ and $B_{k+1}$ do not share, and free their common joint;
         \STATE Compute the position and the radius of the common joint by using the two-bones energy.
     \ENDFOR
     \STATE Compute the length of the last bone and the radius of the last sphere. 
\STATE \textbf{Forward and Backward registration loop:}
   \REPEAT
     \STATE  Reverse the order of the bones in the chain;
     \FOR {$k:=1$ \TO $K-1$}
         \STATE Consider the pair of bones $B_k,B_{k+1}$ :
         \STATE Fix the positions of the 2 joints that $B_k$ and $B_{k+1}$ do not share;
         \STATE Compute the position and the radius of the common joint by using the two-bones energy.
     \ENDFOR
     \STATE Compute the length of the last bone and the radius of the last sphere with the one-bone energy.
 \UNTIL{convergence}
\end{algorithmic}
\label{algo:full_algo}
\end{algorithm}

\paragraph*{Optimization of a single bone}
As stated above, the optimization of the one-bone energy is only used for estimating the parameters of the extremities of a chain, or for the very first forward pass in Algorithm \ref{algo:full_algo}. 

In the single-bone case, the aim is to estimate the 3D rotation of the bone, its length and the radius of its free extremity by minimizing the one-bone energy $E_k(P_k,B_k(l_k,\bf{r}_k),\bm{\theta}_k)$, where $\bm{\theta}_k$ are the angles of rotation with respect to the predecessor bone. This optimization is performed using the Levenberg-Marquardt algorithm for each parameter. In particular, the rotation can be decomposed into two rotations around two axes that are orthogonal to $c_{k}c_{k+1}$, indeed the rotation around $c_{k}c_{k+1}$ is not considered since it leaves the bone unchanged.
To optimize $\bm \theta_k$, we iteratively look for the best angle $\bm\theta_k+\delta \bm\theta_k$.
At a minimum, $\nabla_{\delta \bm \theta_k} E_k(P_k,B_k(l_r,\bm r_k),\bm\theta_k +\delta\bm\theta_k) = \bm 0$, and the value for $\delta\bm\theta_k$ follows. The details for the damped least-squares estimation are provided in the supplementary material for all parameters.

\begin{figure}[ht]
        \begin{subfigure}[ht]{0.45\linewidth}
            \includegraphics[trim = 0mm 130mm 0mm 0mm, clip, width=\linewidth]{images/2bones_1}
            \caption{Rotate $B_k$ $\cup$  $B_{k+1}$ }
            \label{fig:2b_1}
        \end{subfigure}
        \begin{subfigure}[ht]{0.45\linewidth}
            \includegraphics[trim = 0mm 130mm 0mm 0mm, clip,width=\linewidth]{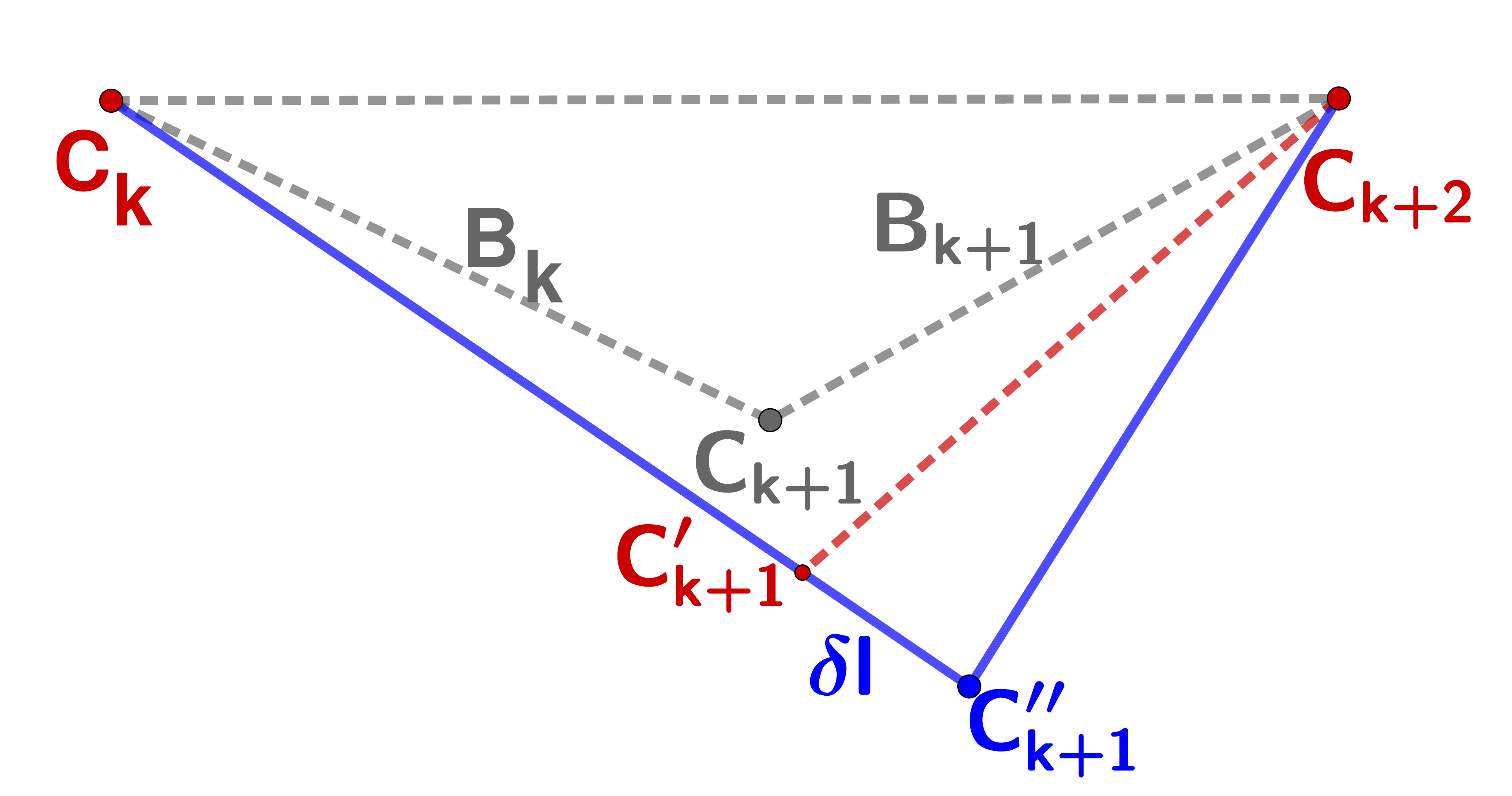}
            \caption{Refine length $l_k$ of $B_k$}
            \label{fig:2b_2}
        \end{subfigure}
    
        \begin{subfigure}[ht]{0.45\linewidth}
            \includegraphics[trim = 0mm 55mm 0mm 0mm, clip,width=\linewidth]{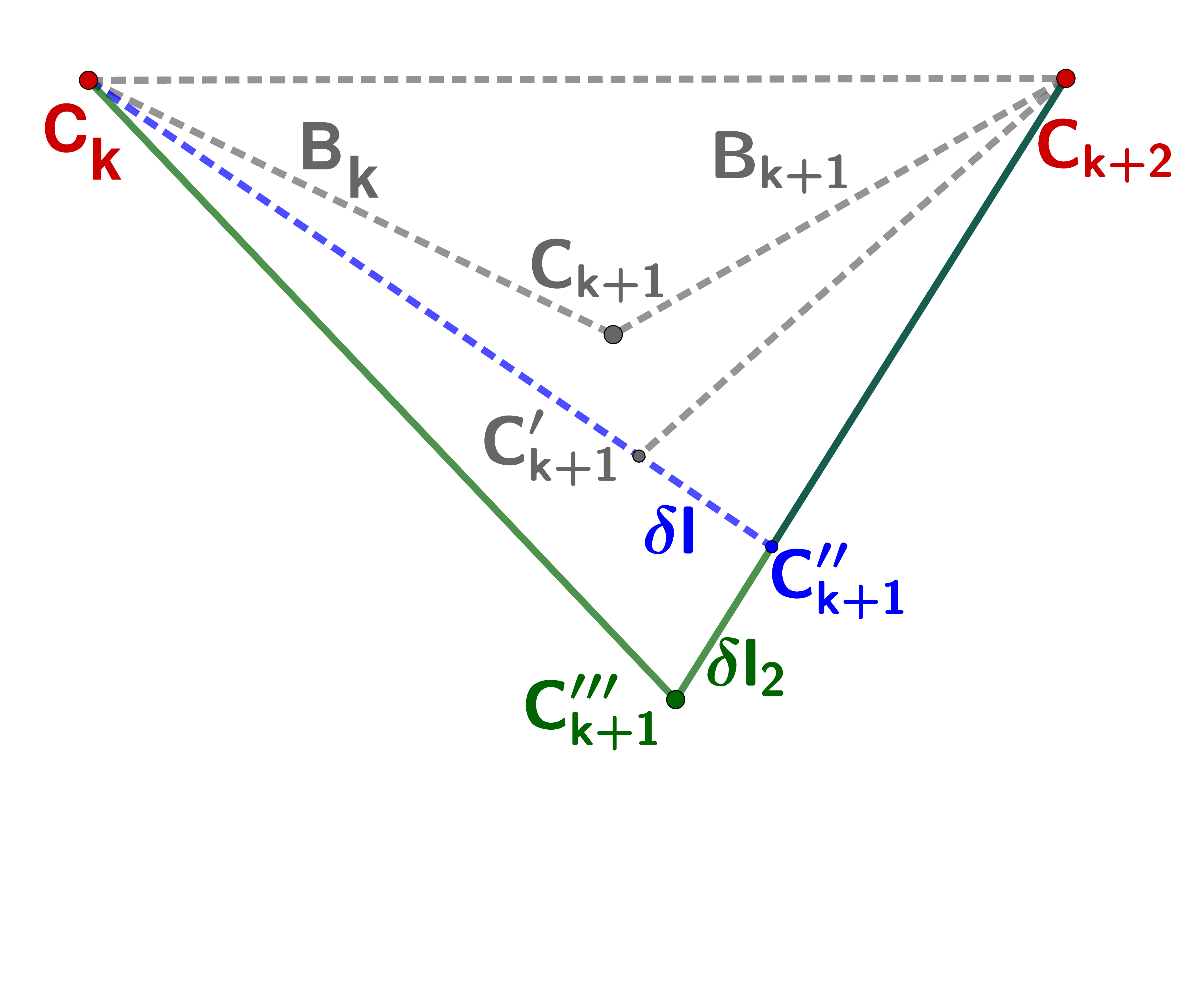}
            \caption{Refine length $l_{k+1}$ of $B_{k+1}$}
            \label{fig:2b_3}
        \end{subfigure}
         \begin{subfigure}[ht]{0.45\linewidth}
            \includegraphics[trim = 0mm 55mm 0mm 0mm, clip,width=\linewidth]{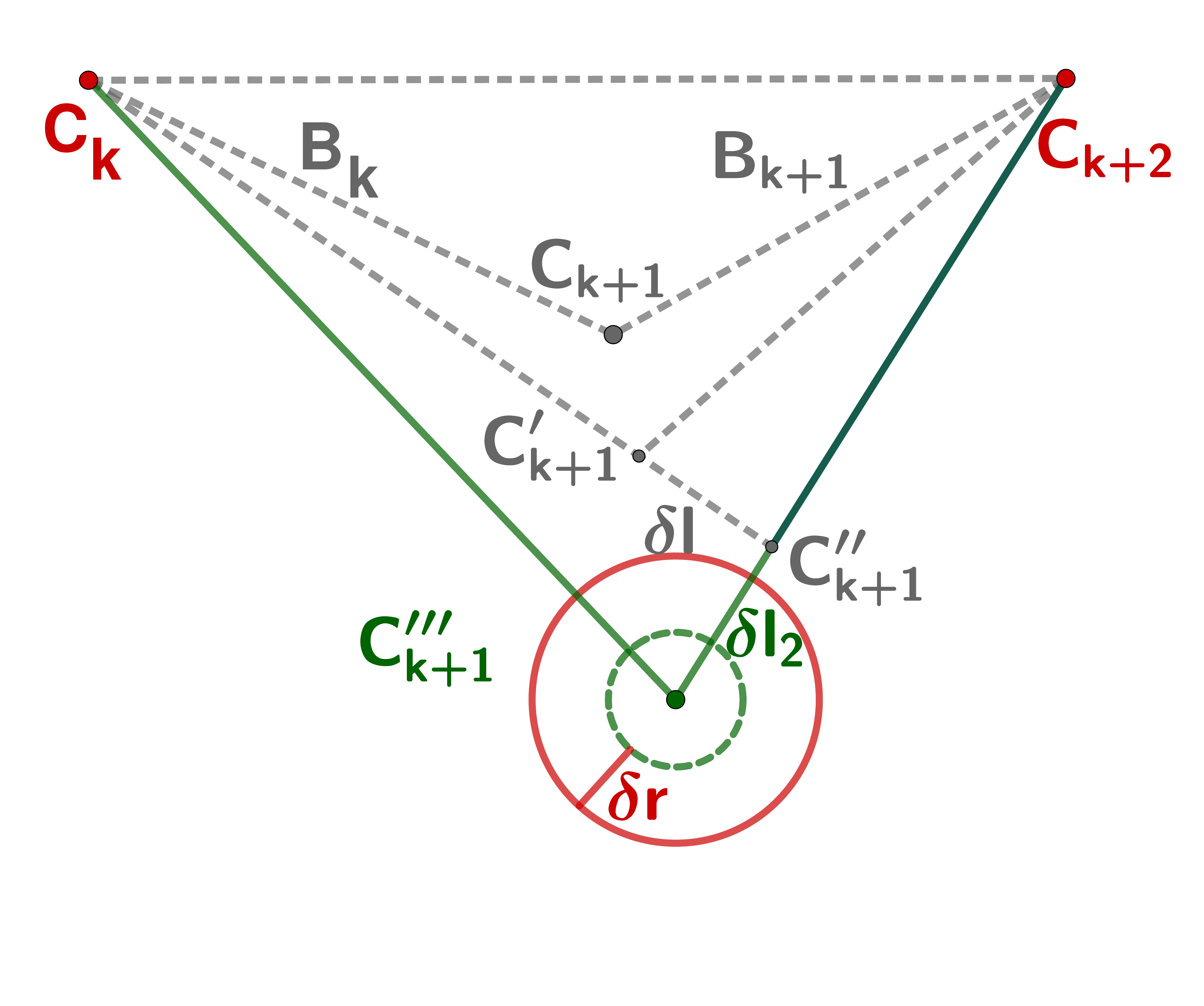}
            \caption{Refine radius $r_{k+1}$}
            \label{fig:2b_4}
        \end{subfigure}
    \caption{Pairwise Optimization. With fixed extremities $c_{k}$ and $c_{k+2}$, the pair of bones $B_k$ and $B_{k+1}$ is first rotated around axis $c_{k}c_{k+2}$. Then the lengths of the bones $B_k$ and $B_{k+1}$ and their common radius $r_{k+1}$ are optimized successively. After these updates, the point-to-bone assignment is recomputed. As the process is repeated the distances are more accurate since the point-to-bone assignment becomes more meaningful.}
    \label{fig:2bones}
\end{figure}

\paragraph*{Optimization for the joint between two consecutive bones.}
The optimization of the position and radius of the joint between two consecutive bones $(B_k,B_{k+1})$ is performed by optimizing a set of four parameters in a loop (an angle, two lengths and a radius) minimizing the two-bones energy. The two end-sphere centers being fixed ($c_k$ and $c_{k+2}$ in Figure \ref{fig:2bones}), we first compute the optimal rotation of the two bones around axis $c_kc_{k+2}$. We then optimize the bone lengths $\hat l_k = l_k + \delta l_k$ and $\hat l_{k+1} = l_{k+1} + \delta l_{k+1}$ and, finally, the radius of the common joint is computed as $\hat{r}_{k+1}=r_{k+1}+\delta r$. The parameters optimization alternates with a recomputation of point sets $P_k$ and $P_{k+1}$, which refines the point-to-bone assignment.
The optimization is also performed using the Levenberg-Marquardt algorithm (see the supplementary material for details).
  
\begin{figure}[ht]
        \centering
        \begin{subfigure}{0.155\textwidth}
            \includegraphics[width=\textwidth]{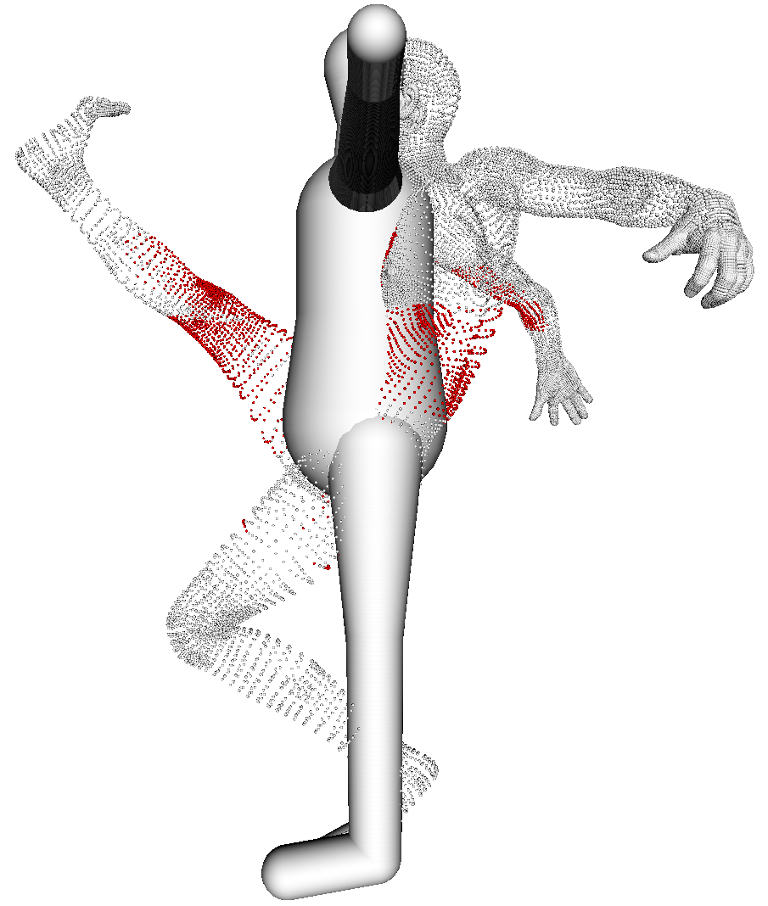}
            \caption{\scriptsize{Initialization}}
            \label{fig:for_0}
        \end{subfigure}
        \begin{subfigure}[ht]{0.155\textwidth}
            \includegraphics[width=\textwidth]{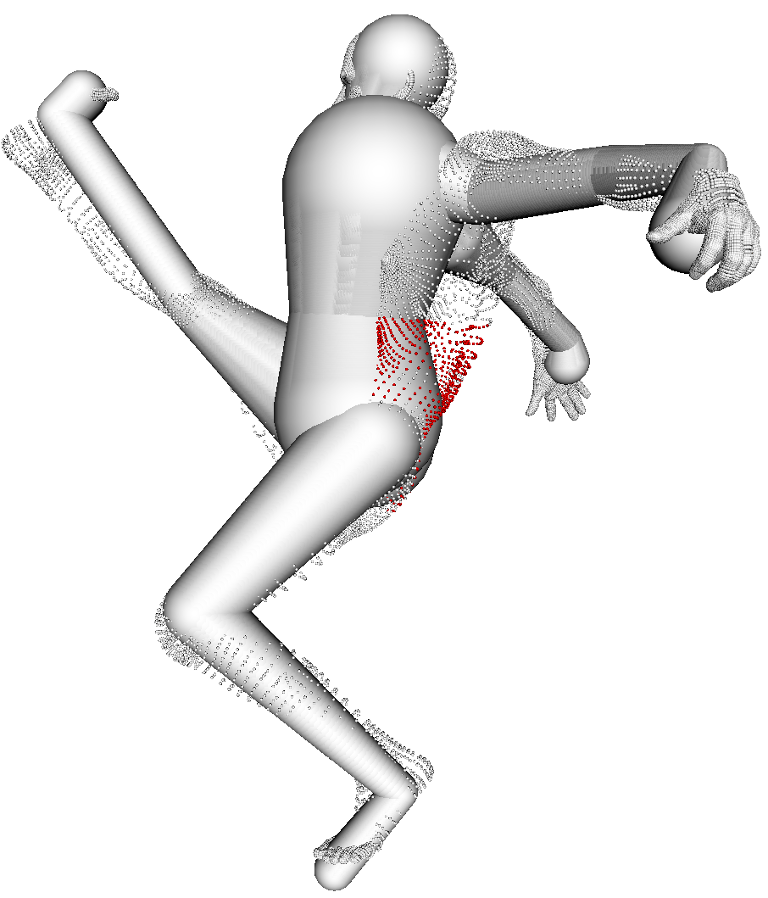}
            \caption{\scriptsize{First forward passage}}
            \label{fig:for_1}
        \end{subfigure}
        \begin{subfigure}{0.155\textwidth}
            \includegraphics[width=\textwidth]{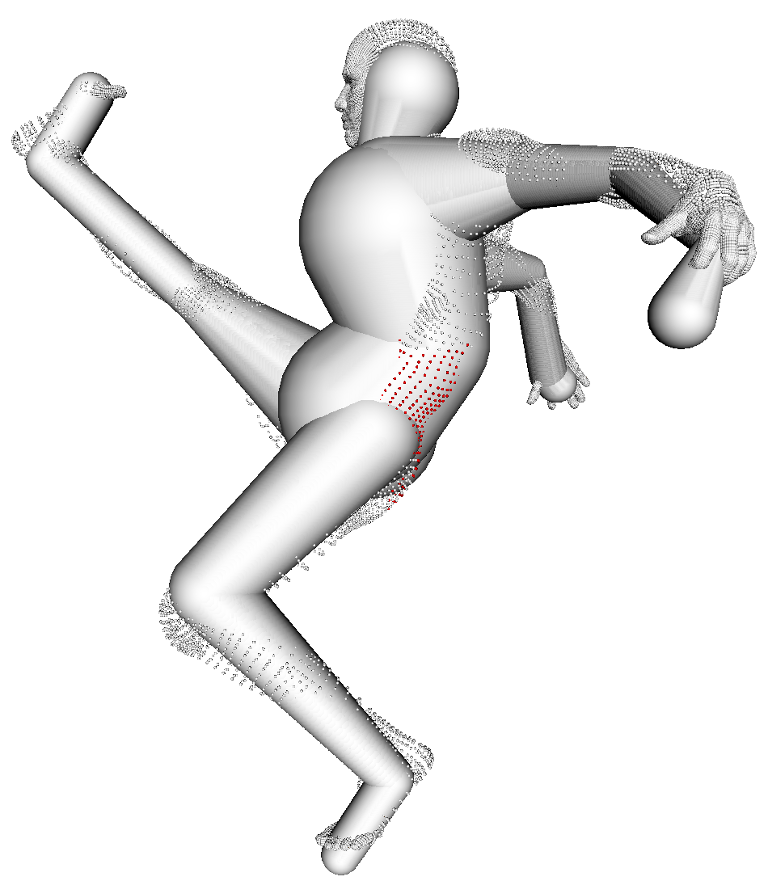}
            \caption{\scriptsize{First backward passage}}
            \label{fig:back_1}
        \end{subfigure}
    \caption{The assigned points for the first spine is showed in red. The spine is first registered with points that are not yet assigned to the legs, which distorts its position. Once the legs are registered, the registration backtracks to the spine and its position can be corrected. The final result is shown on Figure \ref{fig:exp_tosca}.}
    \label{fig:forward_backward}
\end{figure}

\paragraph*{Full Skeleton Registration} 
The full model corresponds to a tree whose branches are composed of chains. The process of registering each of the chains must be done in such an order that it gradually ensures the relevance of the data attached to each chain. Thus, previously registered chains can be questioned again if their attached points are reassigned to other chains or if they catch new points during the process (see Figure \ref{fig:forward_backward}).

Our system depends on the initial position of 
a joint chosen as the root 
of the skeleton. For example, for human or quadrupeds models, we assume that the pelvis part of our model is initialized near the corresponding part of the point set, which is done manually through a single point and click. Each chain is then registered in turn using FAKIR yielding a registered skeleton both in terms of intrinsic parameters and pose in only a couple of iterations.

For human models, the registration order is the following: first the spine chain is registered, refining the pelvis position and scale during the process, followed by each of the two leg chains and each of the two arm chains. When registering the arms and legs chains, the position for the joint attached to the spine or the pelvis remains fixed. 
However, after one arm is registered, changing the spine-arm joint position, the spine chain is updated accordingly (and similarly for legs and head),  thus the forward and backward chain registration extends to the whole model.
This leads to a calibrated and accurately positioned articulated model.

\section{Results}

In this section, we show the performance of FAKIR both on synthetic data and on point sets resulting from statue digitization. We developed our algorithm in C++, using OpenMP for computing point to bone distances in parallel. All experiments are run on an Intel Core i7-4790K CPU @ 4.00GHz. Normals were computed by using the state of the art approach of Hoppe et al \cite{Hoppe92}.

\subsection{Experiments on synthetic data}

We first tested our algorithm on synthetic data to provide a quantitative evaluation of the FAKIR performances. We considered a point set of $5k$ points sampled on a sphere-mesh of a 4-bone chain in a specific pose and registered a generic 4-bone chain to it. Although the point set and the initial chain are quite distant from each other, providing an approximate initial position of a single anchor point (one of the extremity) is enough to register accurately the chain.
The accuracy of the registration is evaluated as the average distance between the point set and the model.

\begin{figure}[ht]
    \centering
    \begin{subfigure}[ht]{0.235\textwidth}
        \includegraphics[width=\textwidth]{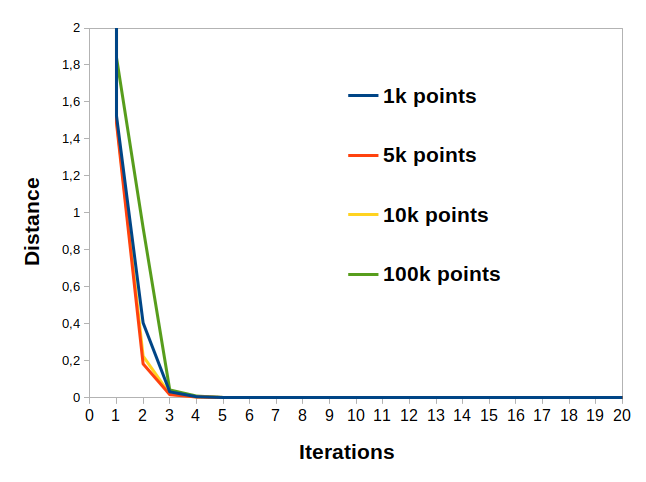}
    \end{subfigure}
    \begin{subfigure}[ht]{0.235\textwidth}
        \includegraphics[width=\textwidth]{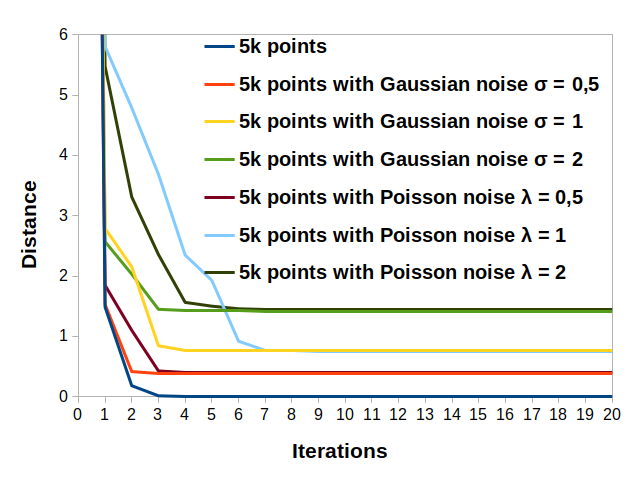}
    \end{subfigure}
    \caption{Evolution of the registration distance with the iterations for different number of points in the point set (left image), different levels of Gaussian noise and Poisson noise (right image).} 
    \label{fig:iteration_curve}
\end{figure}

\begin{figure}[ht]
     \begin{subfigure}[ht]{0.11\textwidth}
        \includegraphics[width=\textwidth]{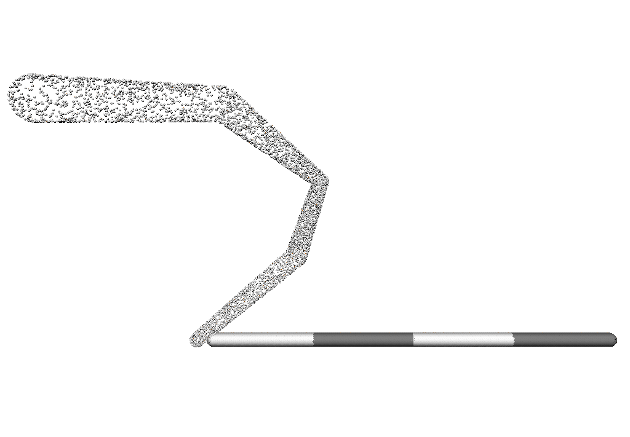}
    \end{subfigure}
    \begin{subfigure}[ht]{0.11\textwidth}
        \includegraphics[width=\textwidth]{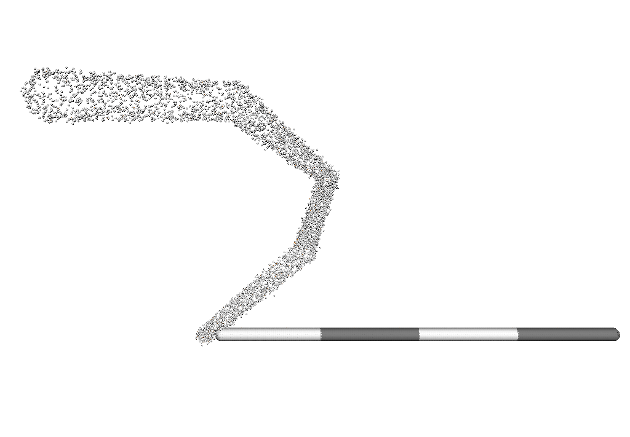}
    \end{subfigure}
    \begin{subfigure}[ht]{0.11\textwidth}
        \includegraphics[width=\textwidth]{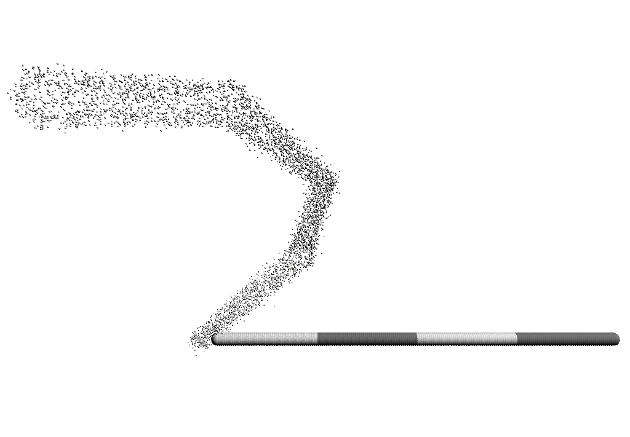}
    \end{subfigure}
    \begin{subfigure}[ht]{0.11\textwidth}
        \includegraphics[width=\textwidth]{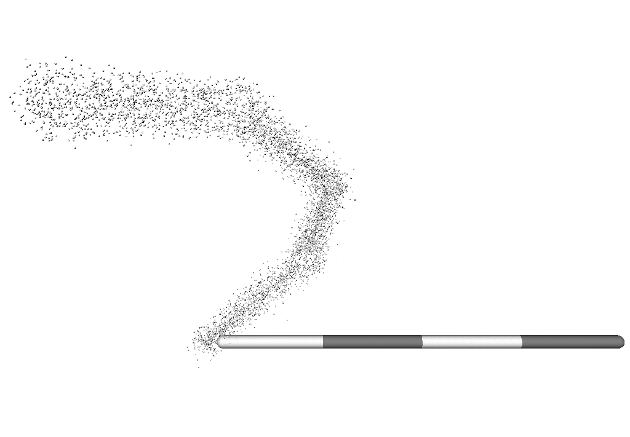}
    \end{subfigure}
    
     \begin{subfigure}[ht]{0.11\textwidth}
        \includegraphics[width=\textwidth]{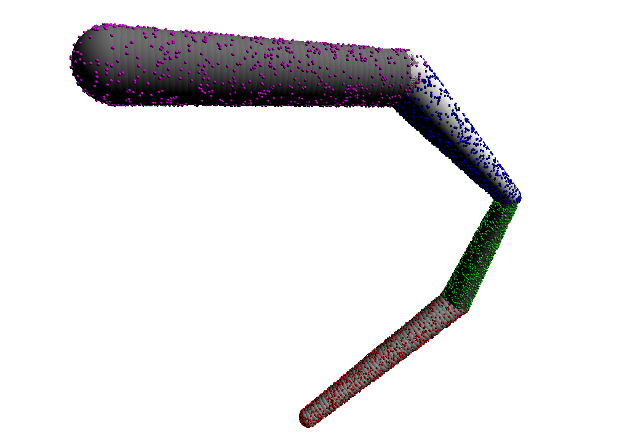}
    \end{subfigure}
    \begin{subfigure}[ht]{0.11\textwidth}
        \includegraphics[width=\textwidth]{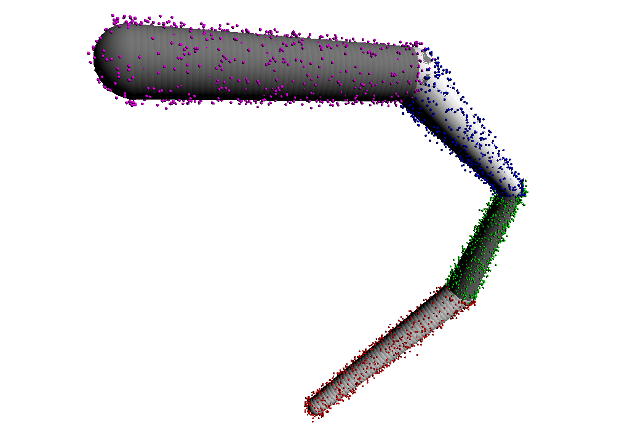}
    \end{subfigure}
    \begin{subfigure}[ht]{0.11\textwidth}
        \includegraphics[width=\textwidth]{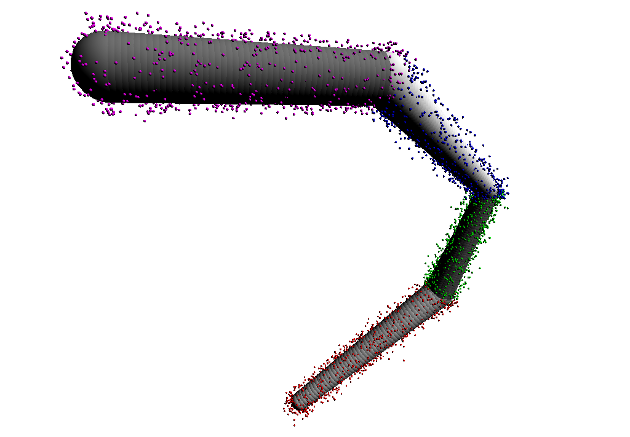}
    \end{subfigure}
    \begin{subfigure}[ht]{0.11\textwidth}
        \includegraphics[width=\textwidth]{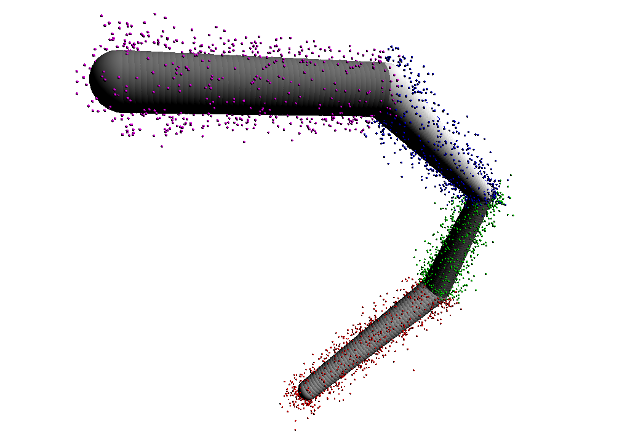}
    \end{subfigure}
    \caption{Evaluation of FAKIR with respect to increasing Gaussian noise after $20$ iterations.
    The first row shows the initial point set and the bottom row shows the registered bone chain. From left to right: without noise, $\sigma=0.5$, $\sigma=1$, and $\sigma = 2$. The total groundtruth model length is 140 (All values are given in length units).}
    \label{fig:synthetic_noise}
\end{figure}

\begin{figure}[ht]
     \begin{subfigure}[ht]{0.11\textwidth}
        \includegraphics[width=\textwidth]{bruit_i_0}
    \end{subfigure}
    \begin{subfigure}[ht]{0.11\textwidth}
        \includegraphics[width=\textwidth]{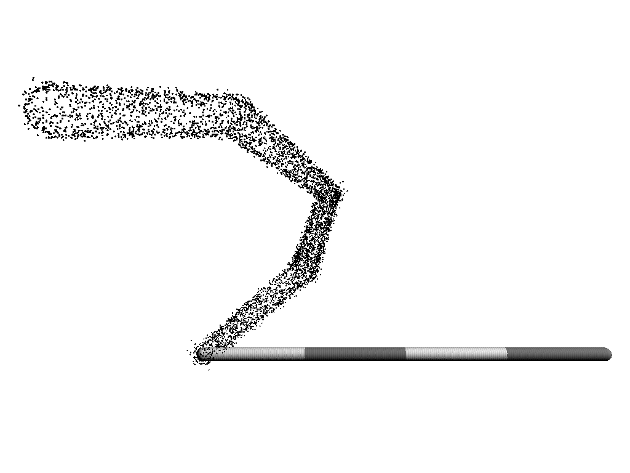}
    \end{subfigure}
    \begin{subfigure}[ht]{0.11\textwidth}
        \includegraphics[width=\textwidth]{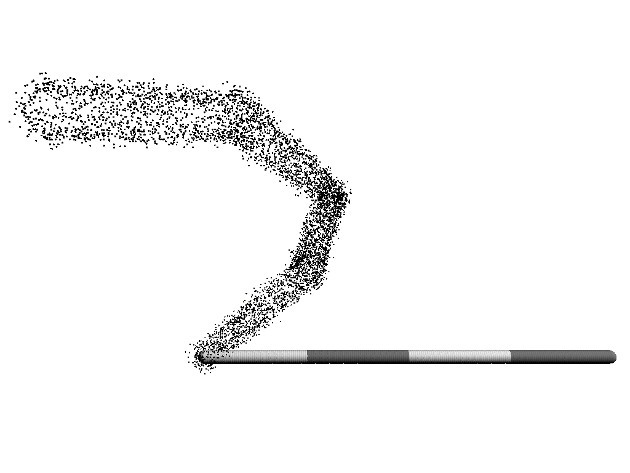}
    \end{subfigure}
    \begin{subfigure}[ht]{0.11\textwidth}
        \includegraphics[width=\textwidth]{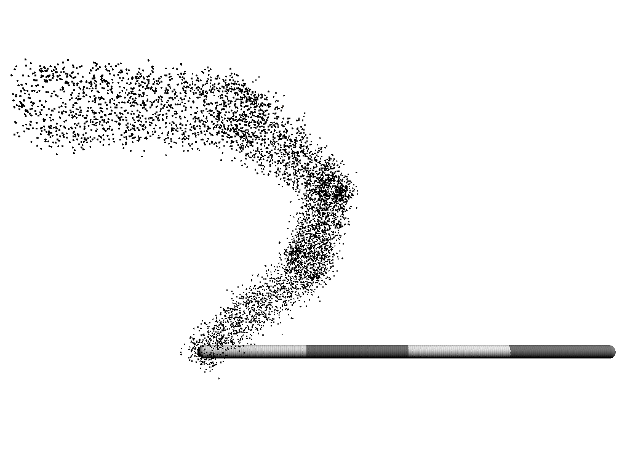}
    \end{subfigure}
    
     \begin{subfigure}[ht]{0.11\textwidth}
        \includegraphics[width=\textwidth]{bruit_r_0}
    \end{subfigure}
    \begin{subfigure}[ht]{0.11\textwidth}
        \includegraphics[width=\textwidth]{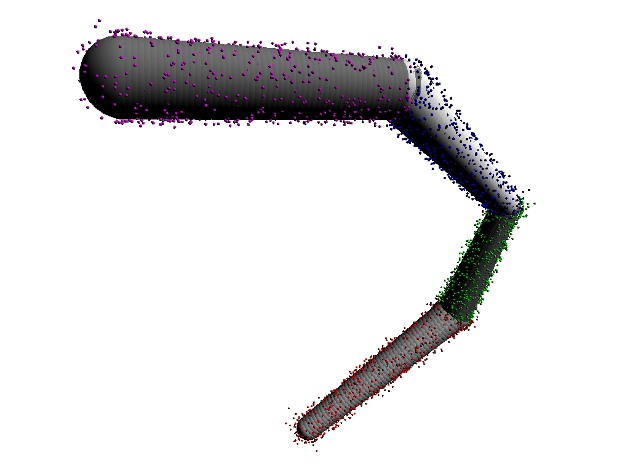}
    \end{subfigure}
    \begin{subfigure}[ht]{0.11\textwidth}
        \includegraphics[width=\textwidth]{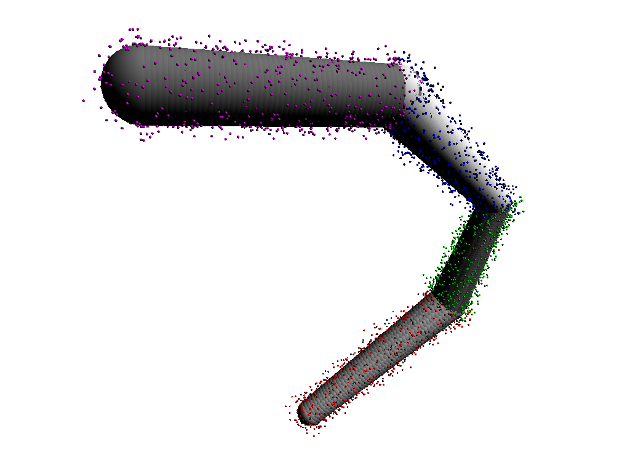}
    \end{subfigure}
    \begin{subfigure}[ht]{0.11\textwidth}
        \includegraphics[width=\textwidth]{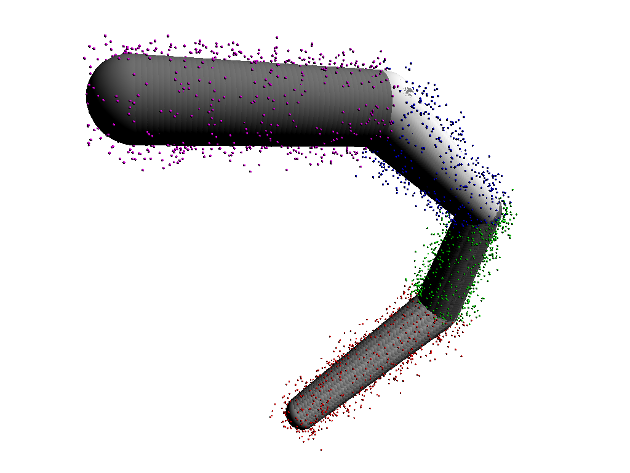}
    \end{subfigure}
        \caption{Evaluation of FAKIR with respect to increasing Poisson noise after $20$ iterations.
    The first row shows the initial point set and the bottom row shows the registered bone chain. From left to right: without noise, $\lambda=0.5$, $\lambda=1$, and $\lambda = 2$. The total groundtruth model length is 140 (All values are given in length units).}
    \label{fig:synthetic_noise_p}
\end{figure}

\begin{figure}[ht]
    \begin{subfigure}[ht]{0.11\textwidth}
        \includegraphics[width=\textwidth]{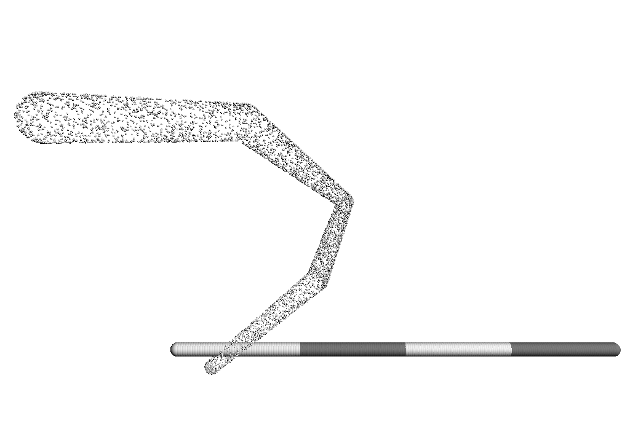}
    \end{subfigure}
    \begin{subfigure}[ht]{0.11\textwidth}
        \includegraphics[width=\textwidth]{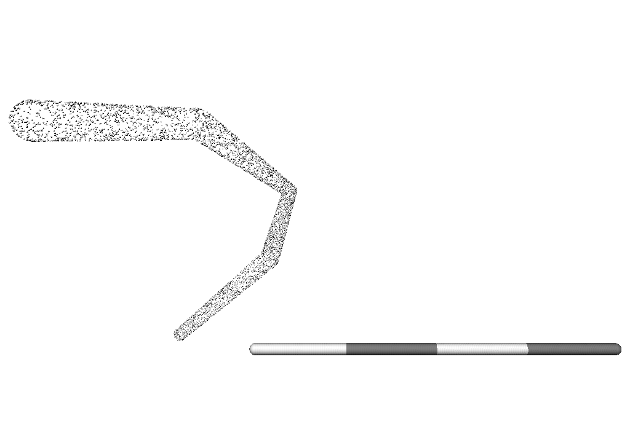}
    \end{subfigure}
    \begin{subfigure}[ht]{0.11\textwidth}
        \includegraphics[width=\textwidth]{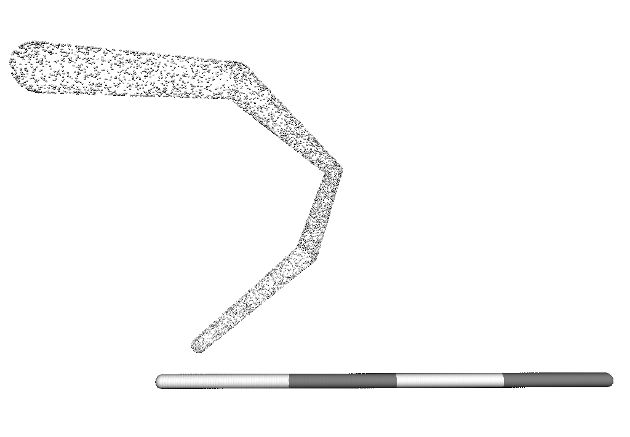}
    \end{subfigure}
    \begin{subfigure}[ht]{0.11\textwidth}
        \includegraphics[width=\textwidth]{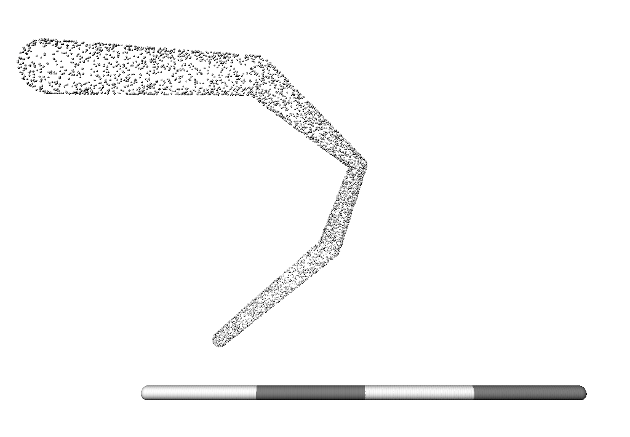}
    \end{subfigure}
    
     \begin{subfigure}[ht]{0.11\textwidth}
        \includegraphics[width=\textwidth]{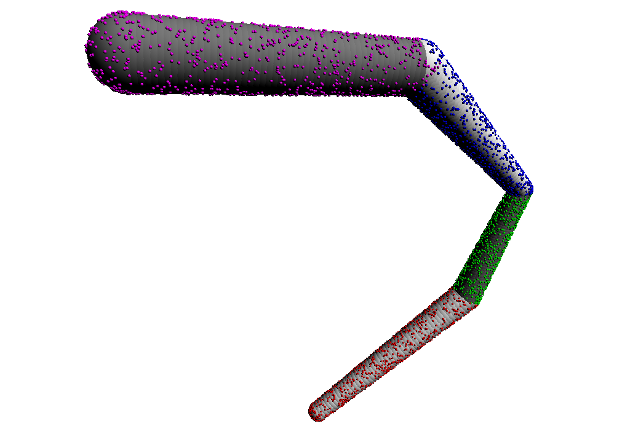}
    \end{subfigure}
    \begin{subfigure}[ht]{0.11\textwidth}
        \includegraphics[width=\textwidth]{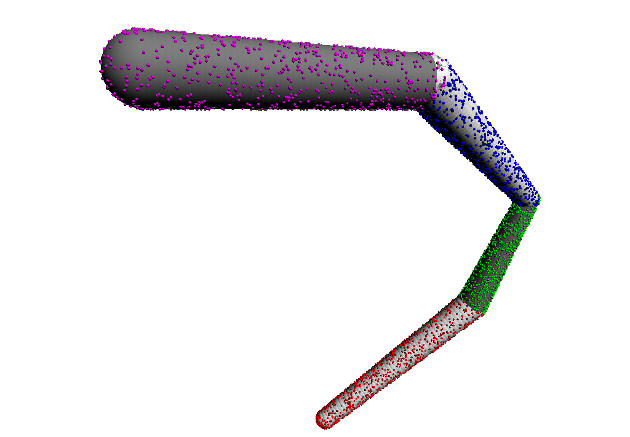}
    \end{subfigure}
    \begin{subfigure}[ht]{0.11\textwidth}
        \includegraphics[width=\textwidth]{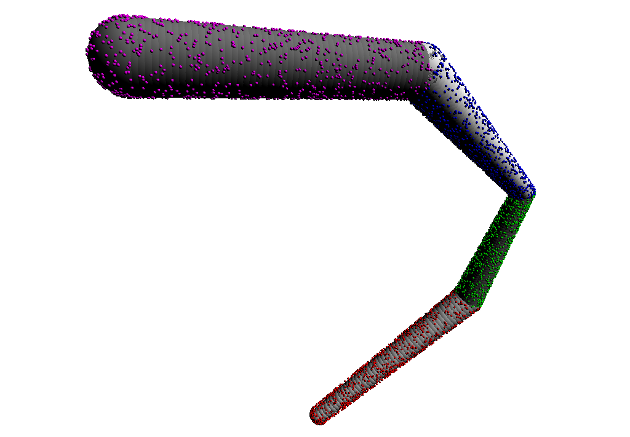}
    \end{subfigure}
    \begin{subfigure}[ht]{0.11\textwidth}
        \includegraphics[width=\textwidth]{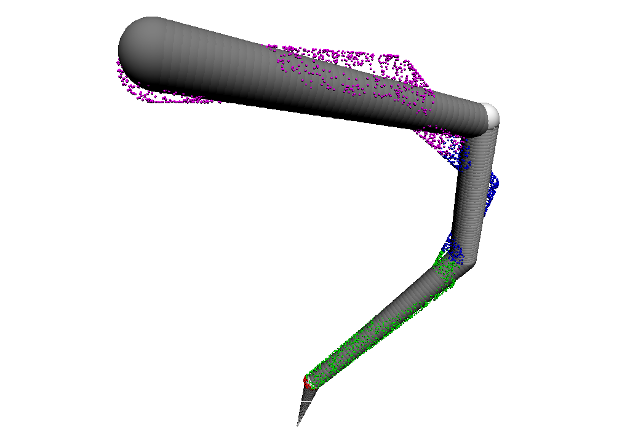}
    \end{subfigure}
        \caption{Evaluation of FAKIR with respect to a bad initial anchor point position after $20$ iterations. The first row shows the initial point set and the bottom row shows the registered bone chain. The last column shows that due to a bad initialization, the points (plotted in red) that are affected to the first bone do not bring enough information for the one-bone energy to move the chain extremity. Then, not enough bones remain to approximate the whole point set.} 
    \label{fig:synthetic_anchor}
\end{figure}

\begin{figure}[ht]
    \begin{subfigure}[ht]{0.11\textwidth}
        \includegraphics[width=\textwidth]{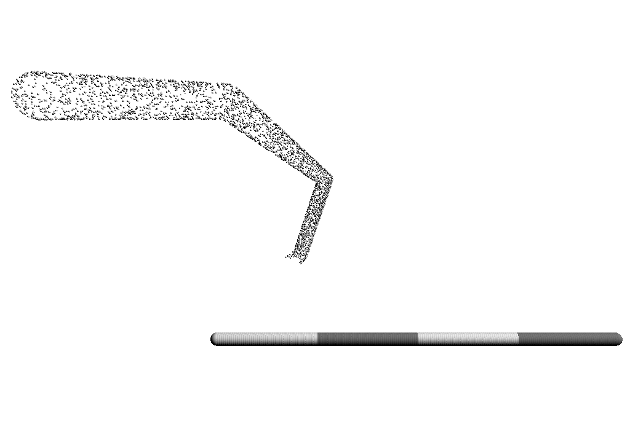}
    \end{subfigure}
    \begin{subfigure}[ht]{0.11\textwidth}
        \includegraphics[width=\textwidth]{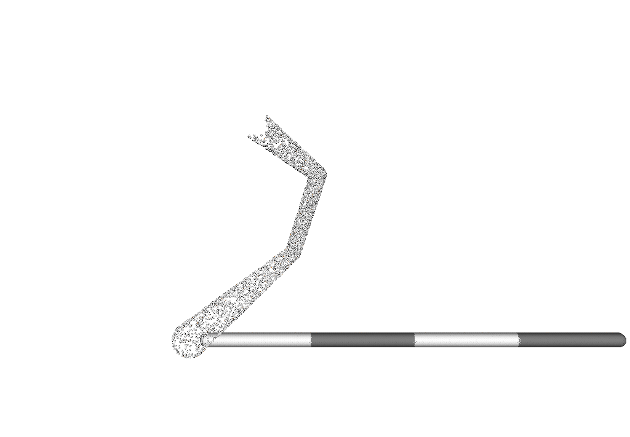}
    \end{subfigure}
    \begin{subfigure}[ht]{0.11\textwidth}
        \includegraphics[width=\textwidth]{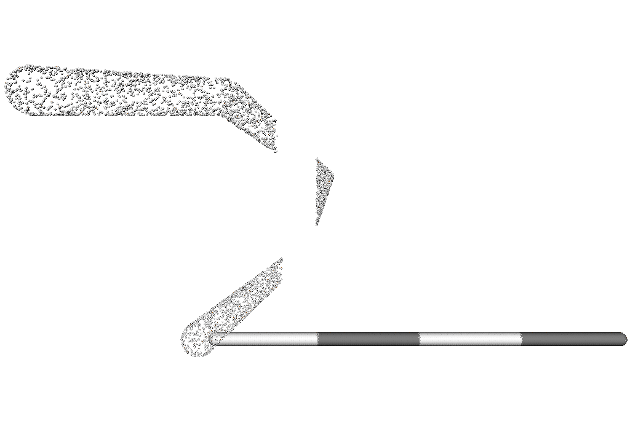}
    \end{subfigure}
    \begin{subfigure}[ht]{0.11\textwidth}
        \includegraphics[width=\textwidth]{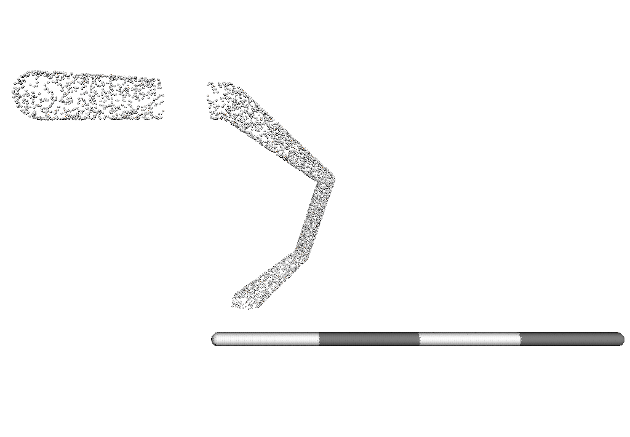}
    \end{subfigure}
    
     \begin{subfigure}[ht]{0.11\textwidth}
        \includegraphics[width=\textwidth]{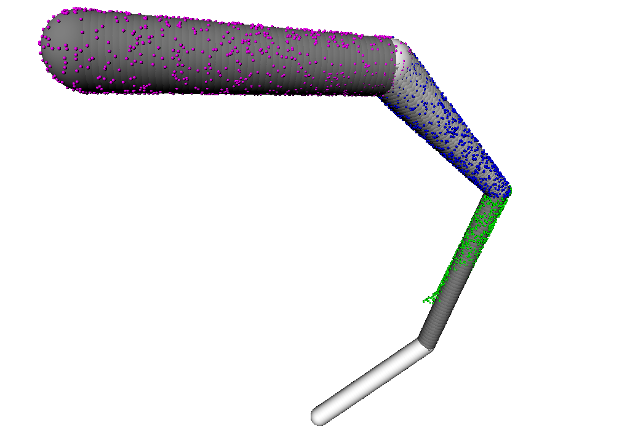}
    \end{subfigure}
    \begin{subfigure}[ht]{0.11\textwidth}
        \includegraphics[width=\textwidth]{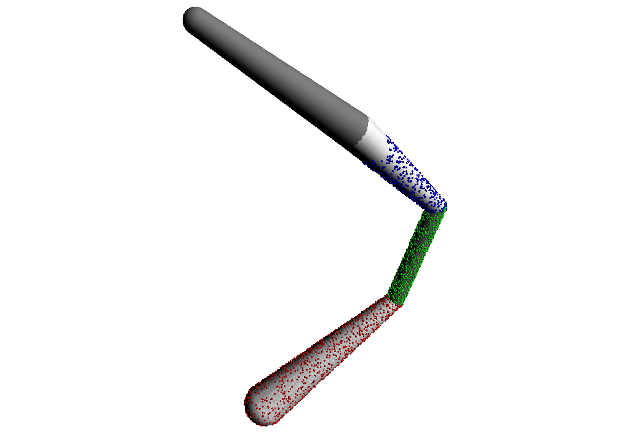}
    \end{subfigure}
    \begin{subfigure}[ht]{0.11\textwidth}
        \includegraphics[width=\textwidth]{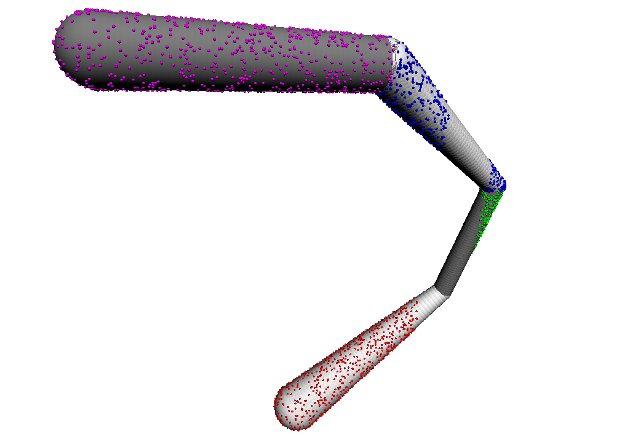}
    \end{subfigure}
    \begin{subfigure}[ht]{0.11\textwidth}
        \includegraphics[width=\textwidth]{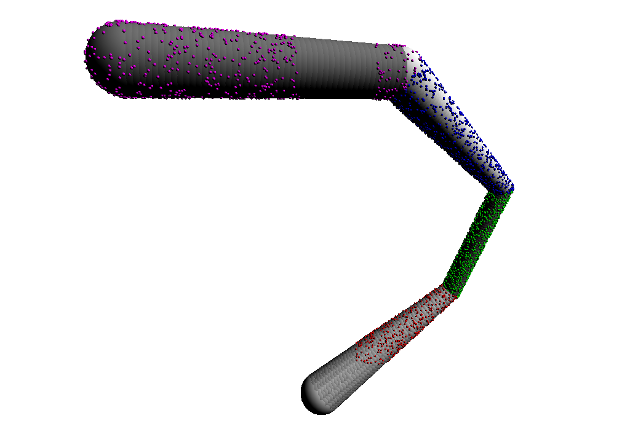}
    \end{subfigure}
    \caption{Evaluation of the FAKIR algorithm with respect to missing data after $20$ iterations. The first row shows the initial point set and the bottom row shows the registered bone chain.}
    \label{fig:synthetic_holes}
\end{figure}

In the noiseless case, our algorithm takes $7$ iterations to converge to a $0$ distance in $2.37s$, including $0.62s$ for the first forward pass.
The distance of the point set to the model with respect to the iterations for larger point sets and increasing noise is shown on Figure \ref{fig:iteration_curve}: the number of points has only a moderate impact on the number of iterations needed to converge (around $7$). When there is noise in the data, the distance also converges in a few iterations independently of the noise, however the distance at convergence is directly correlated to the variance of the noise. As shown by our experiments, FAKIR is rather resilient to even relatively high levels of noise (Gaussian noise in Figure \ref{fig:synthetic_noise} and Poisson noise in Figure \ref{fig:synthetic_noise_p}).
Figure \ref{fig:synthetic_anchor} shows how FAKIR handles an initial position of the anchor point that is not in the vicinity of its optimal position in the point set. FAKIR can handle initial positions that are moderately far from the true position, but in some cases (last column),
the backward optimization of the one-bone energy alone fails to reduce the length of the first bone and the radius of its free extremity degenerates to $0$ instead. This is due to the fact that no point is projected on the spherical free extremity of that bone.
This problem could be avoided by adding a bone occupancy term to the one-bone energy. A preferential alternative would be to modify the one-bone energy of the first and last bones by adding a term corresponding to the distance of the free caps to the data points. However, if the initial point is reasonably close to its true position, this problem does not occur.
FAKIR is also rather robust to missing data thanks to the iterated forward and backward passes (Figure \ref{fig:synthetic_holes}). Naturally when the missing parts are on the first or last bone or when a full bone is missing, the algorithm cannot predict the right length or angle.

\subsection{Skeleton registration results on statues}
We selected some interesting statues from various sources.
\begin{enumerate}
\item Dancer with Crotales, Louvre Museum
\item Dancing Faun, Pompei excavations
\item Aphrodite, Museum of Thorvaldsens
\item Old man walking, Nye Carlsberg Glyptotek
\item Esquiline Venus, Capitoline Museum
\item Old Fisherman Vatican-Louvre, Louvre Museum
\item Venus de Milo: Louvre Museum
\item Mermaid, Royal Bibliotek of Copenhagen
\end{enumerate}

While the 'Dancer with crotales' is a raw point set. The other 7 models are point sets sampled on meshes extracted from the Sketchfab website.
We also use several models from the TOSCA dataset~\cite{Bronstein08}, including nonhuman models to show the pliability of our method.

\begin{figure}[ht]
      \begin{subfigure}[ht]{0.31\linewidth}
      \includegraphics[width=\textwidth]{r_dan_o}
      \end{subfigure}
      \begin{subfigure}[ht]{0.31\linewidth}
          \includegraphics[width=\textwidth]{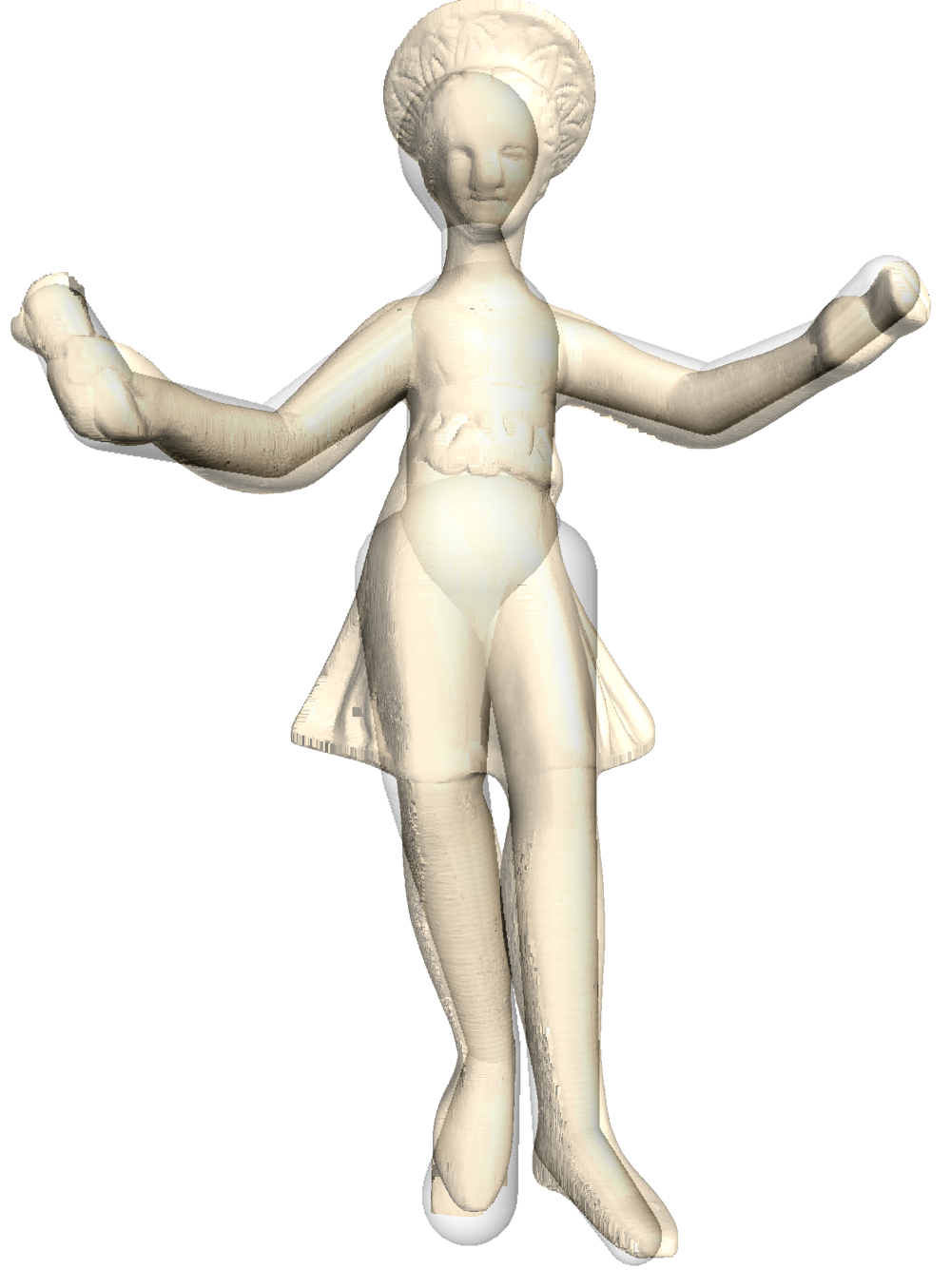}
      \end{subfigure}
      \begin{subfigure}[ht]{0.31\linewidth}
          \includegraphics[width=\textwidth]{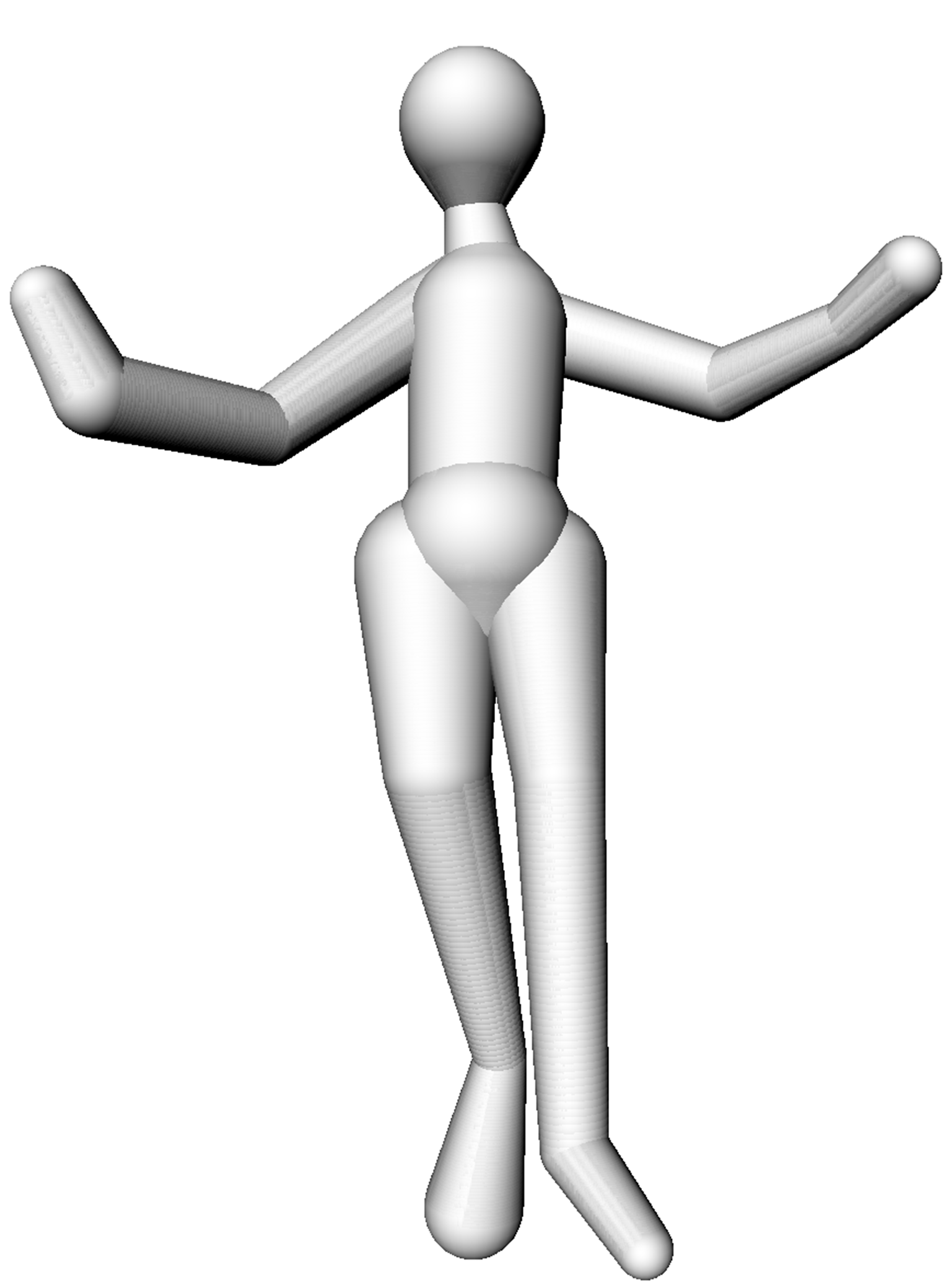}
      \end{subfigure}

      \begin{subfigure}[ht]{0.31\linewidth}
          \includegraphics[width=\textwidth]{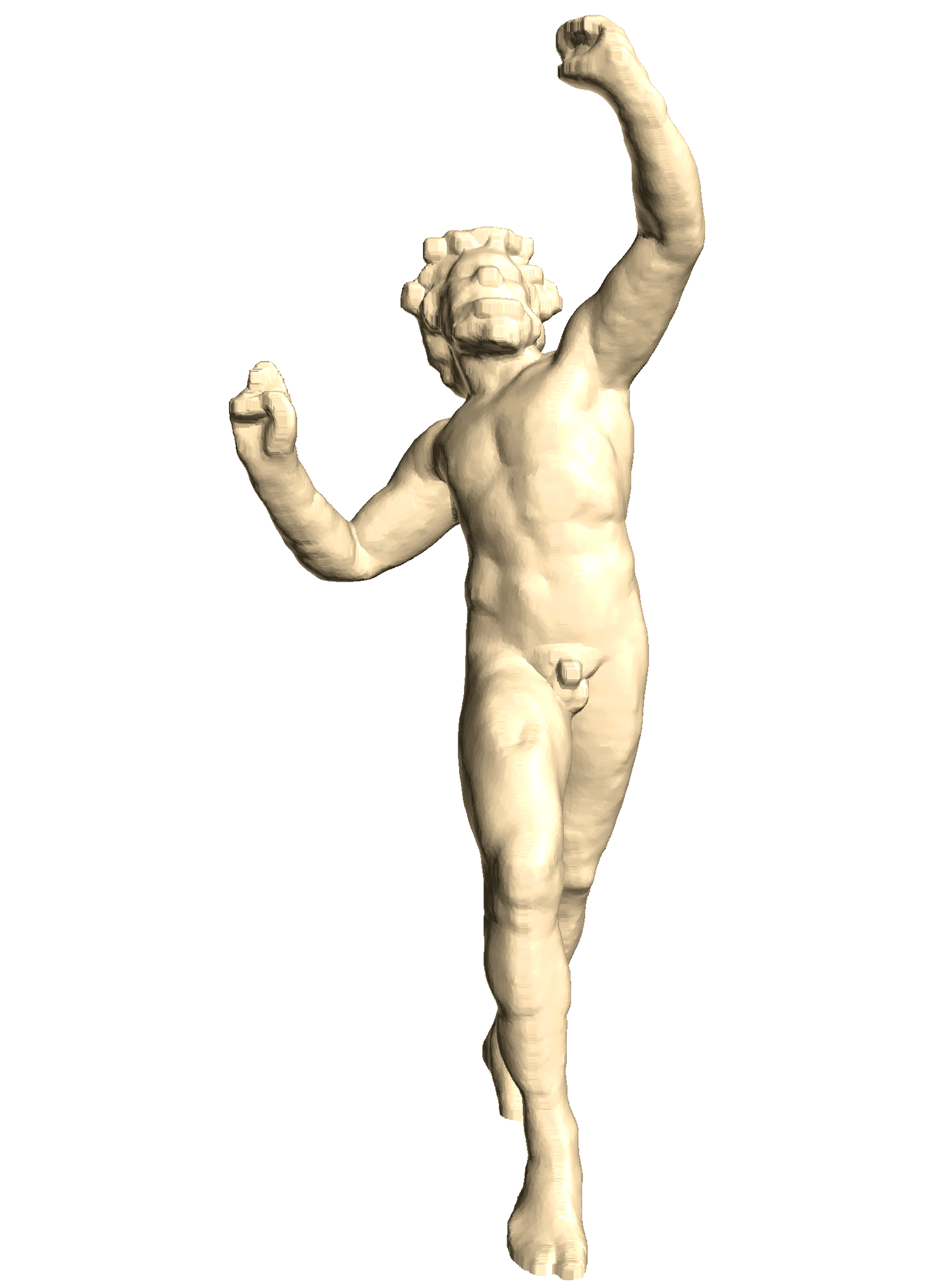}
      \end{subfigure}
      \begin{subfigure}[ht]{0.31\linewidth}
          \includegraphics[width=\textwidth]{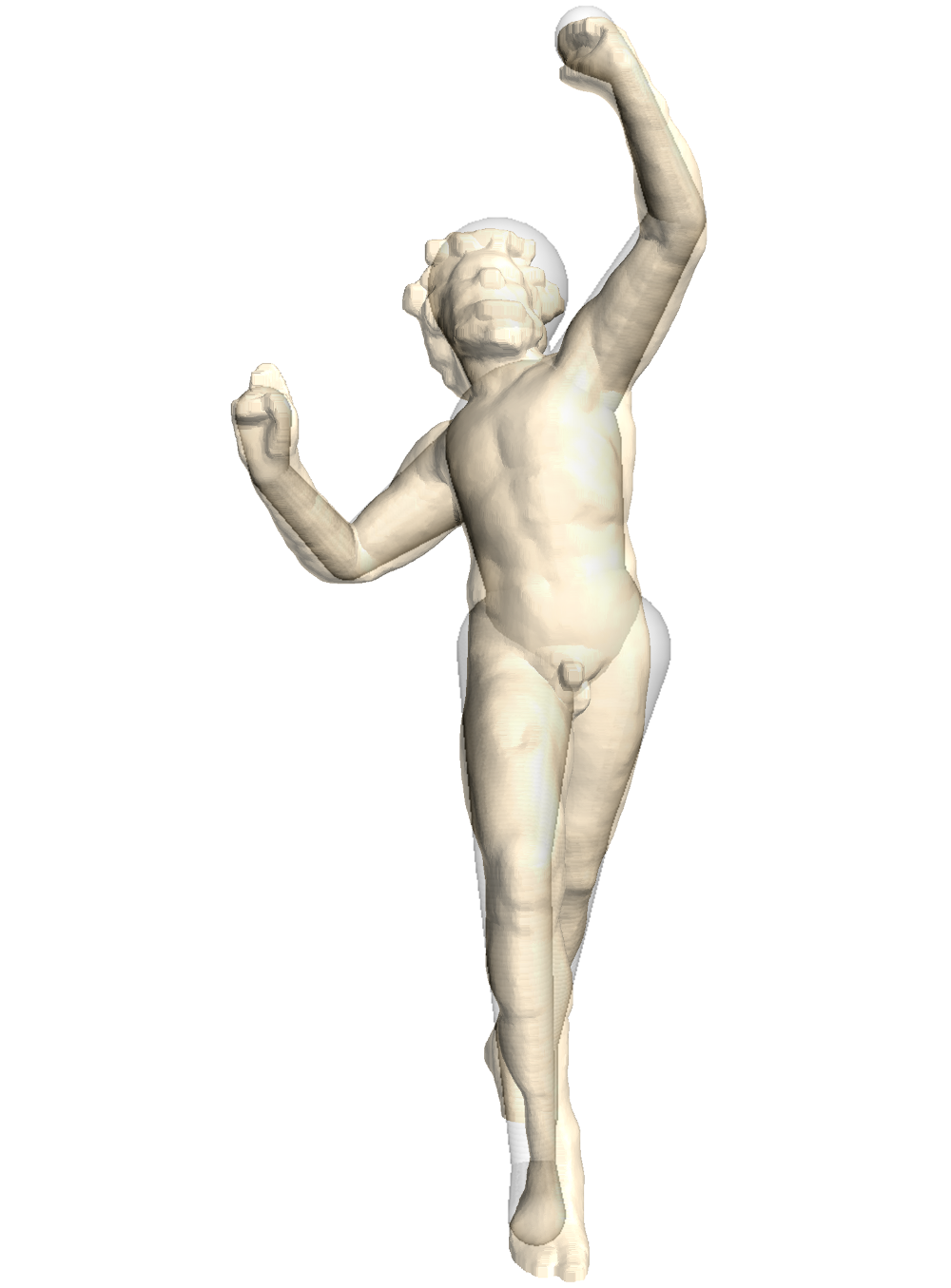}
      \end{subfigure}
      \begin{subfigure}[ht]{0.31\linewidth}
          \includegraphics[width=\textwidth]{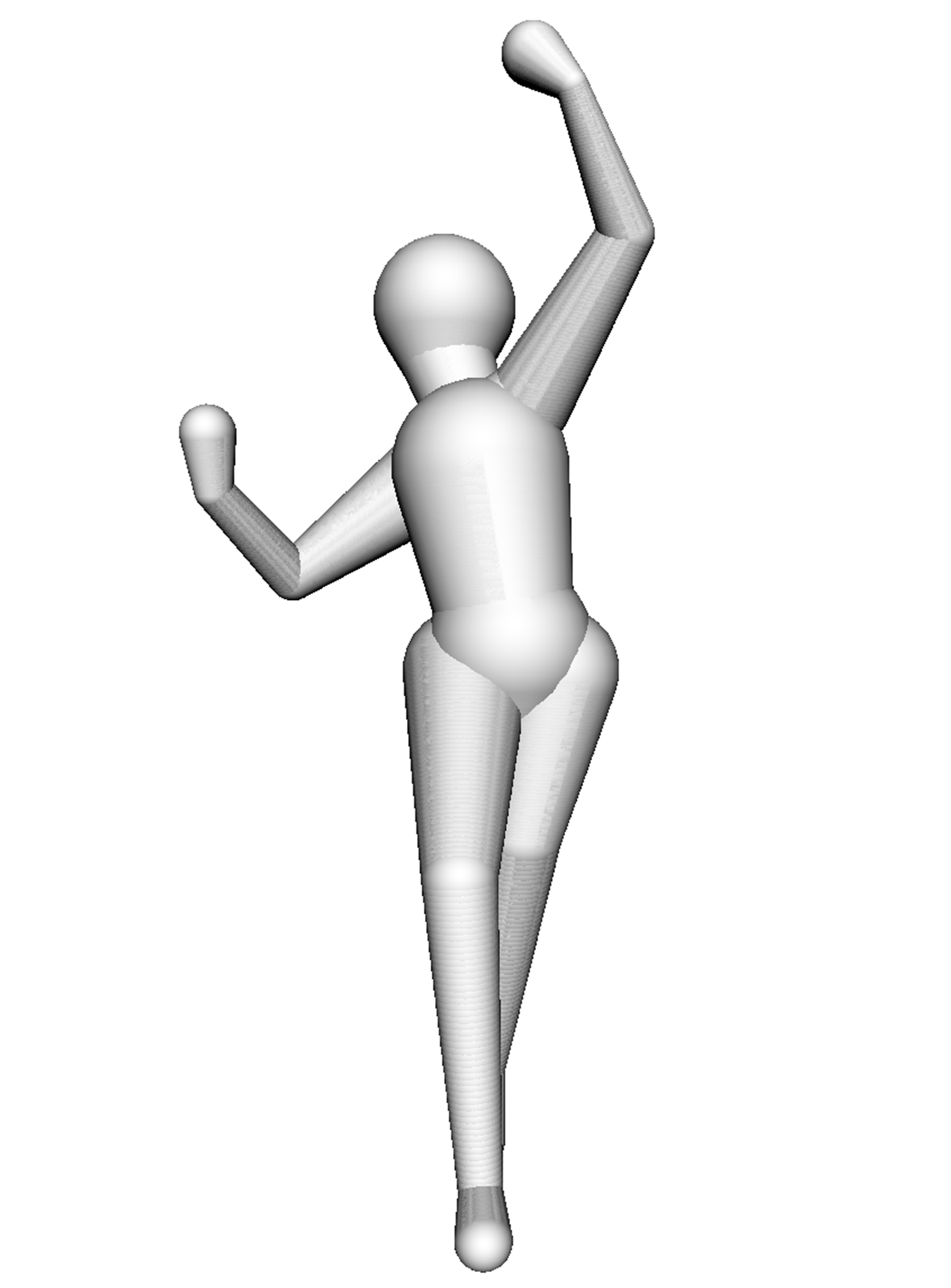}
      \end{subfigure}

      \begin{subfigure}[ht]{0.31\linewidth}
          \includegraphics[width=\textwidth]{r_aph_o}
      \end{subfigure}
      \begin{subfigure}[ht]{0.31\linewidth}
          \includegraphics[width=\textwidth]{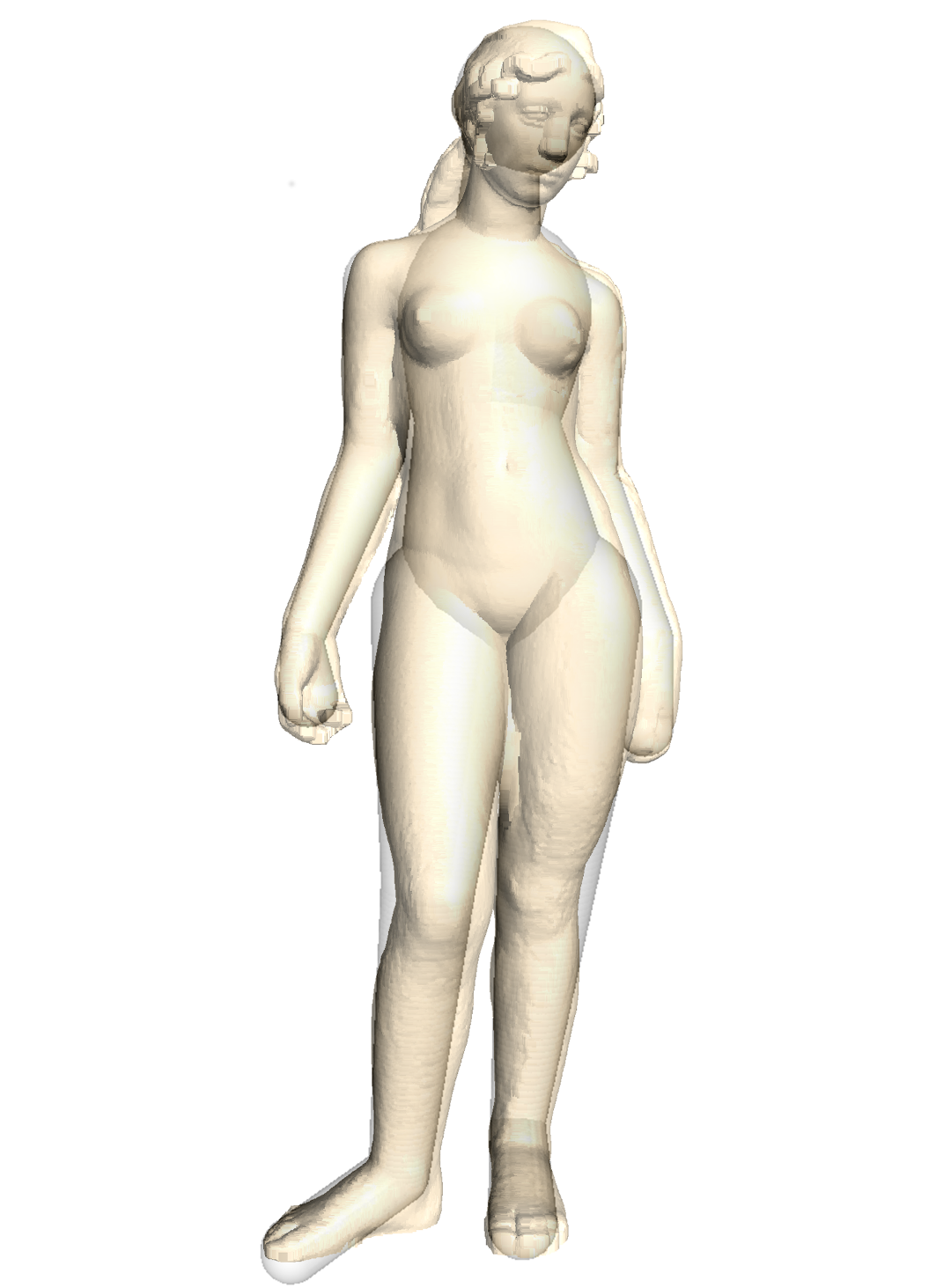}
      \end{subfigure}
      \begin{subfigure}[ht]{0.31\linewidth}
          \includegraphics[width=\textwidth]{r_aph_m}
      \end{subfigure}

      \begin{subfigure}[ht]{0.31\linewidth}
          \includegraphics[ width=\textwidth]{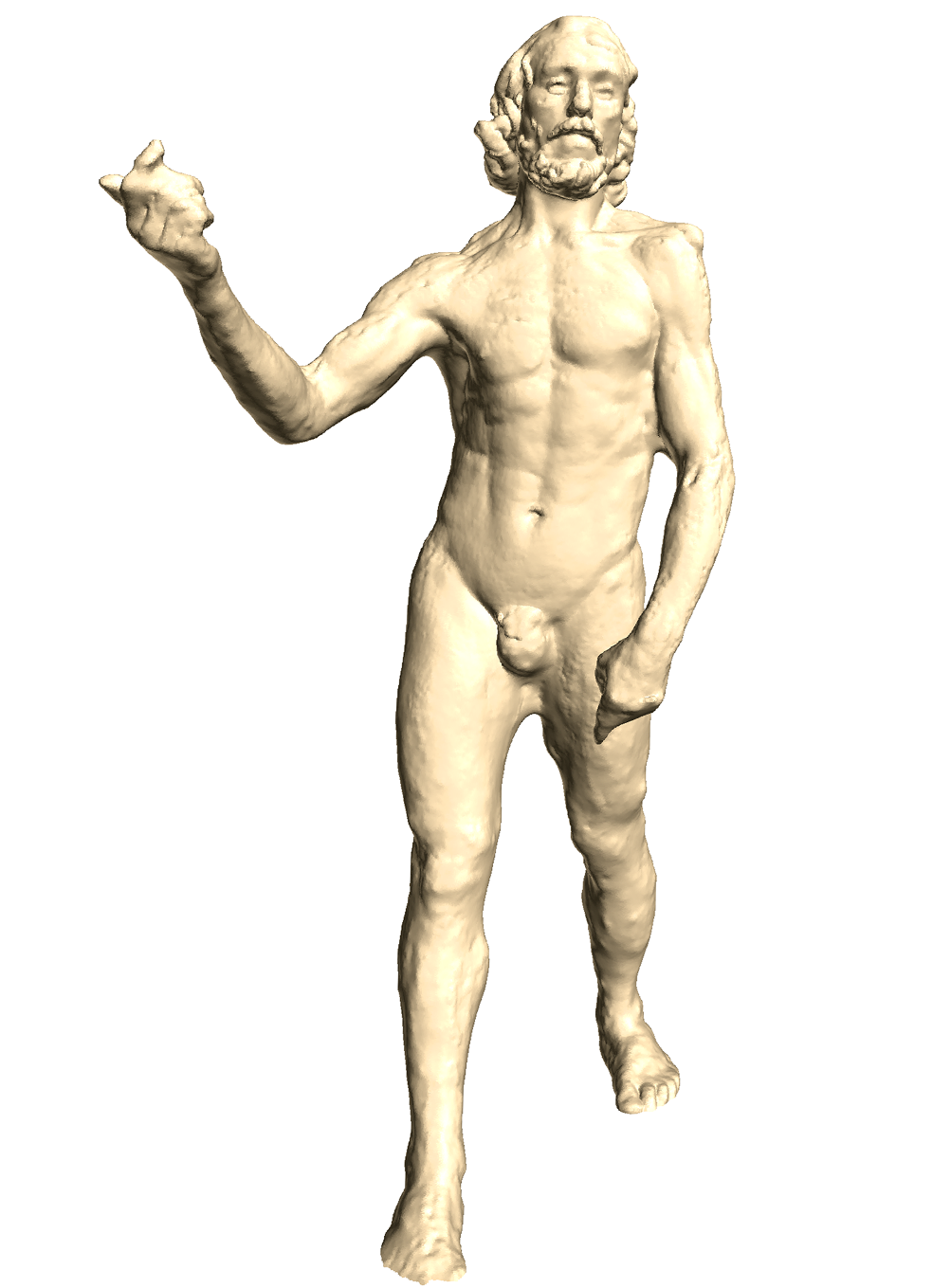}
      \end{subfigure}
      \begin{subfigure}[ht]{0.31\linewidth}
          \includegraphics[width=\textwidth]{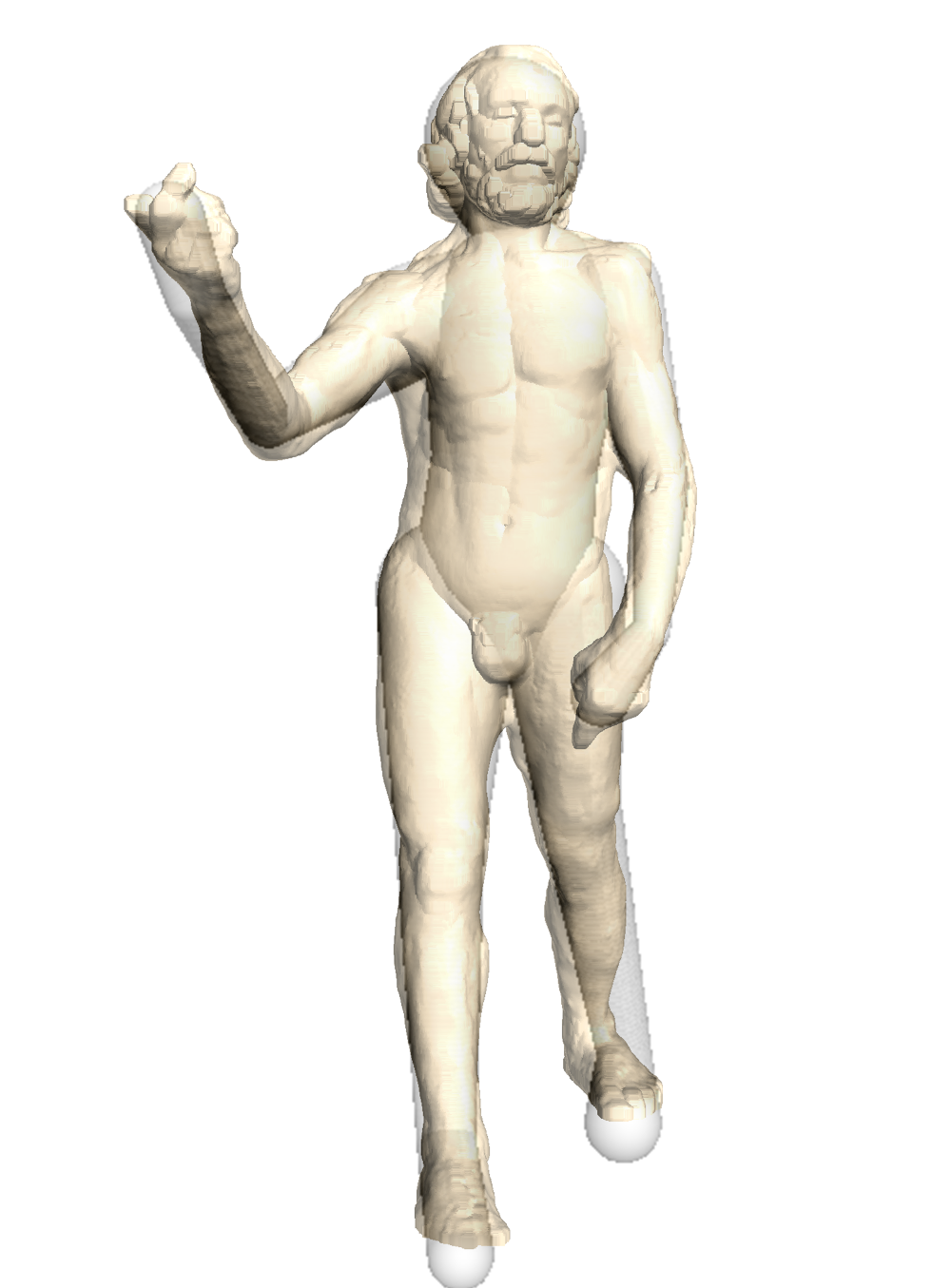}
      \end{subfigure}
      \begin{subfigure}[ht]{0.31\linewidth}
          \includegraphics[width=\textwidth]{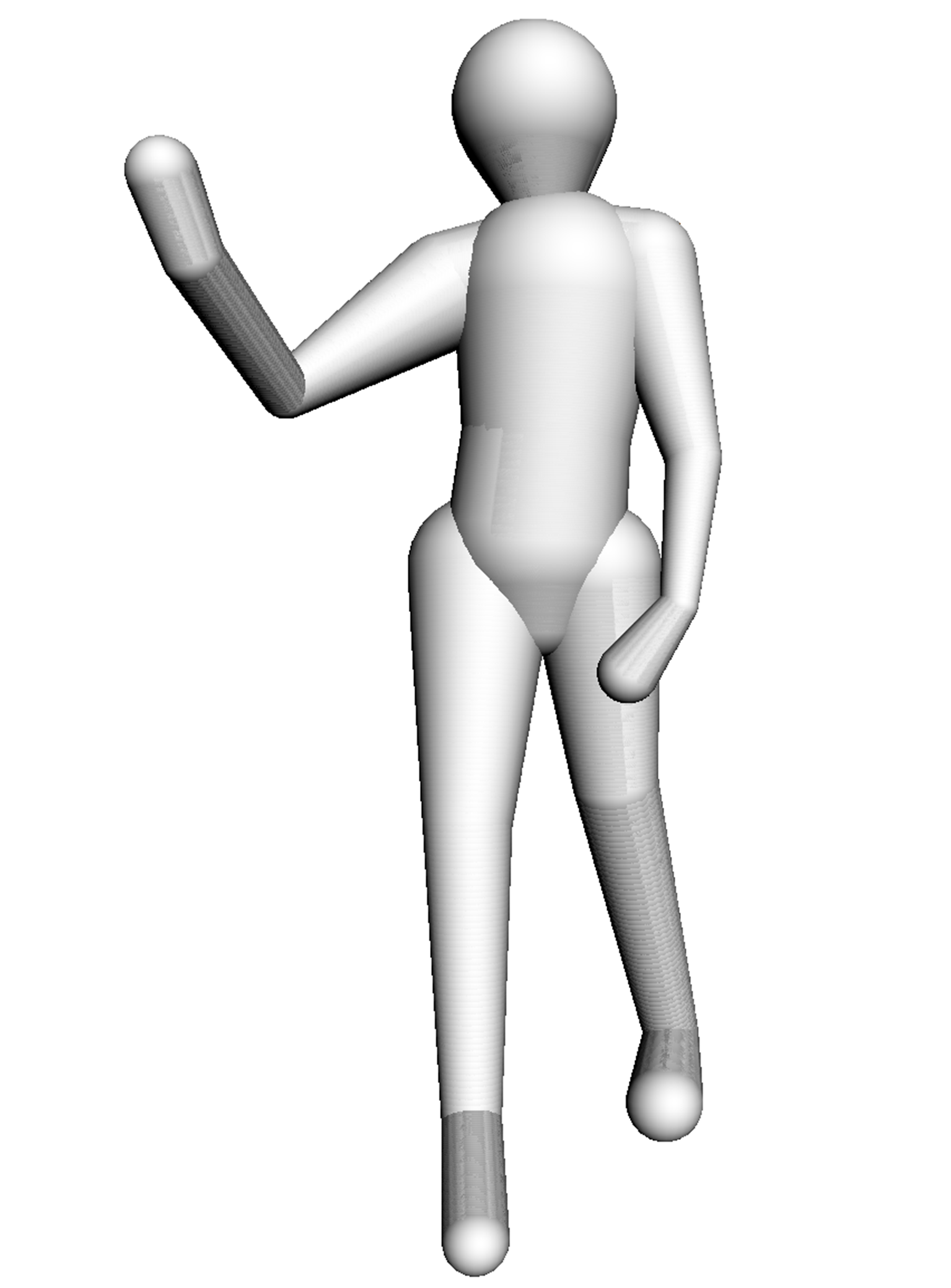}
      \end{subfigure}

  \caption{Registration of 4 statues: the Dancer with Crotales (first row), the Dancing Faun (second row), Aphrodite (third row) and the Old Man Walking (fourth row). First column: initial point set, second column: overlay of the registered model and the point cloud, third column: registered model.}
  \label{fig:result}
\end{figure}

\begin{figure}[ht]
    \centering
        \begin{subfigure}[ht]{0.32\linewidth}
            \includegraphics[width=\textwidth]{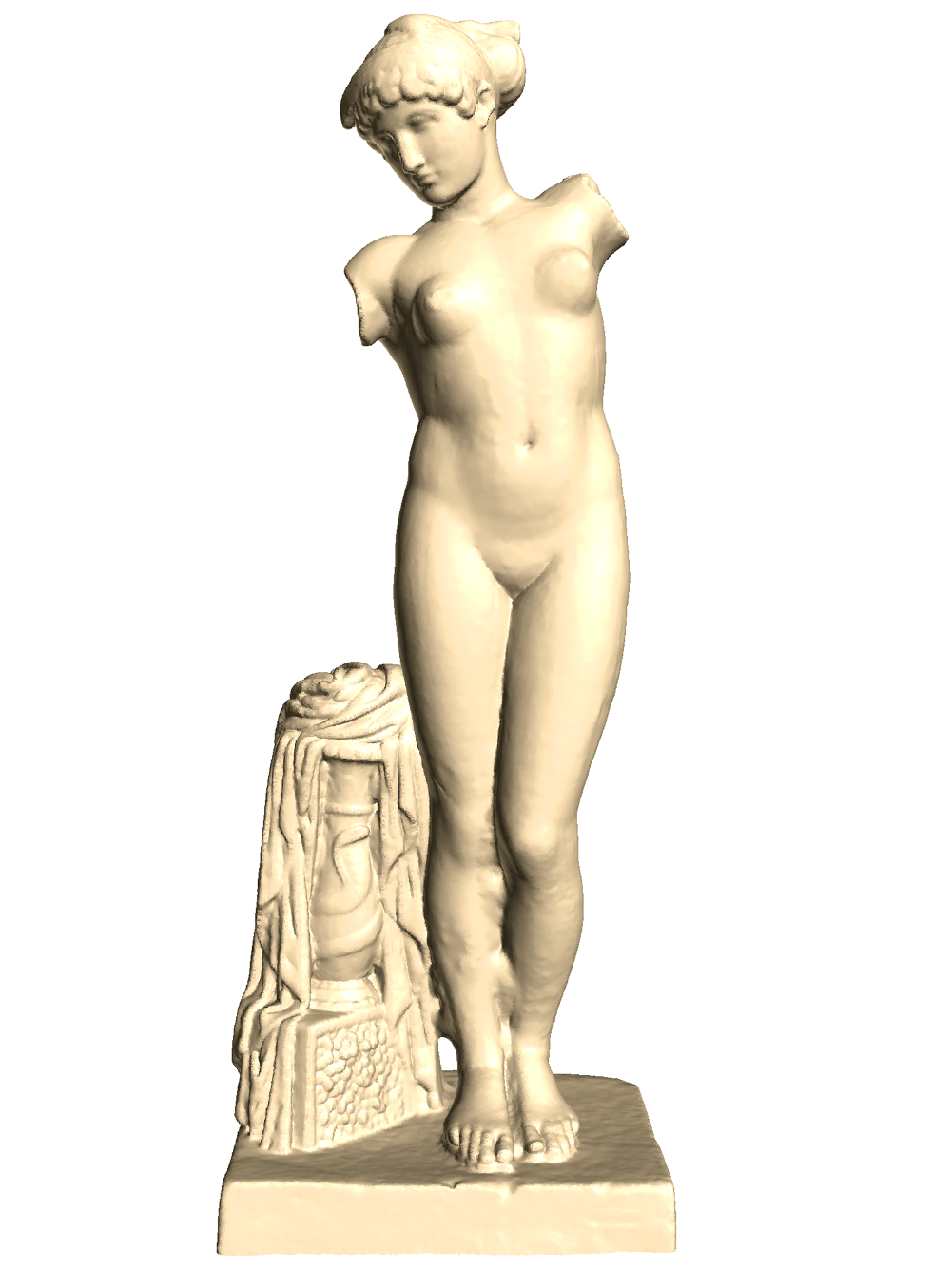}
        \end{subfigure}
        \begin{subfigure}[ht]{0.32\linewidth}
            \includegraphics[width=\textwidth]{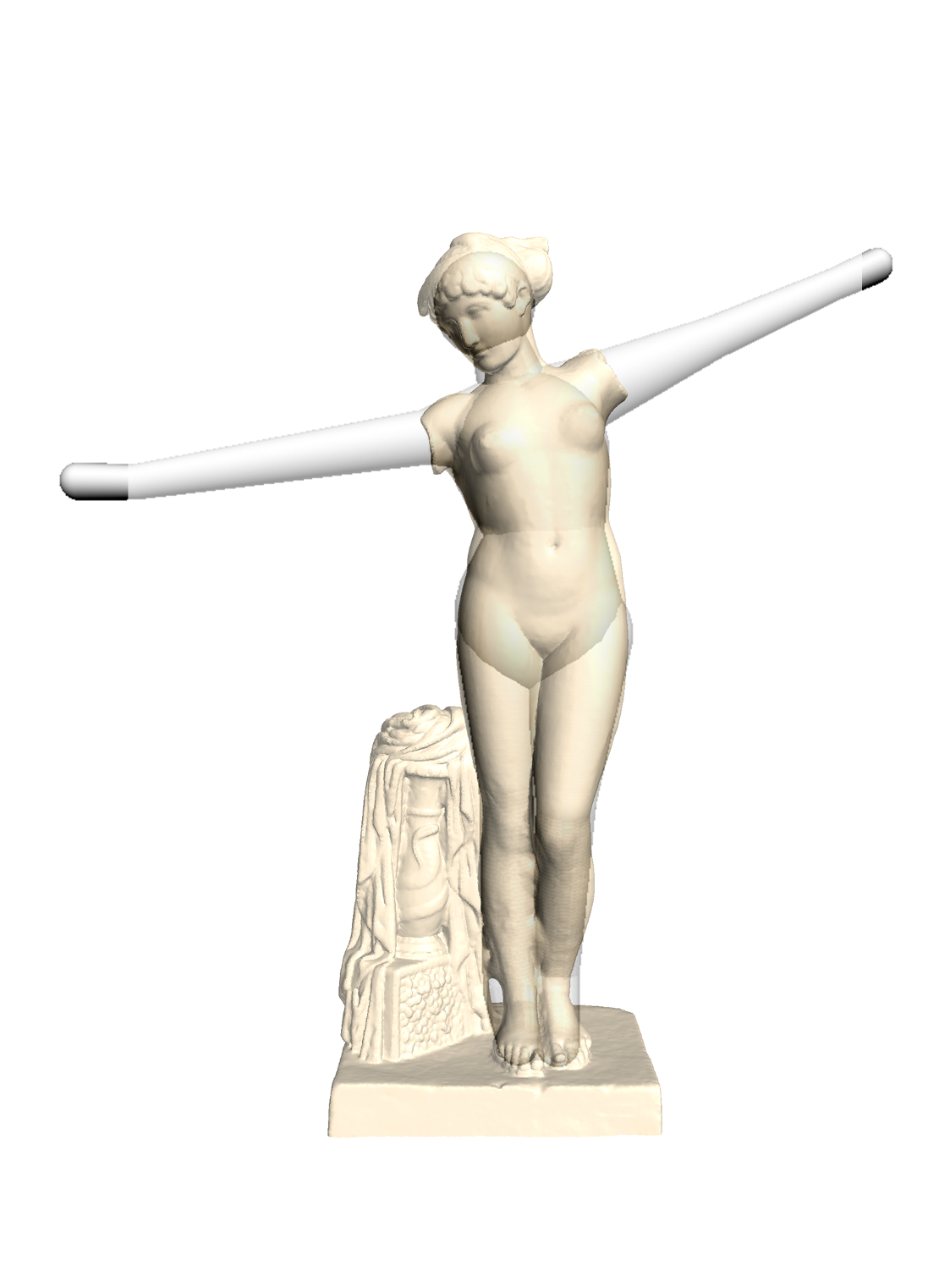}
        \end{subfigure}
        \begin{subfigure}[ht]{0.32\linewidth}
            \includegraphics[width=\textwidth]{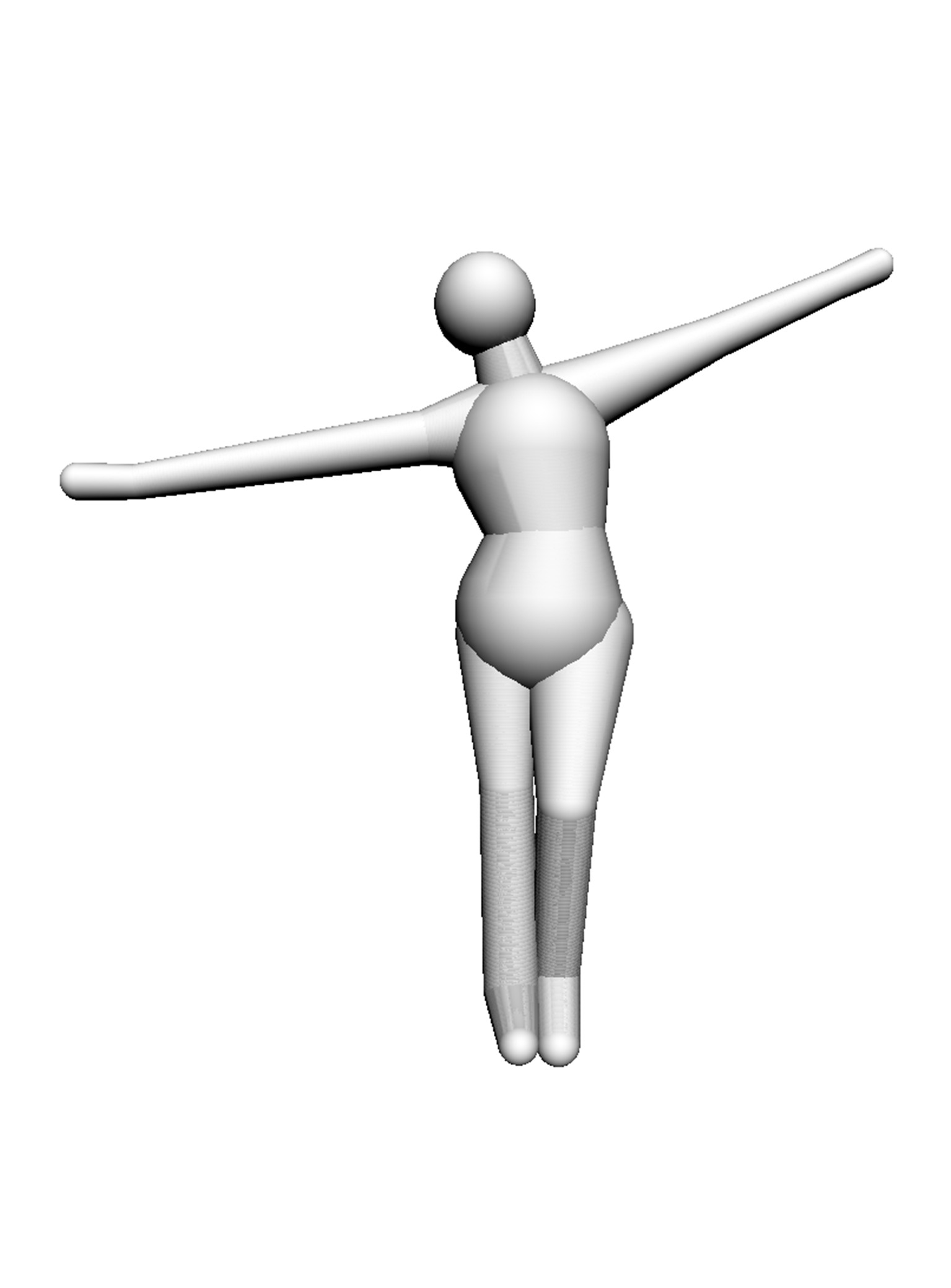}
        \end{subfigure}

        \begin{subfigure}[ht]{0.32\linewidth}
            \includegraphics[width=\textwidth]{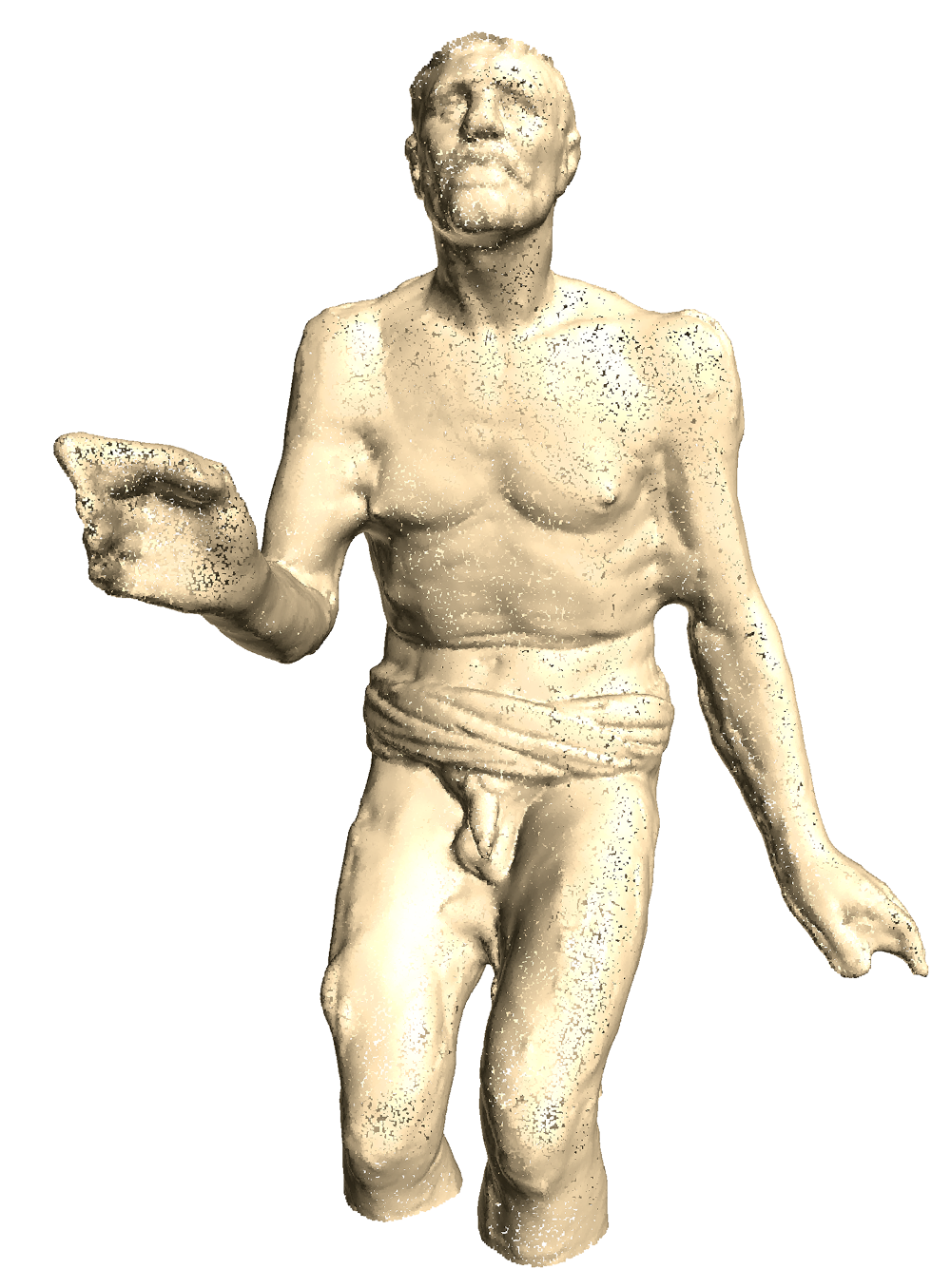}
        \end{subfigure}
        \begin{subfigure}[ht]{0.32\linewidth}
            \includegraphics[width=\textwidth]{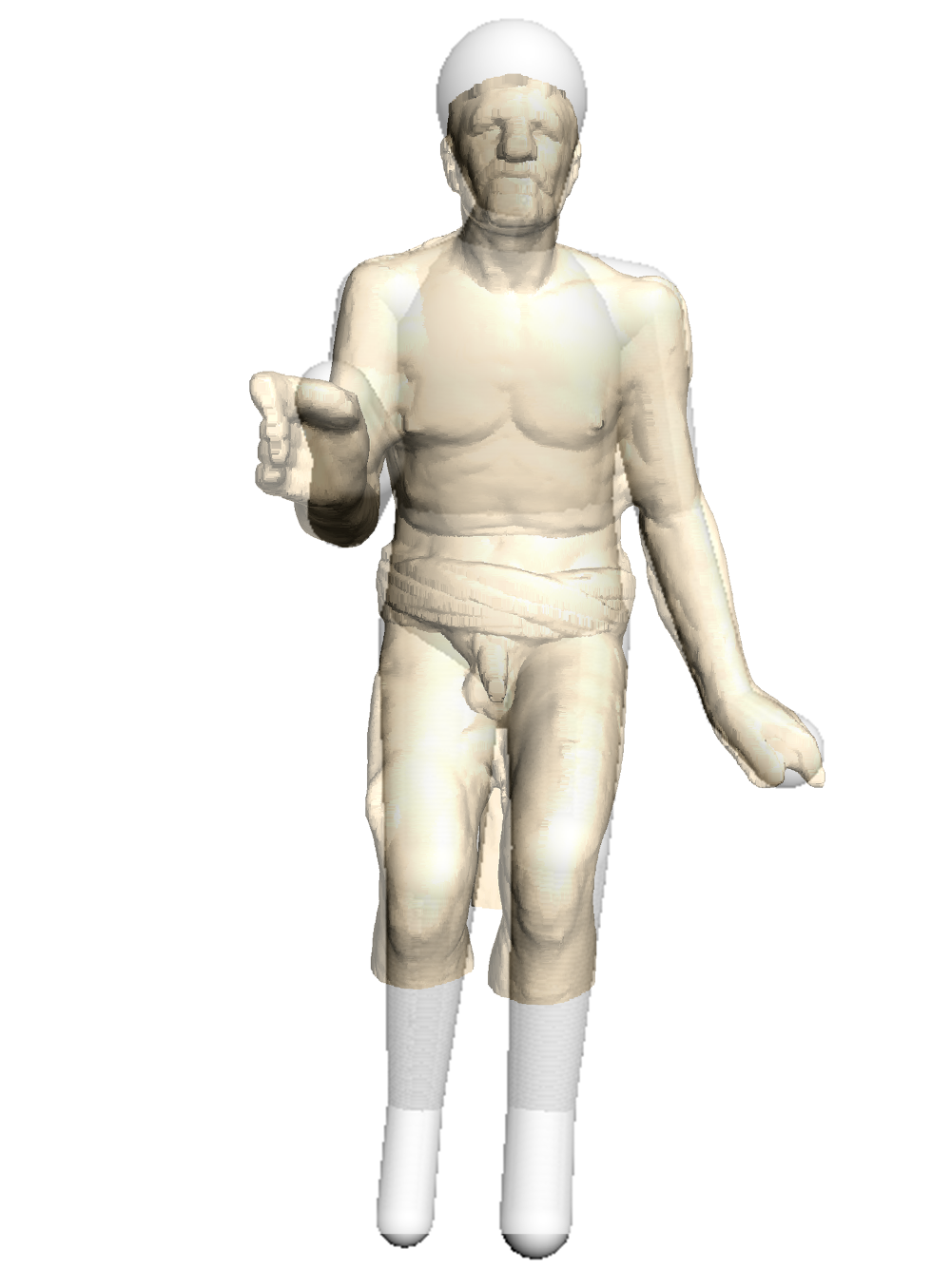}
        \end{subfigure}
        \begin{subfigure}[ht]{0.32\linewidth}
            \includegraphics[width=\textwidth]{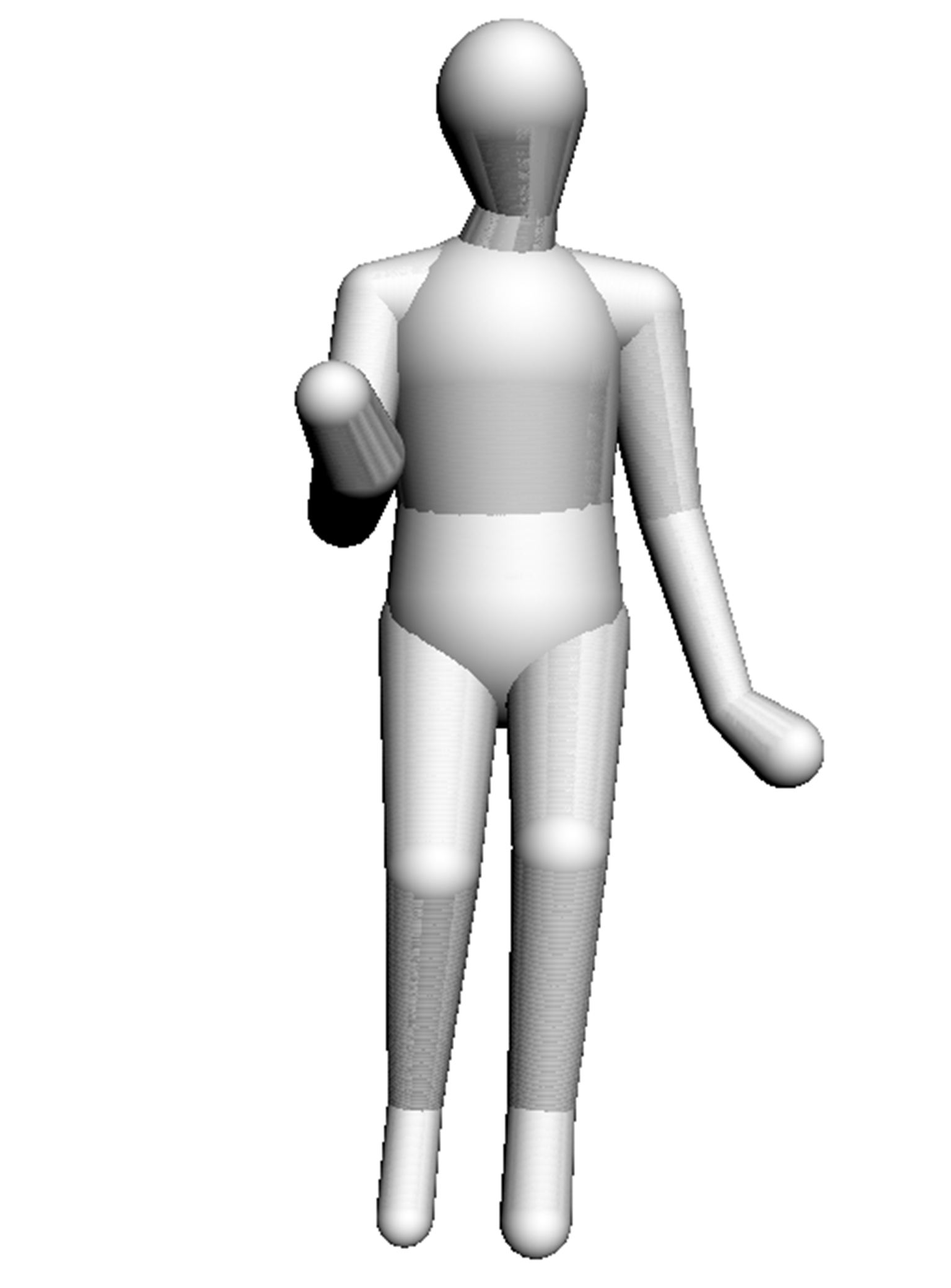}
        \end{subfigure}

        \begin{subfigure}[ht]{0.32\linewidth}
            \includegraphics[width=\textwidth]{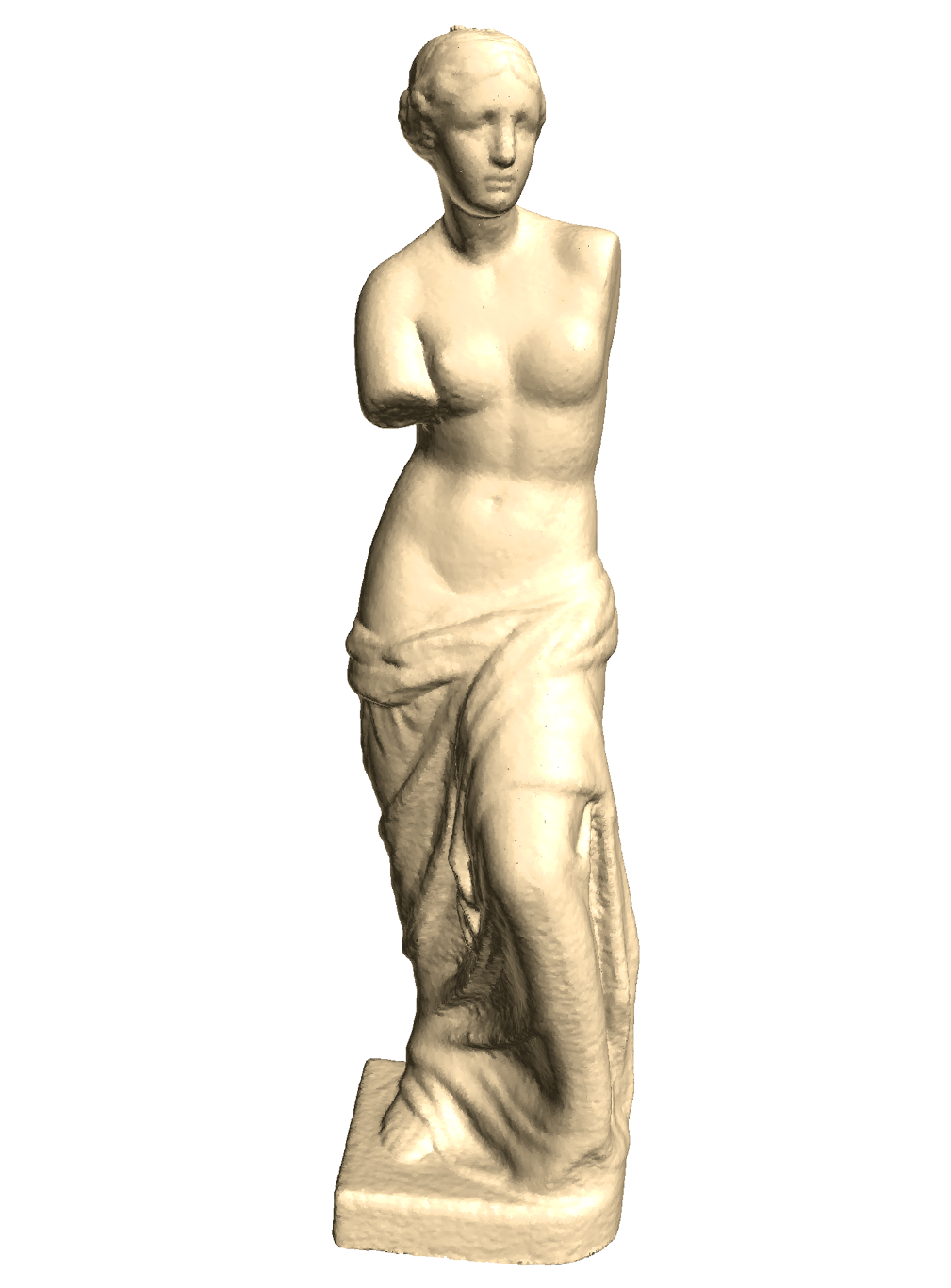}
        \end{subfigure}
        \begin{subfigure}[ht]{0.32\linewidth}
            \includegraphics[width=\textwidth]{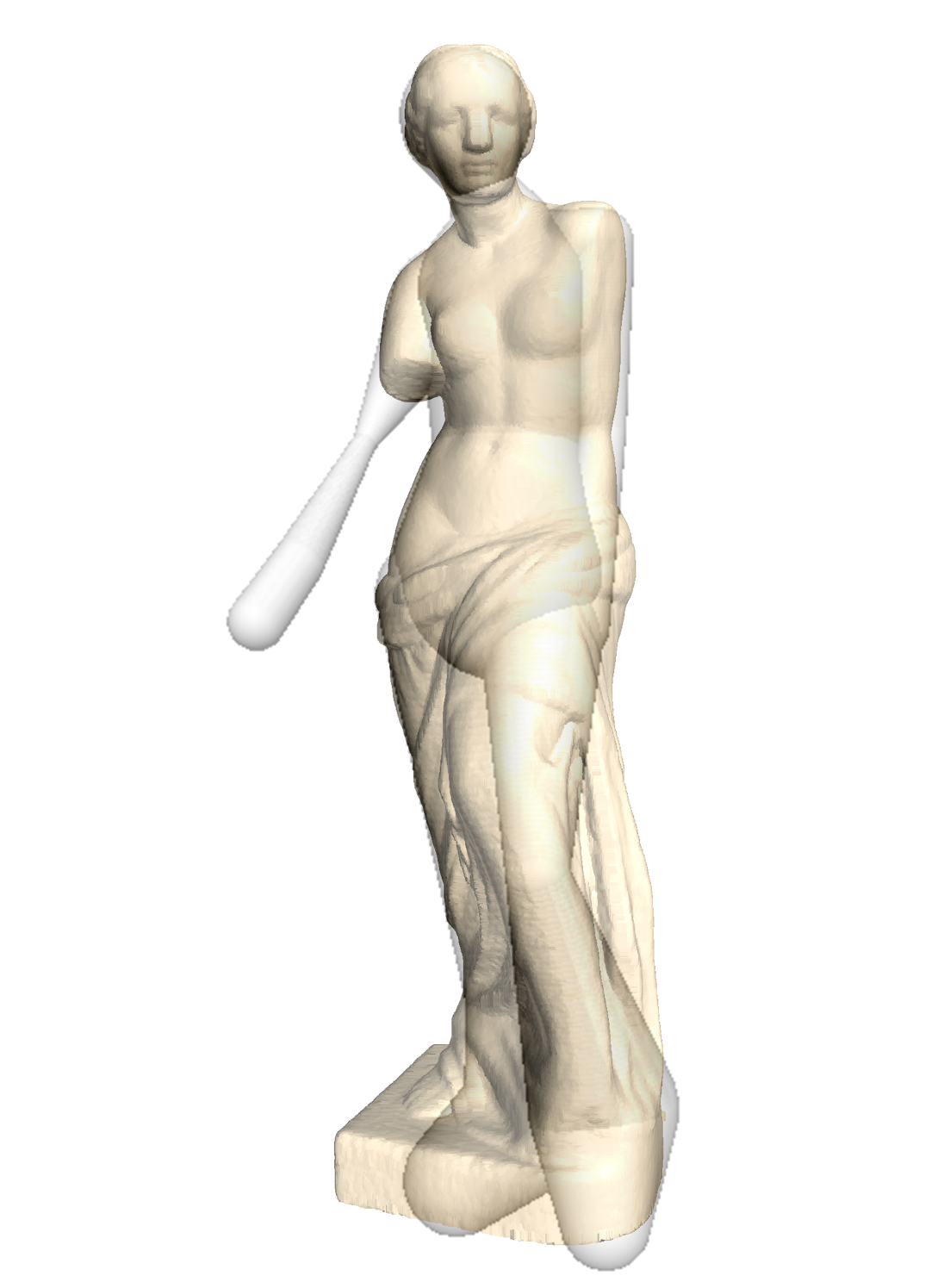}
        \end{subfigure}
        \begin{subfigure}[ht]{0.32\linewidth}
            \includegraphics[width=\textwidth]{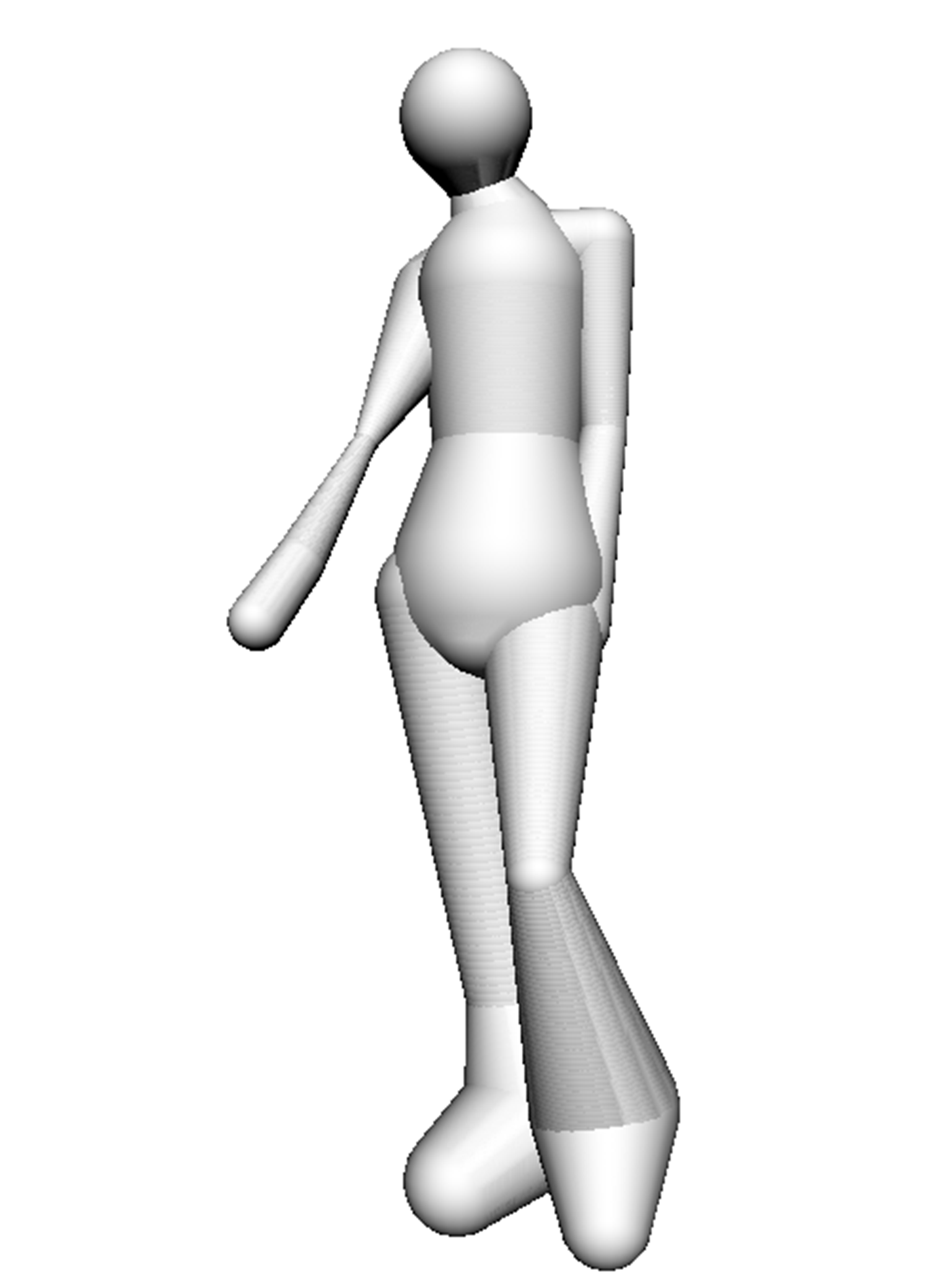}
        \end{subfigure}

    \caption{Registration of 3 incomplete statues, Esquiline Venus (first row), Old Fisherman (second row) and Venus de Milo (third row). First column: initial point set, second column: overlay of the registered model and the point set, third column: registered model alone.}
    \label{fig:incomplete_result}
\end{figure}

\begin{figure}[ht]
  \centering
    \begin{subfigure}[t]{0.32\linewidth}
        \centering
        \includegraphics[width=\textwidth]{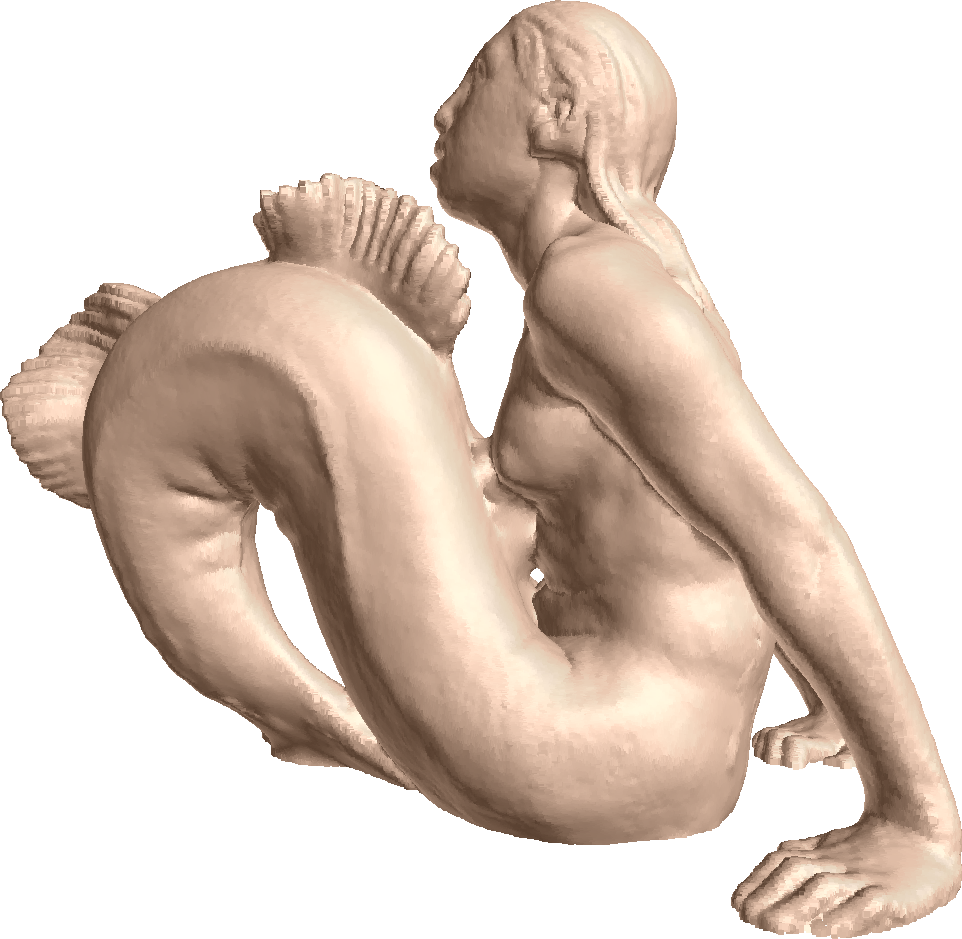}
        \caption{Mermaid statue}
    \end{subfigure}
    \begin{subfigure}[t]{0.32\linewidth}
        \centering
        \includegraphics[width=\textwidth]{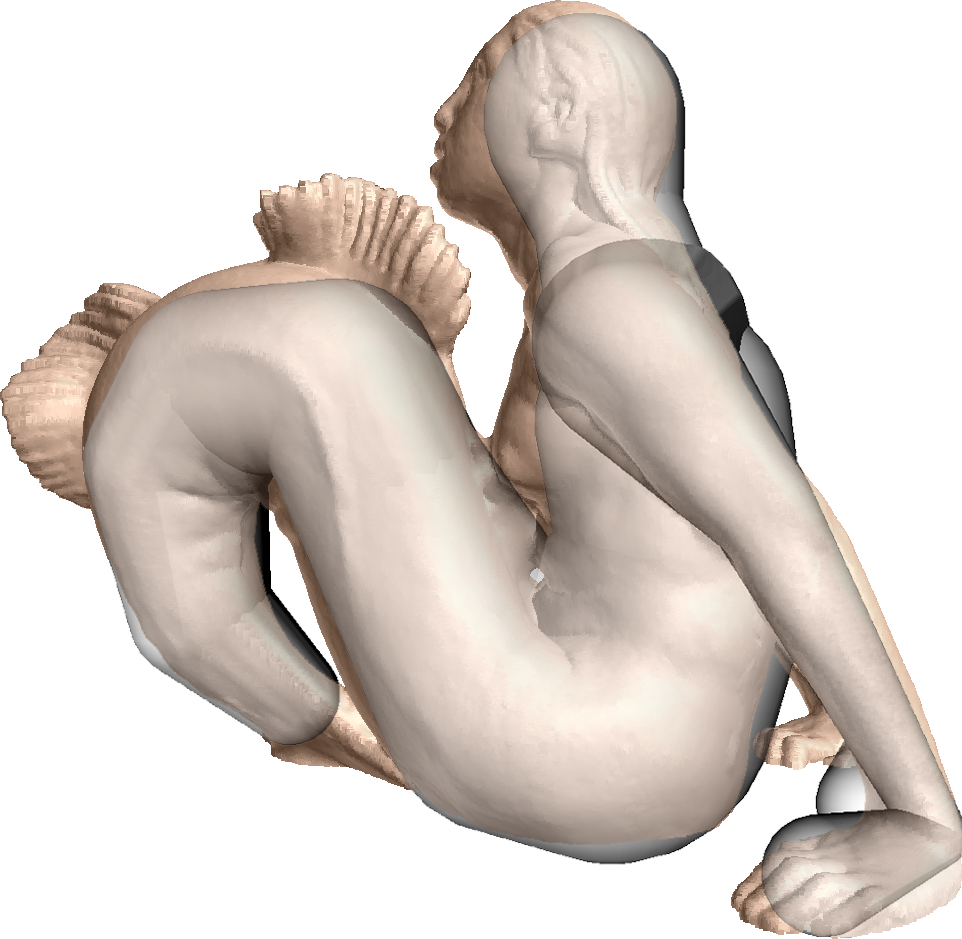}
        \caption{Overlayed registered model}
    \end{subfigure}
    \begin{subfigure}[t]{0.32\linewidth}
        \centering
        \includegraphics[width=\textwidth]{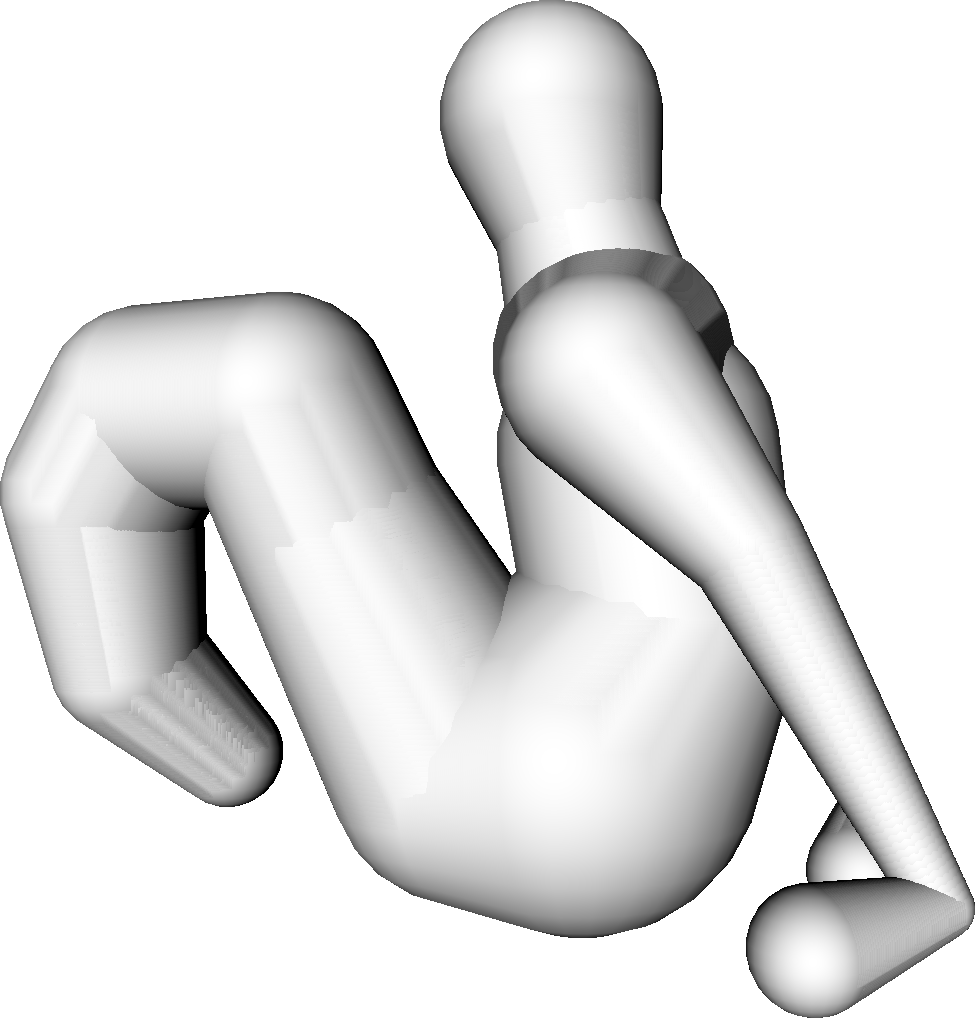}
        \caption{Registered model}
    \end{subfigure}
    \caption{FAKIR result on the Mermaid statue, the skeleton is a human one but with a single chain instead of two legs.}
    \label{fig:mermaid_res}
\end{figure}

\begin{figure}[ht]
    \centering
    \includegraphics[width=0.15\linewidth]{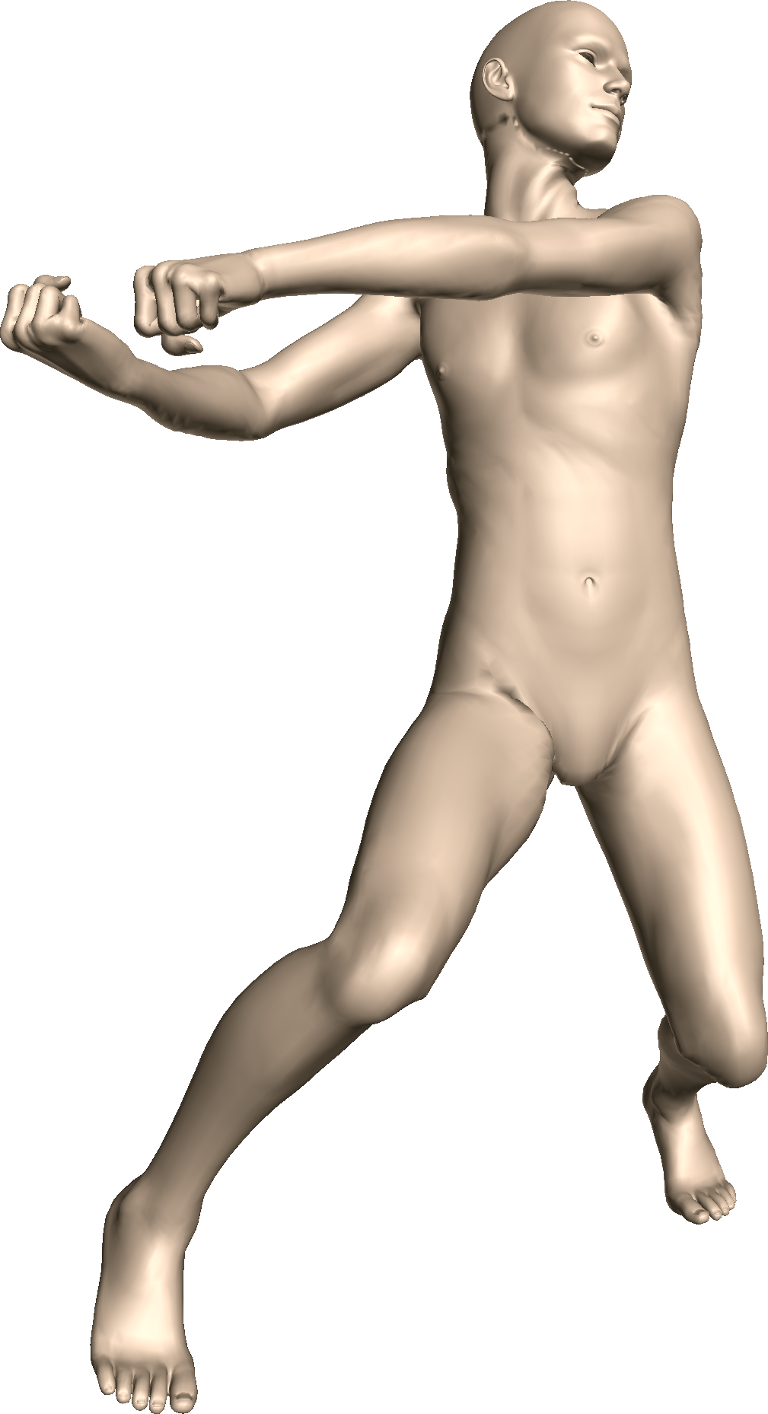}
    \includegraphics[width=0.15\linewidth]{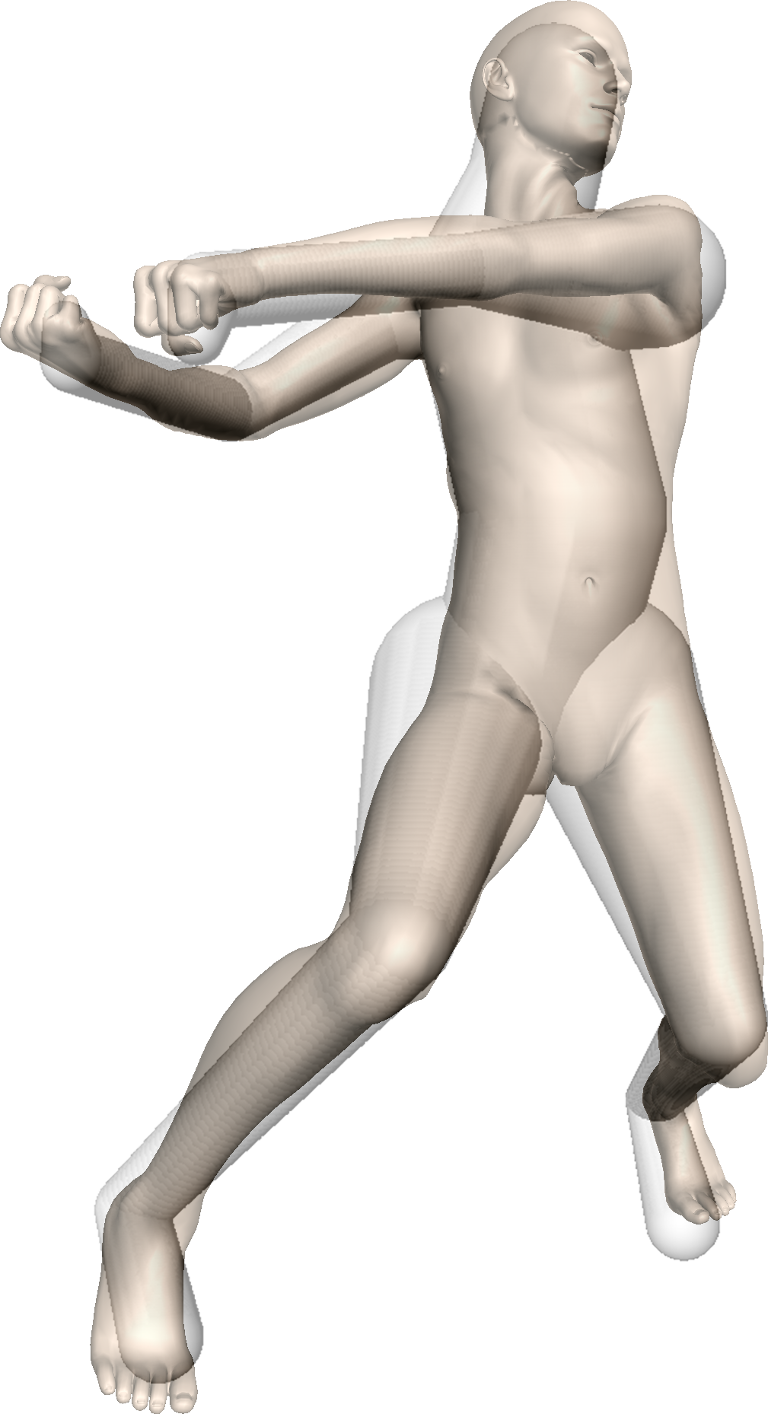}
    \includegraphics[width=0.15\linewidth]{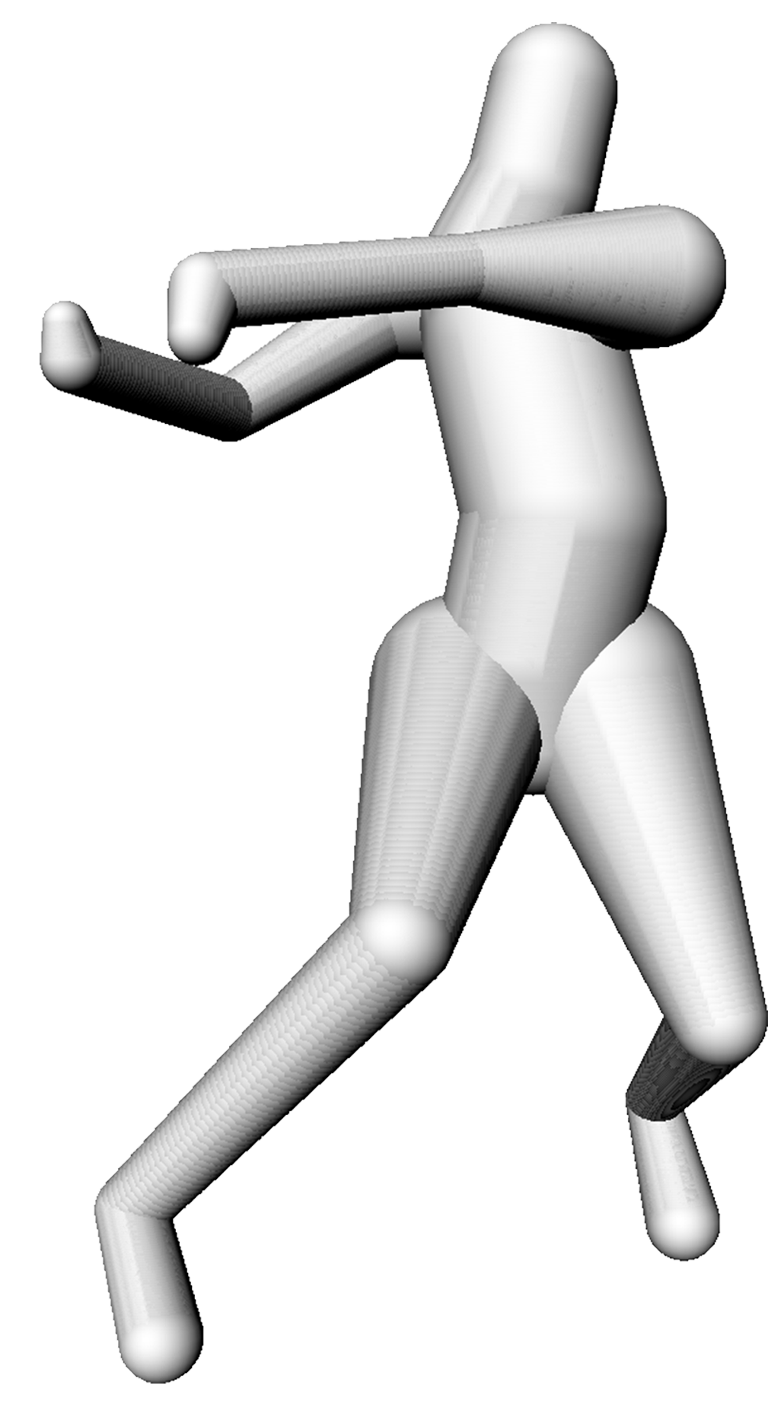}
    ~
    \includegraphics[width=0.15\linewidth]{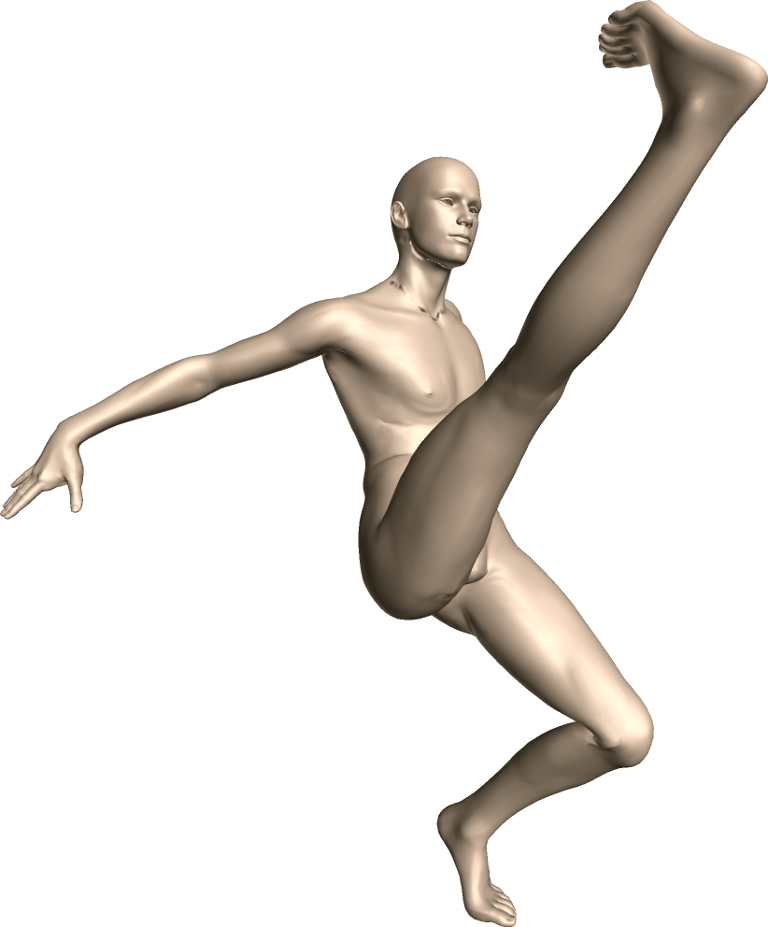}
    \includegraphics[width=0.15\linewidth]{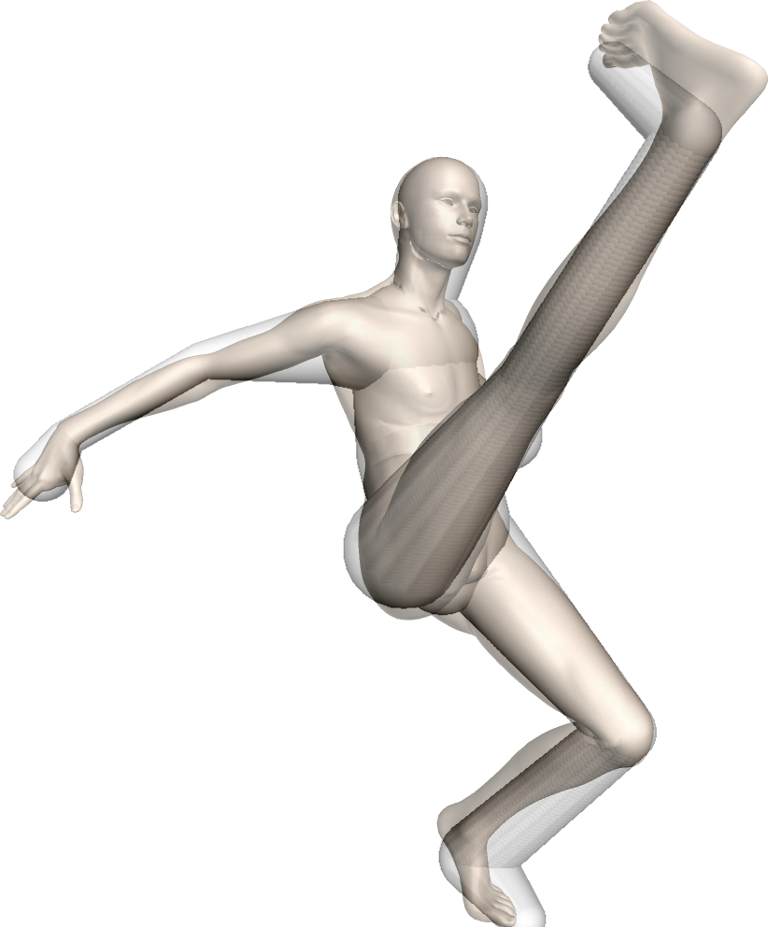}
    \includegraphics[width=0.15\linewidth]{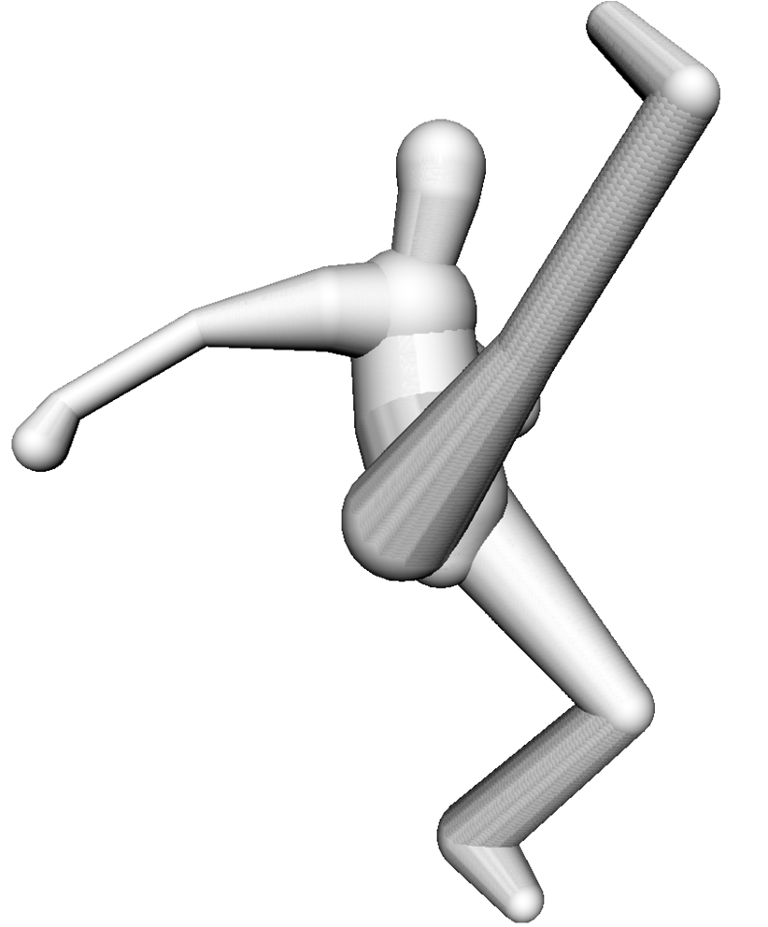}

    \includegraphics[width=0.15\linewidth]{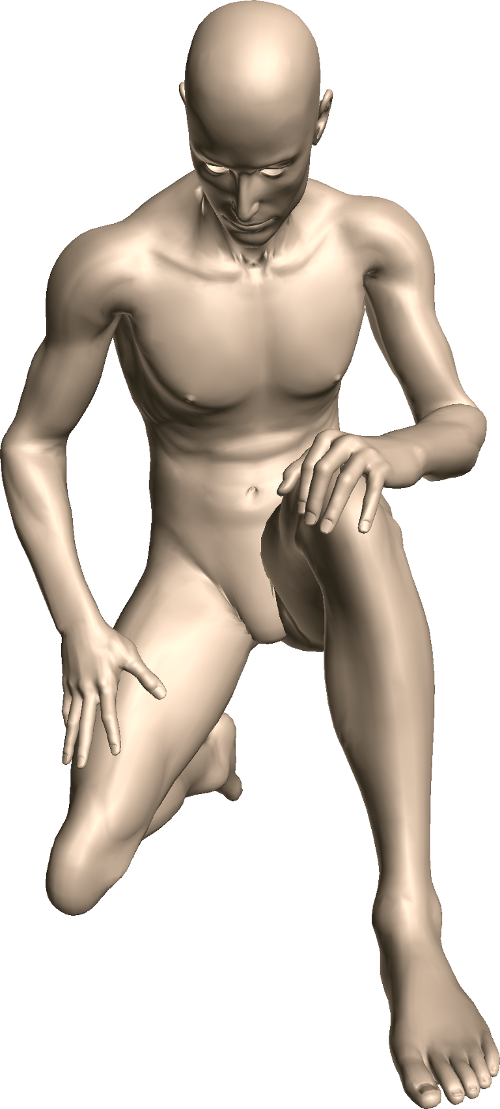}
    \includegraphics[width=0.15\linewidth]{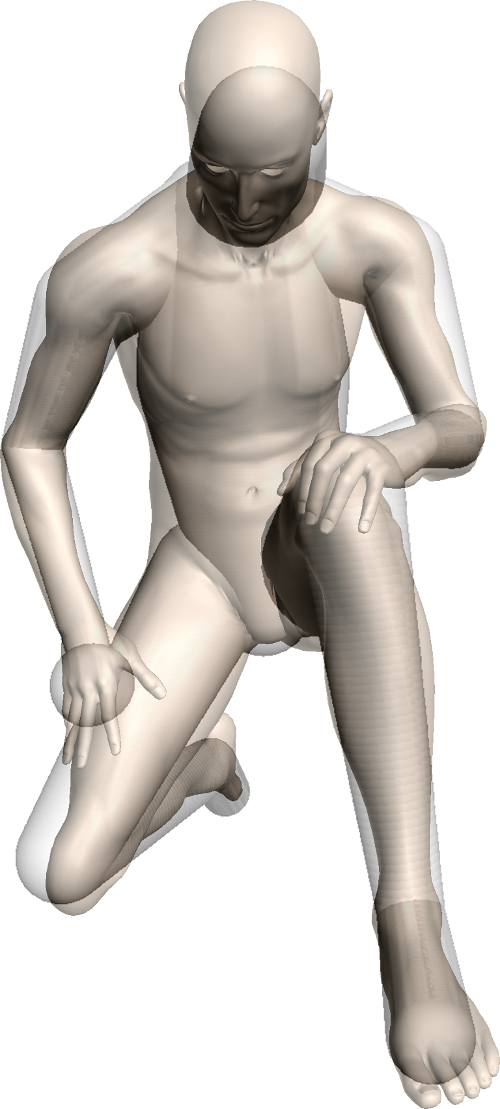}
    \includegraphics[width=0.15\linewidth]{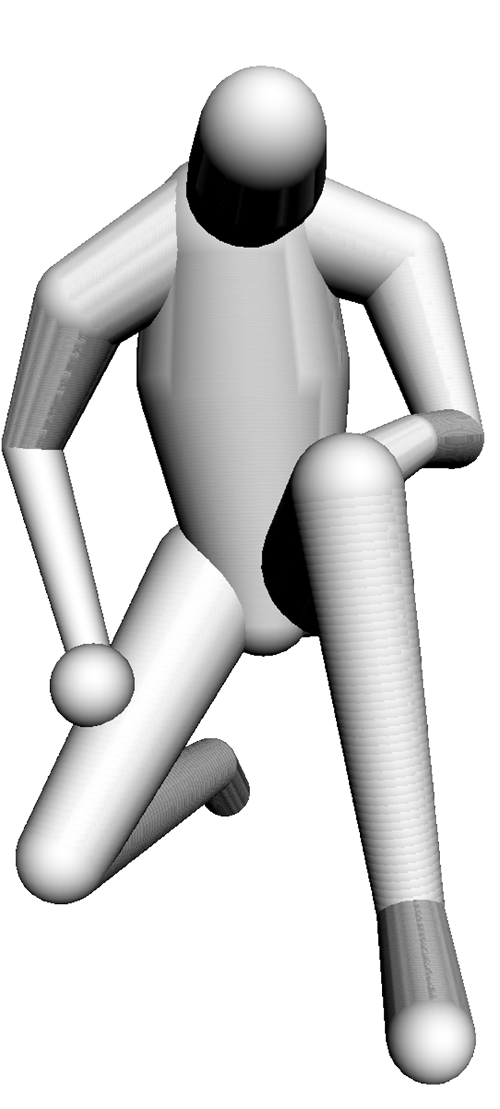}
    ~
    \includegraphics[width=0.15\linewidth]{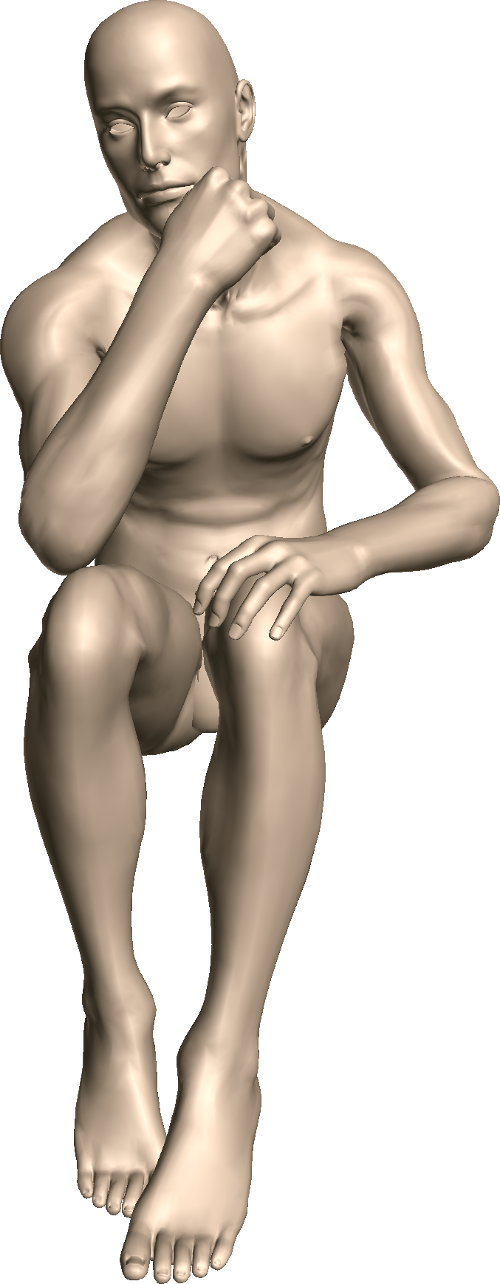}
    \includegraphics[width=0.15\linewidth]{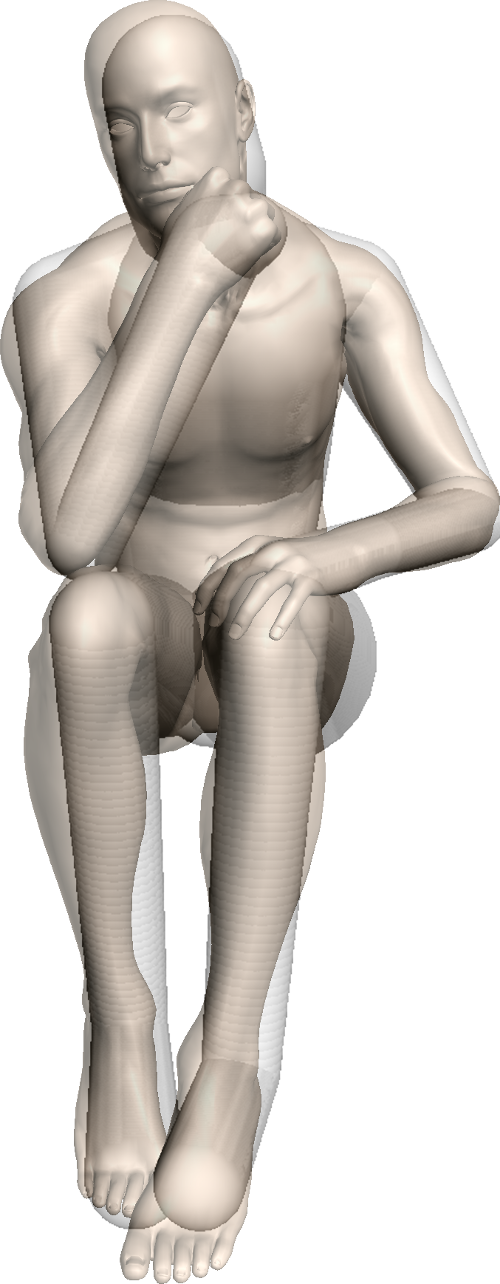}
    \includegraphics[width=0.15\linewidth]{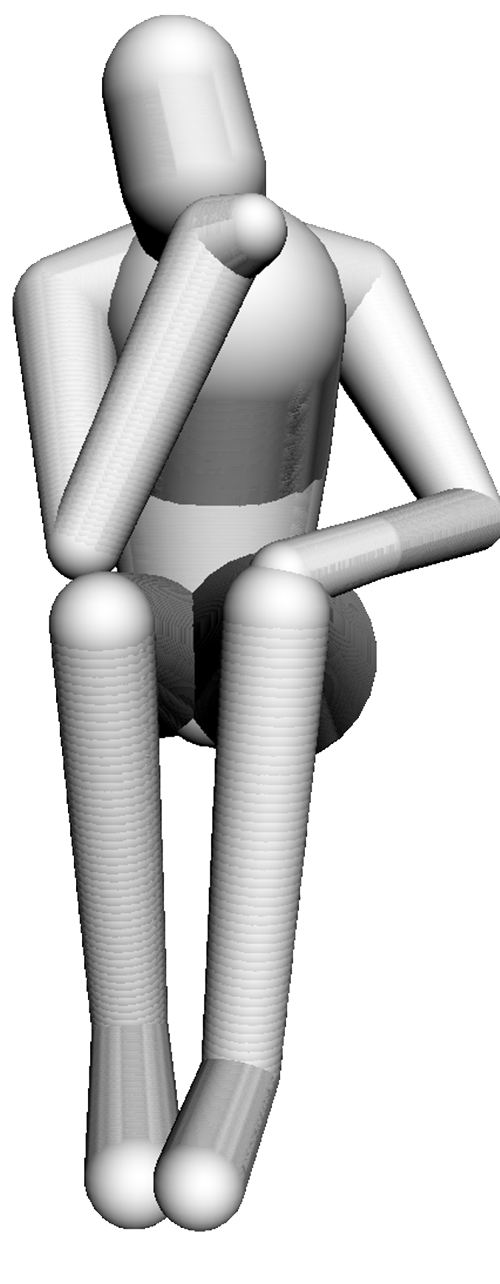}
    \caption{Skeleton registration on human in various poses from the TOSCA dataset.}
    \label{fig:exp_tosca}
\end{figure}

\begin{figure}[ht]
    \centering
    \includegraphics[width=0.32\linewidth]{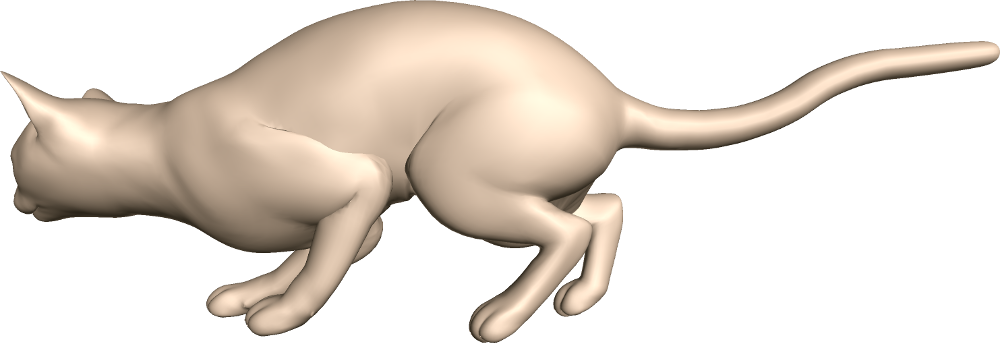}
    \includegraphics[width=0.32\linewidth]{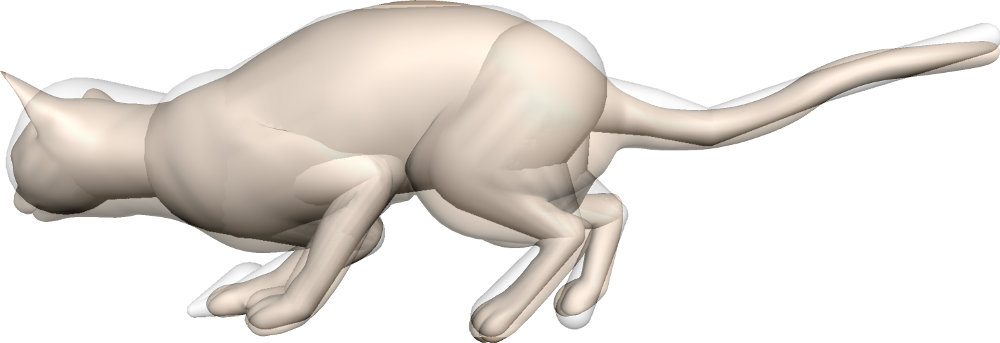}
    \includegraphics[width=0.32\linewidth]{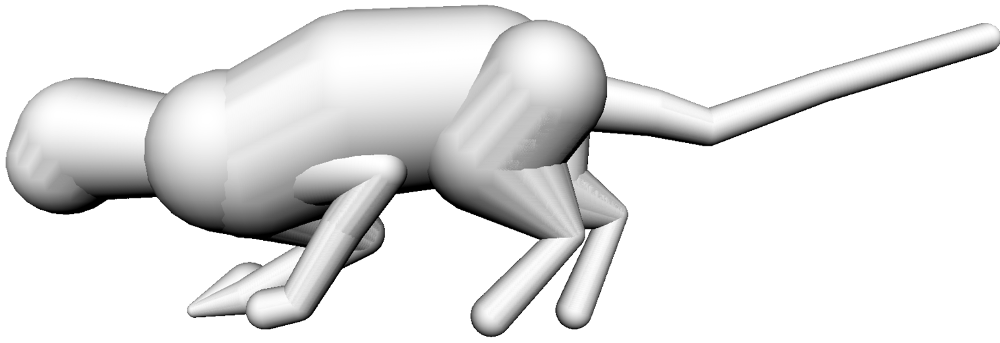}\\
    \includegraphics[width=0.32\linewidth]{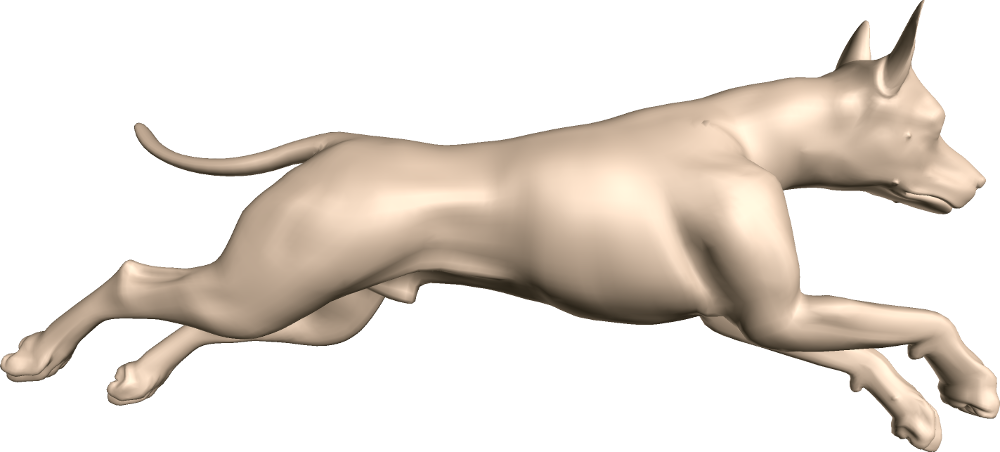}
    \includegraphics[width=0.32\linewidth]{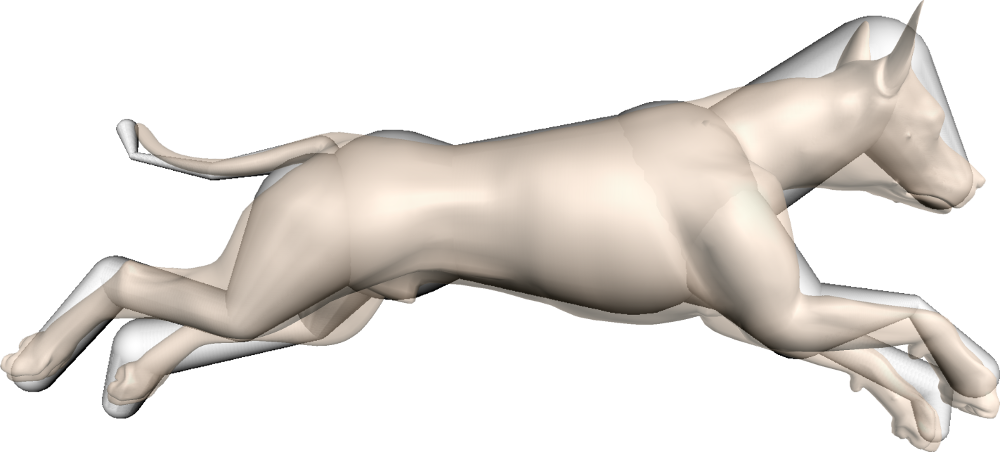}
    \includegraphics[width=0.32\linewidth]{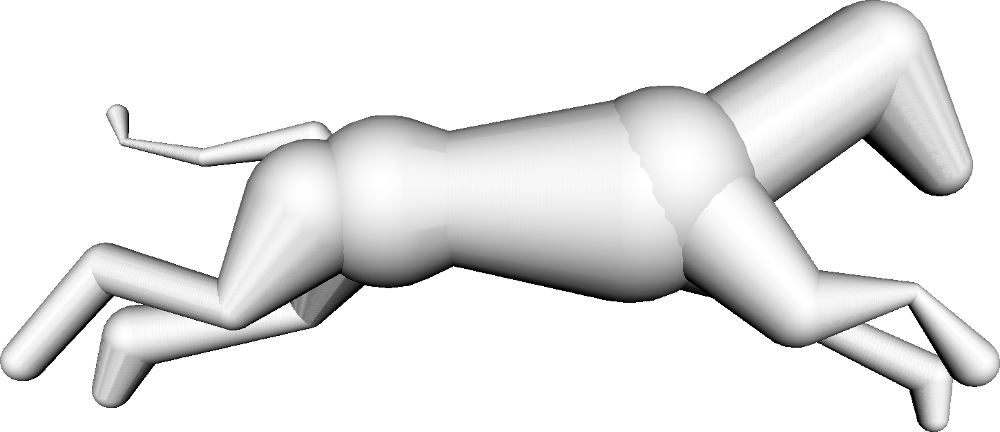}\\
    \includegraphics[width=0.32\linewidth]{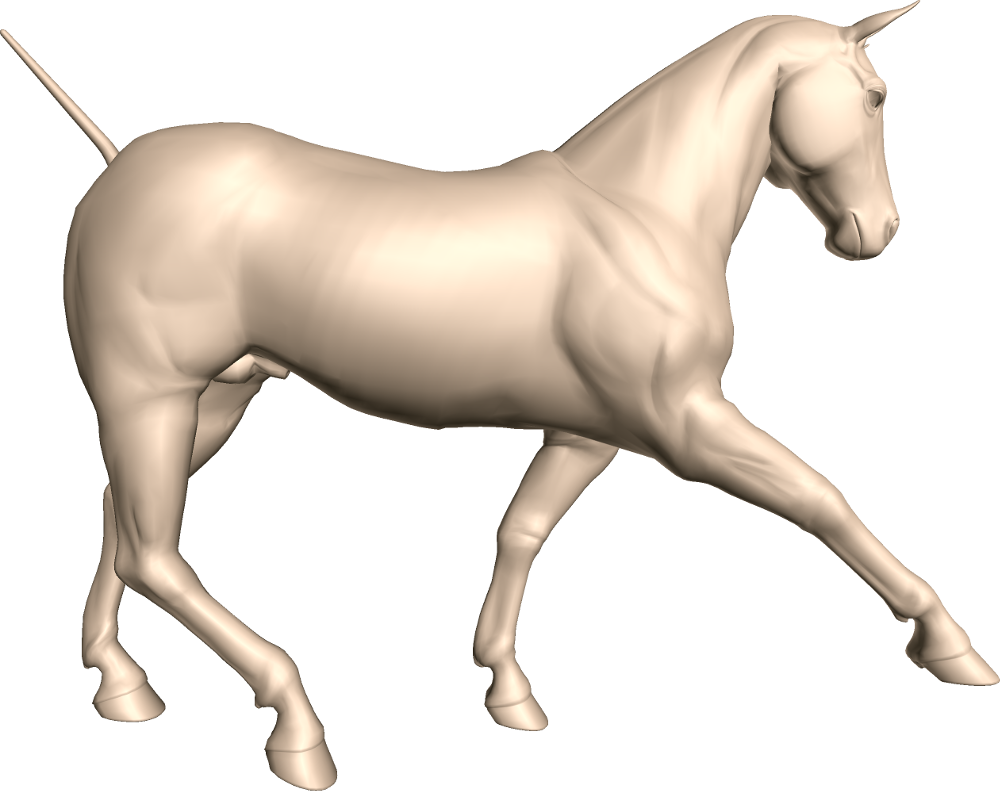}
    \includegraphics[width=0.32\linewidth]{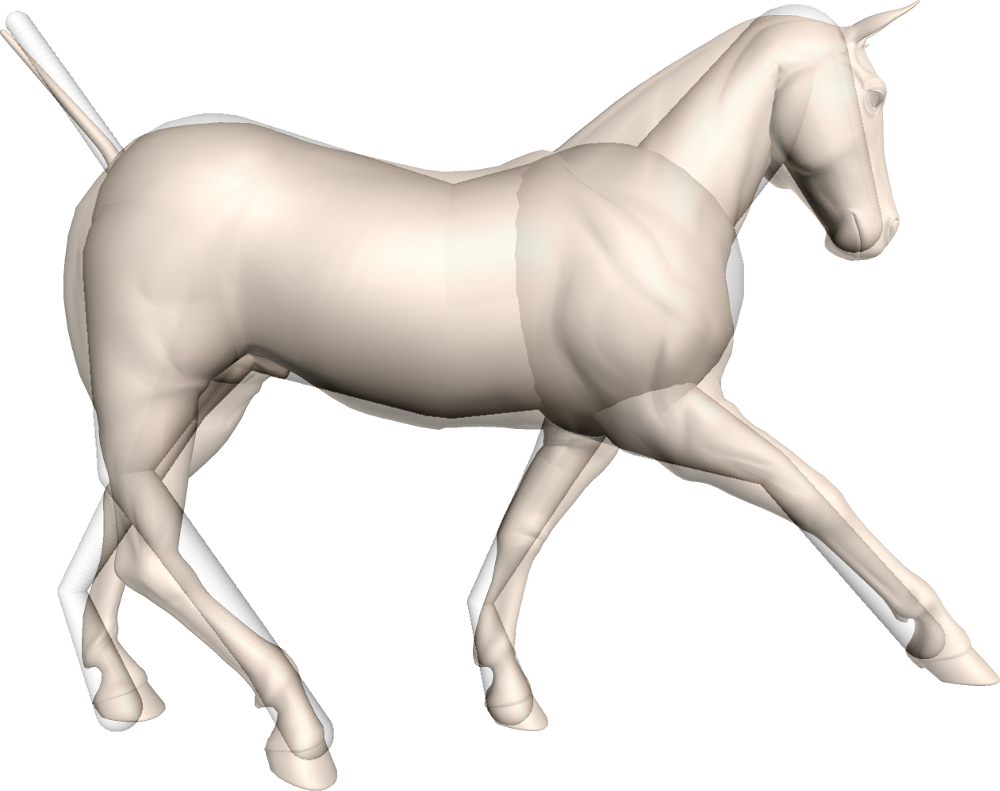}
    \includegraphics[width=0.32\linewidth]{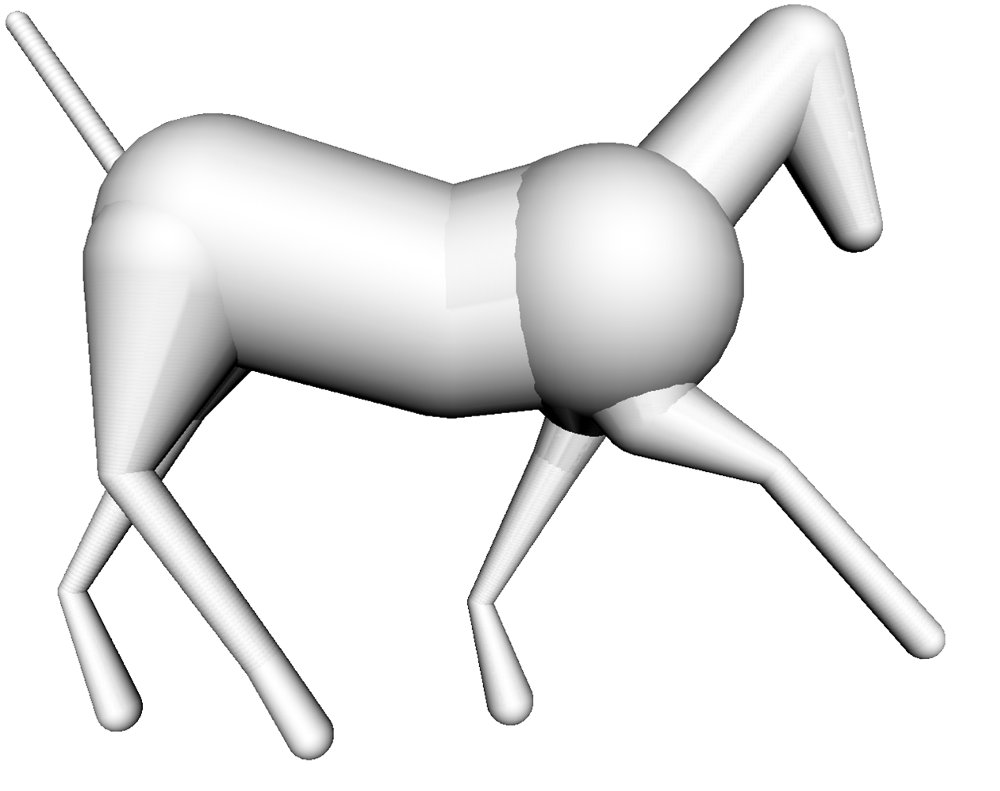}\\
    \includegraphics[width=0.32\linewidth]{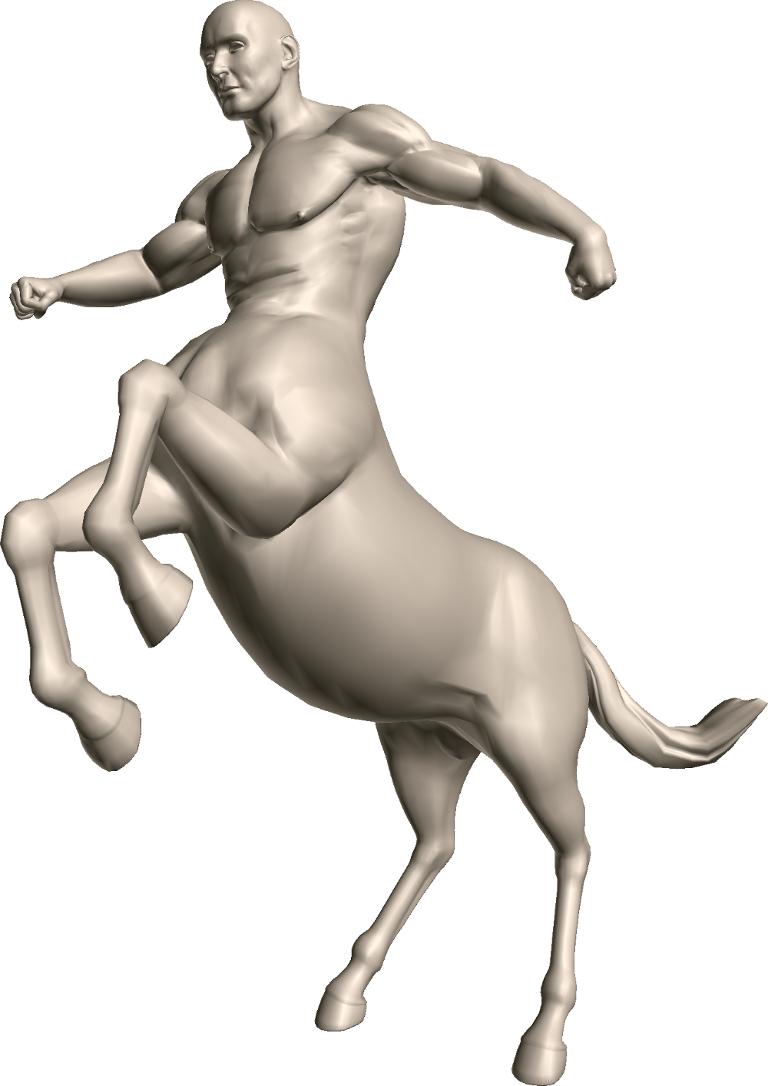}
    \includegraphics[width=0.32\linewidth]{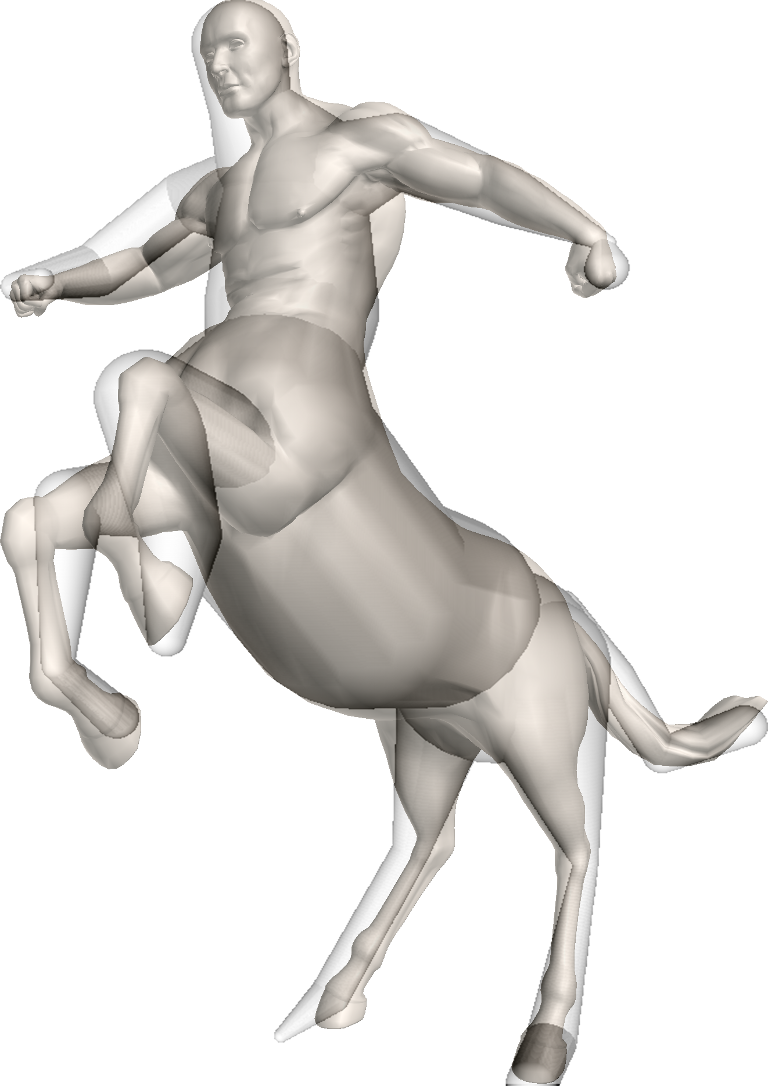}
    \includegraphics[width=0.32\linewidth]{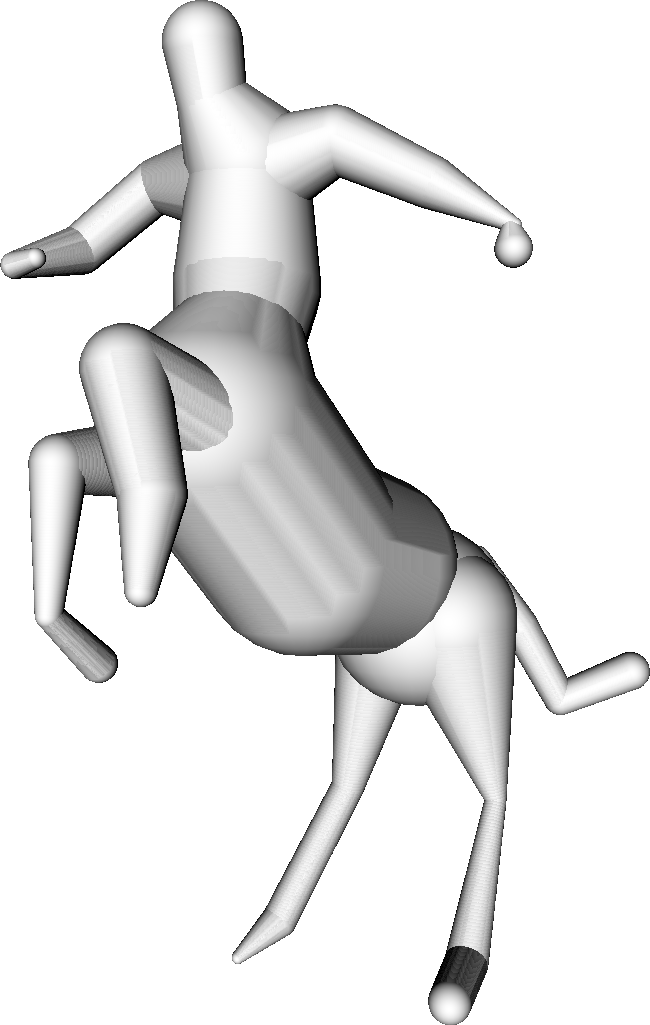}
    \caption{Skeleton registration on different animals from the TOSCA dataset. The skeleton is simply the human skeleton  (Figure \ref{fig:human_model}) supplemented with a bone chain for the tail.}
    \label{fig:tosca_animals}
\end{figure}

Figure \ref{fig:result} shows our registrations on four statues.
The registration algorithm performs well for statues depicting naked characters: in this case, the registration is not hindered by additional clothing or accessories, and the simple sphere-mesh model fits well the data. Even with moderate clothing (Dancer with Crotales) FAKIR recovers the pose of the statue (see also the supplementary for more registration results).
FAKIR can also work on incomplete statues on as shown in Figure \ref{fig:incomplete_result} and on imaginary creature statues (Figure \ref{fig:mermaid_res}). 
We also demonstrate that FAKIR can work with real human bodies in more complex poses (Figure \ref{fig:exp_tosca}) or animals (Figure \ref{fig:tosca_animals}) from the TOSCA dataset.

\begin{figure}[ht]
    \centering
    \begin{subfigure}[t]{0.65\linewidth}
    \includegraphics[width=0.48\linewidth]{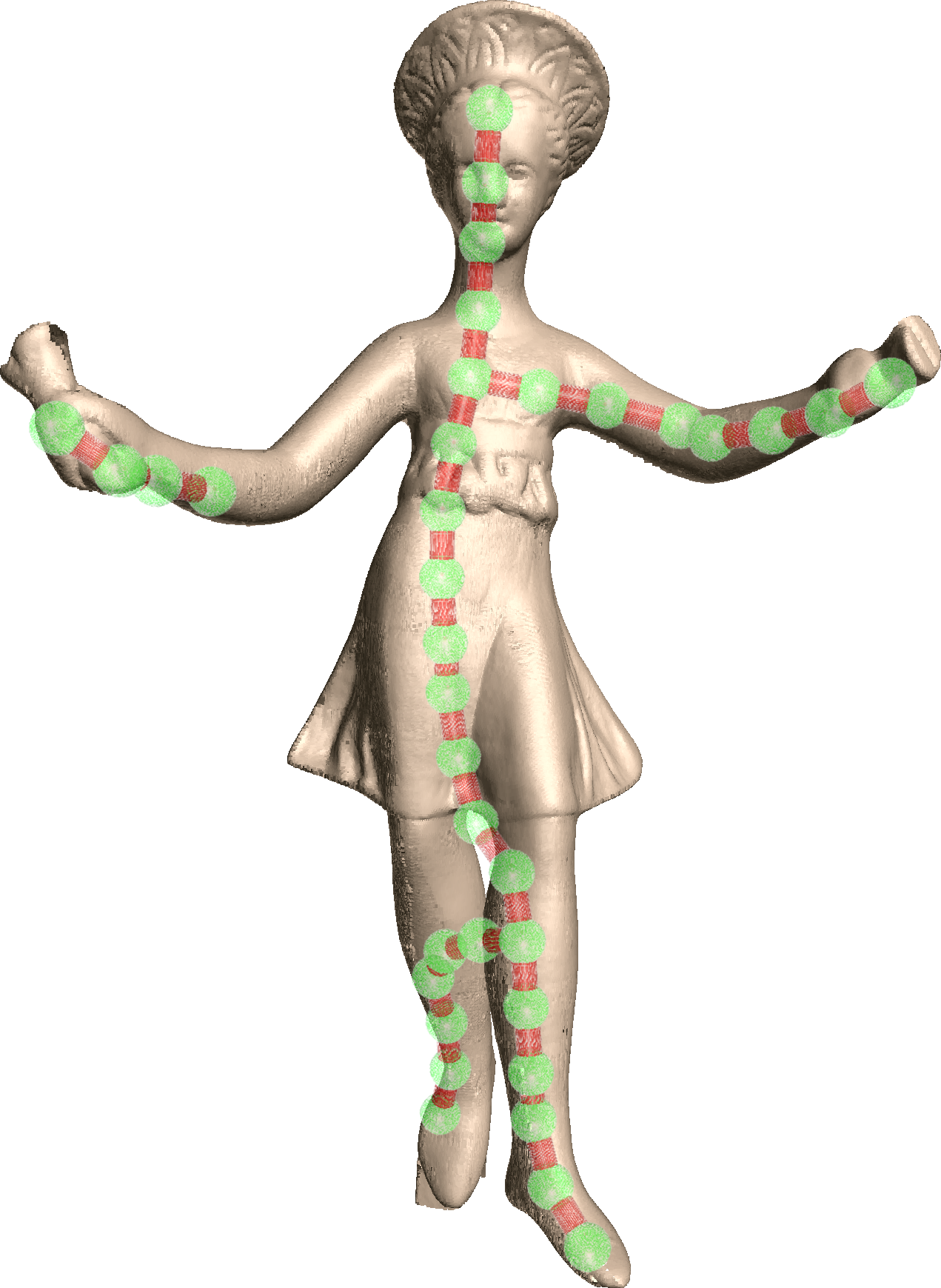}
    \includegraphics[width=0.48\linewidth]{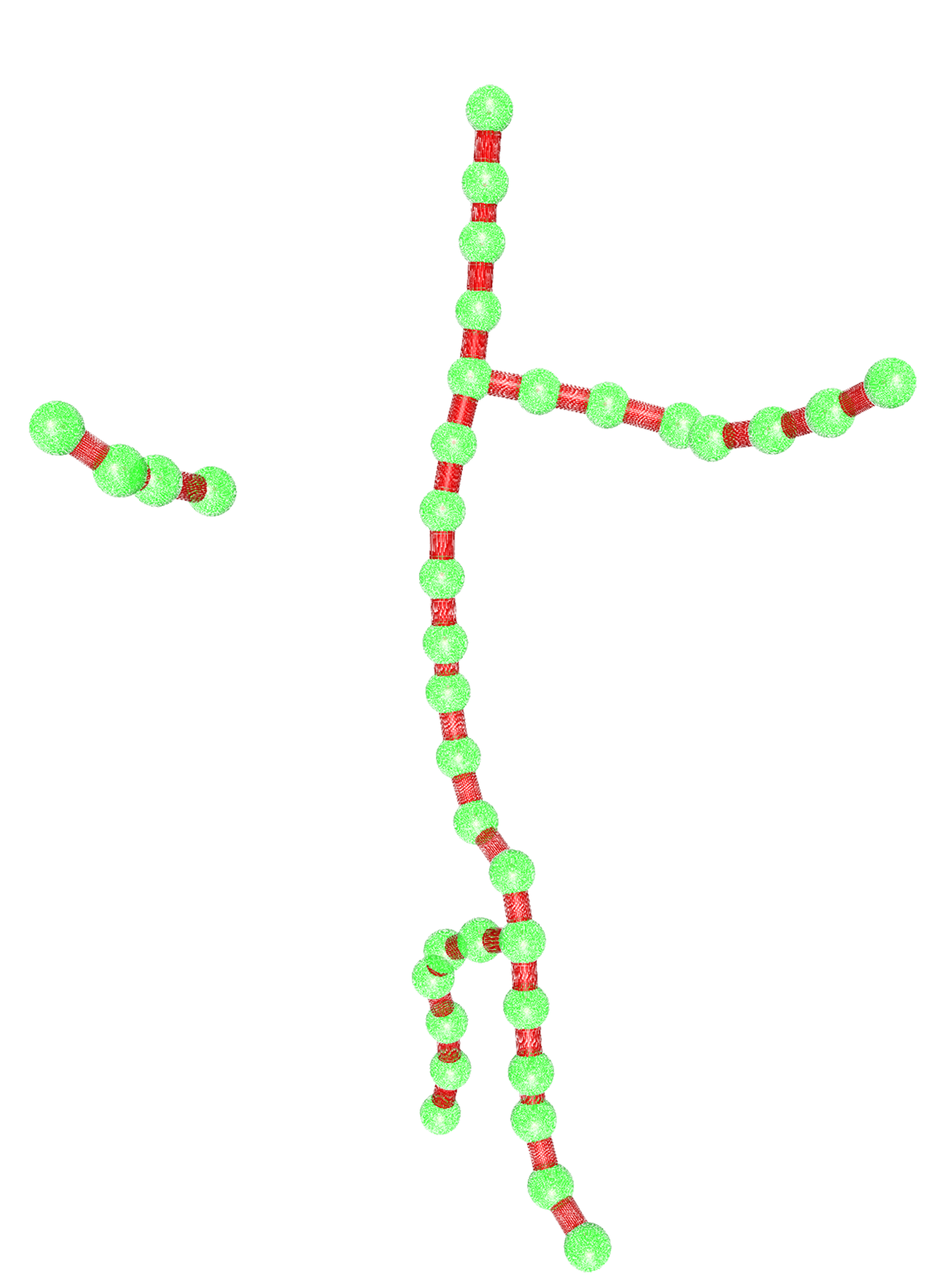}
    \caption{Danseuse with crotales}
    \end{subfigure}
    \begin{subfigure}[t]{0.32\linewidth}
    \includegraphics[width=0.48\linewidth]{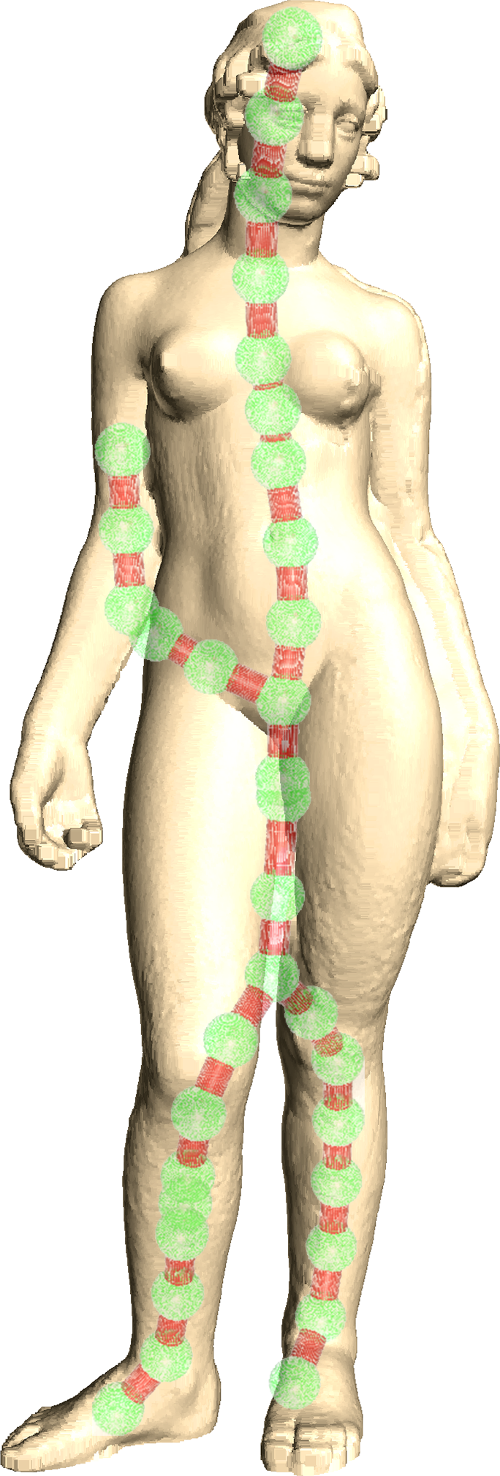}
    \includegraphics[width=0.48\linewidth]{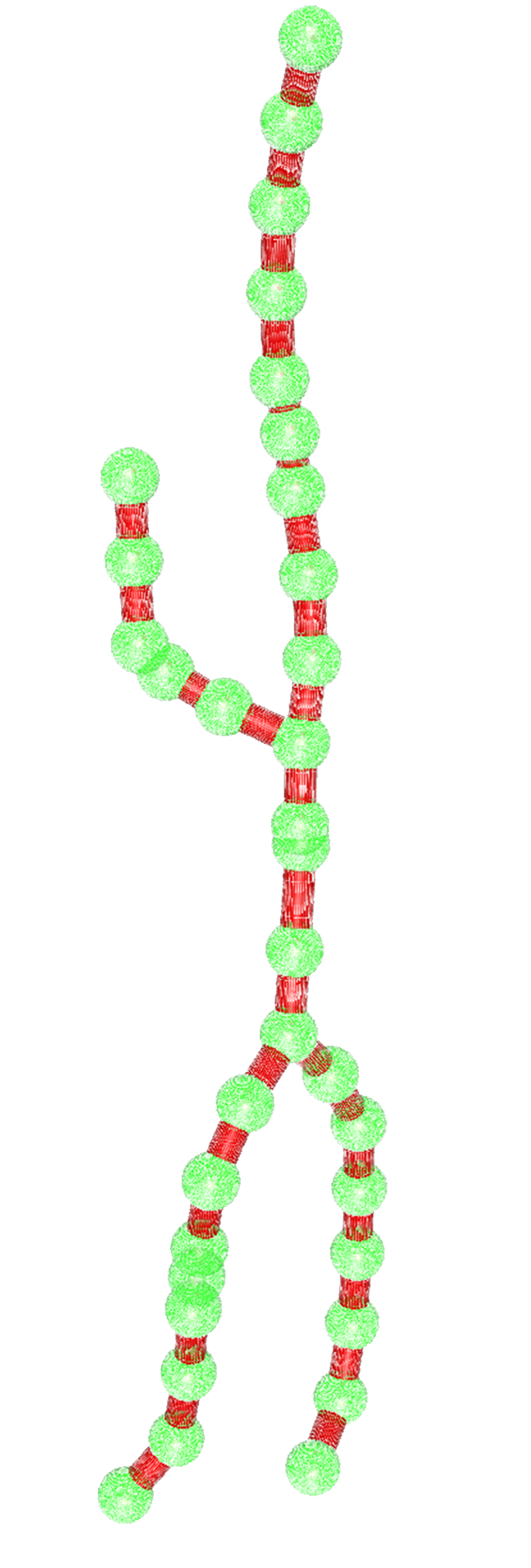}
    \caption{Aphrodite}
    \end{subfigure}
    \caption{$\ell^1$-medial skeletons~\cite{Huang2013} extracted from the Aphrodite and Danseuse with Crotales point sets. To compare to our results on Figures \ref{fig:pinocchio} and \ref{fig:smpl_comp}.}
    \label{fig:compare_huang}
\end{figure}

\begin{figure}[ht]
    \centering
     \begin{subfigure}[t]{0.19\linewidth}
        \centering
        \includegraphics[width=\textwidth]{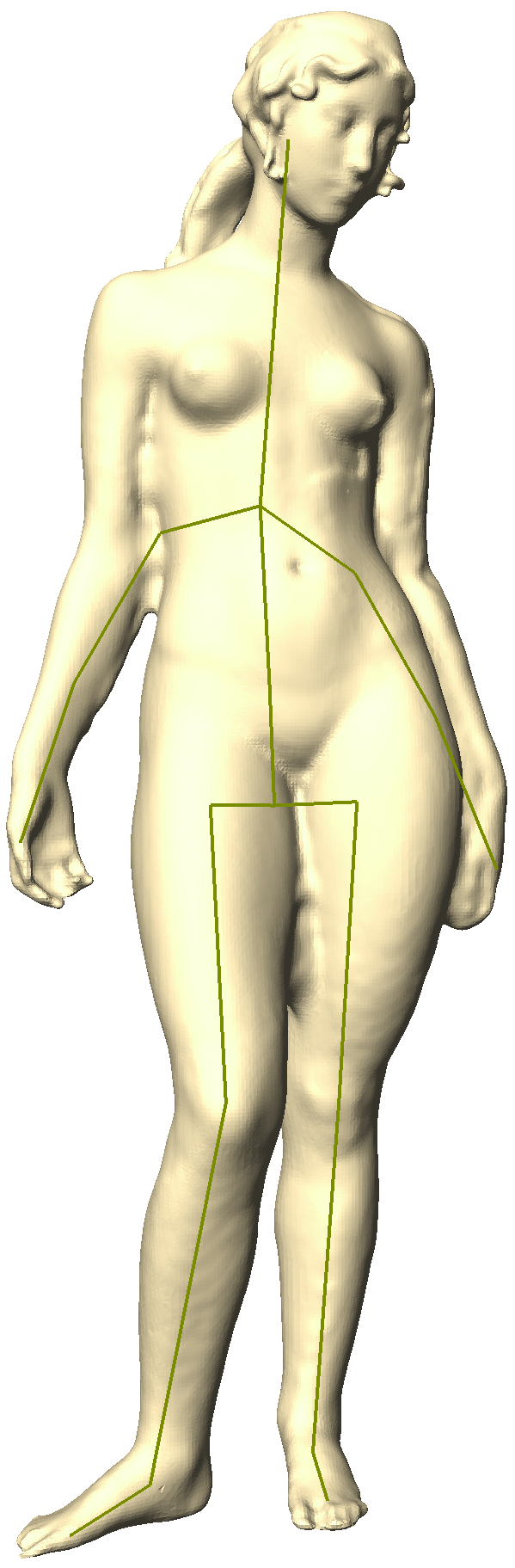}
        \caption{\scriptsize{Pinocchio (vanilla)}}
    \end{subfigure}
    \begin{subfigure}[t]{0.19\linewidth}
        \centering
        \includegraphics[width=\textwidth]{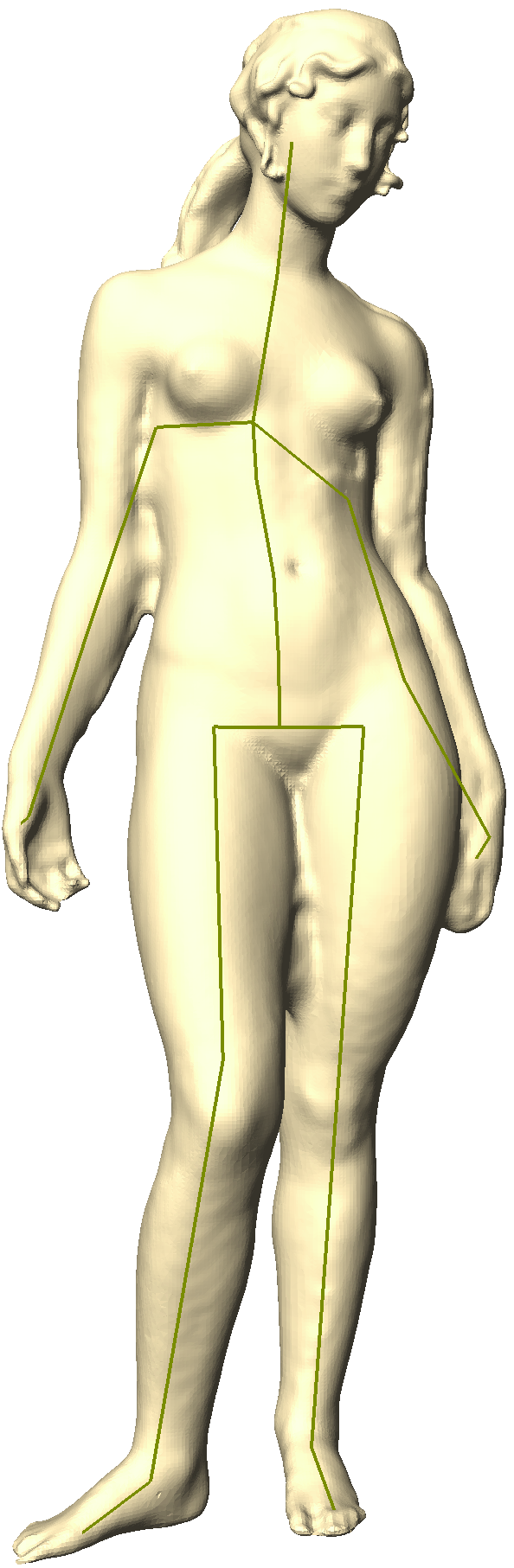}
        \caption{\scriptsize{Pinocchio (our skeleton)}}
    \end{subfigure}
    \begin{subfigure}[t]{0.19\linewidth}
        \centering
        \includegraphics[width=\textwidth]{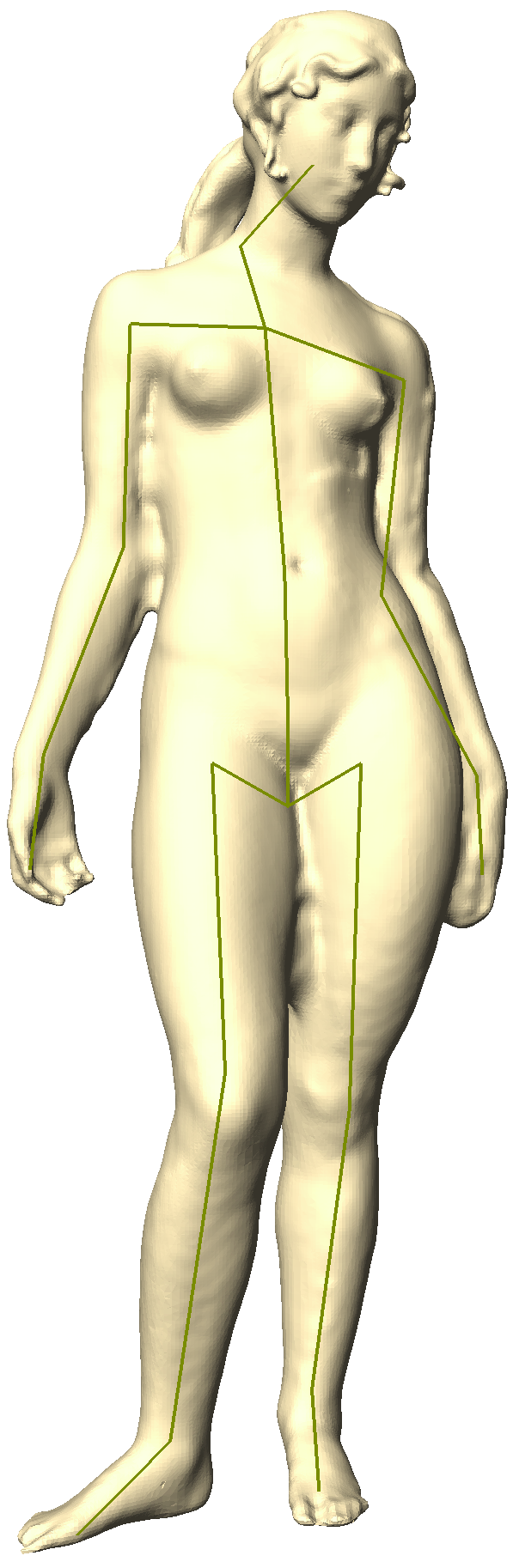}
        \caption{\scriptsize{FAKIR (1 iteration)}}
    \end{subfigure}
    \begin{subfigure}[t]{0.19\linewidth}
        \centering
        \includegraphics[width=\textwidth]{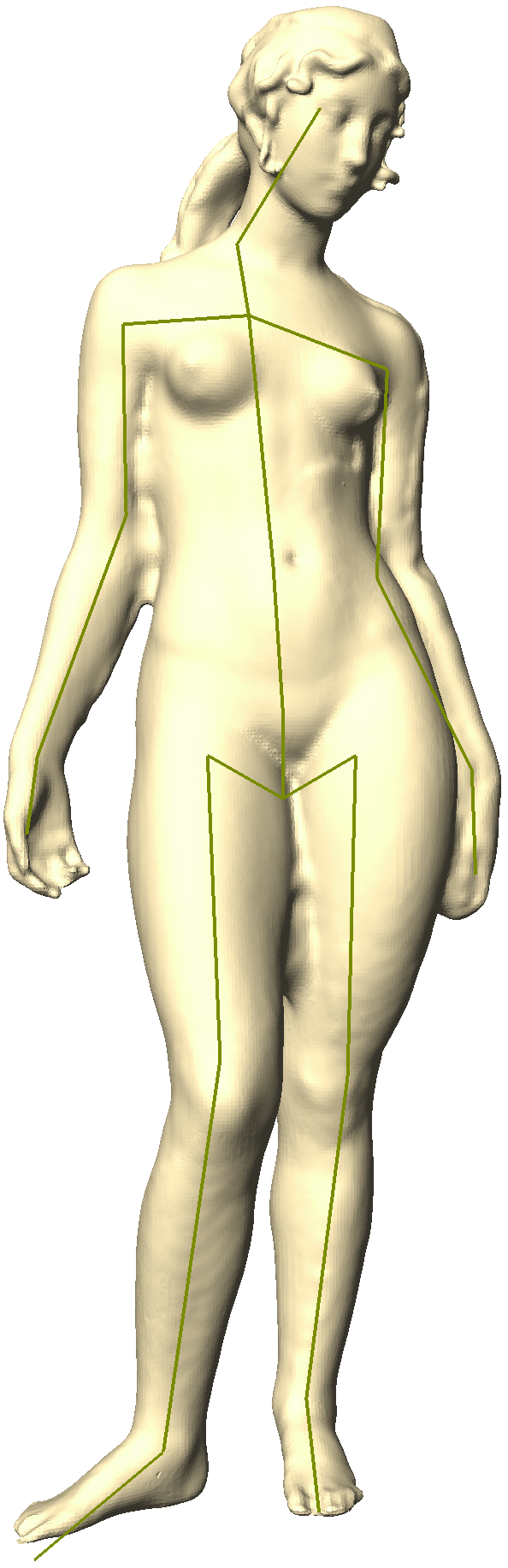}
        \caption{\scriptsize{FAKIR (10 iterations)}}
    \end{subfigure}
    \begin{subfigure}[t]{0.19\linewidth}
        \centering
        \includegraphics[width=\textwidth]{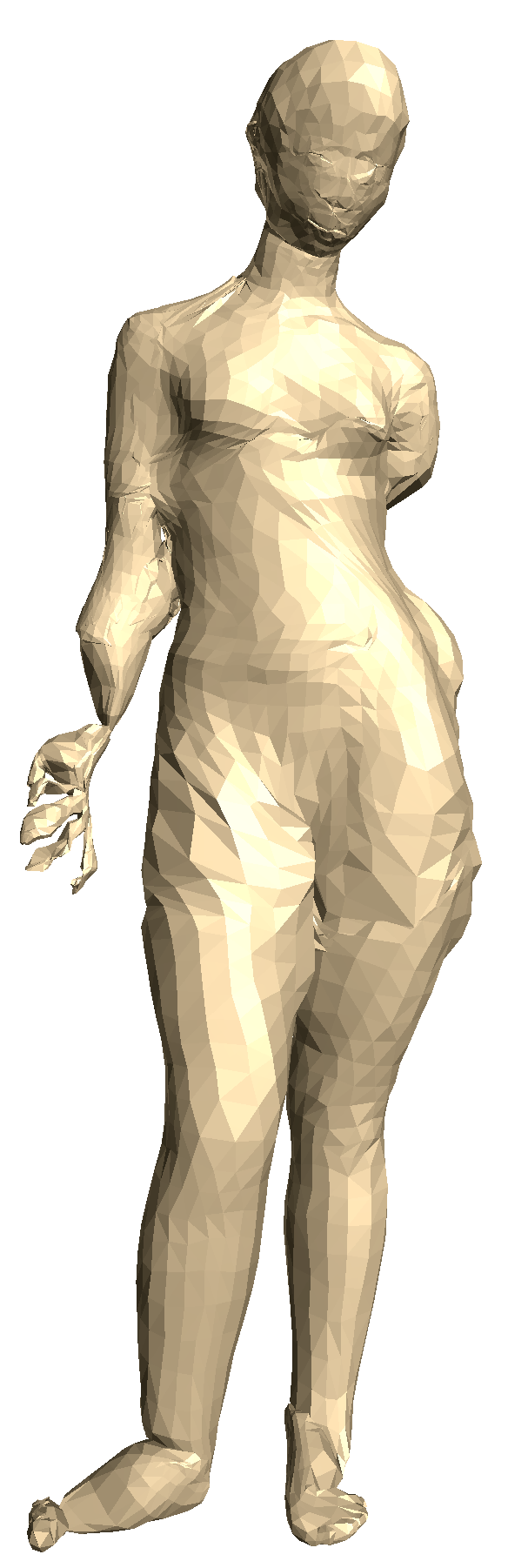}
        \caption{\footnotesize{FARM}}
    \end{subfigure}
    \caption{Comparison with Pinocchio~\cite{Baran07} and FARM~\cite{FARM} algorithm on the Aphrodite statue. From left to right: (a) Pinocchio with the Pinocchio-provided initial skeleton (17 bones); (b) Pinocchio with our initial skeleton (22 bones); (c) FAKIR with our initial skeleton after a single forward iteration; (d) FAKIR with our initial skeleton in 10 iterations. Only the skeleton is displayed since the bone radii are not taken into account by Pinocchio. (e)  Model reconstructed by the FARM method.}
    \label{fig:pinocchio}
\end{figure}

\begin{figure*}[ht]
   \centering
   \begin{tabular}{c | c | c | c | c}
        \includegraphics[width=0.10\textwidth]{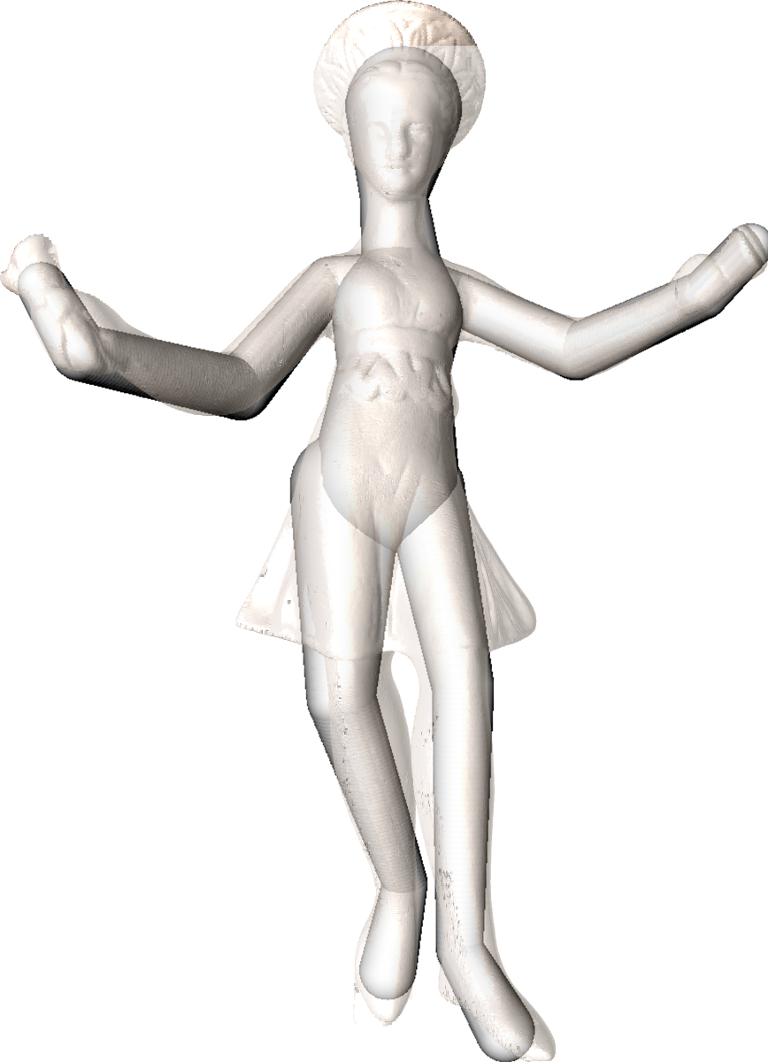}
        \includegraphics[width=0.05\textwidth]{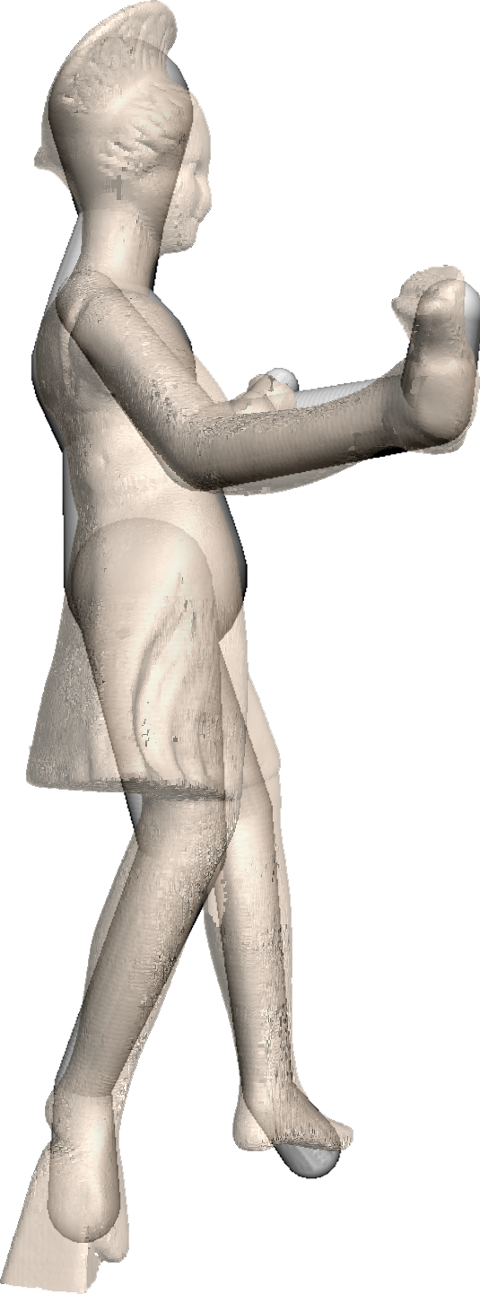} &
        \includegraphics[width=0.10\textwidth]{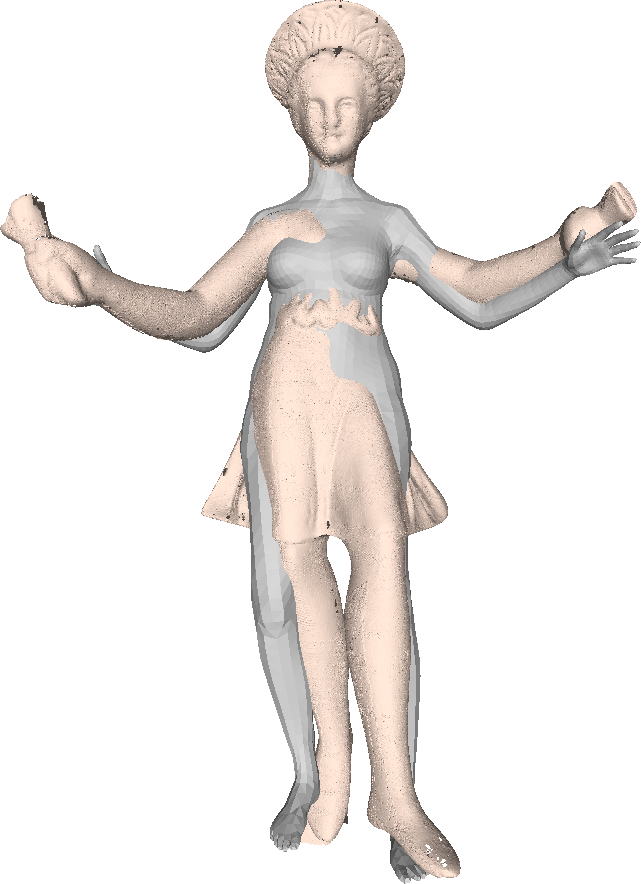}
        \includegraphics[width=0.05\textwidth]{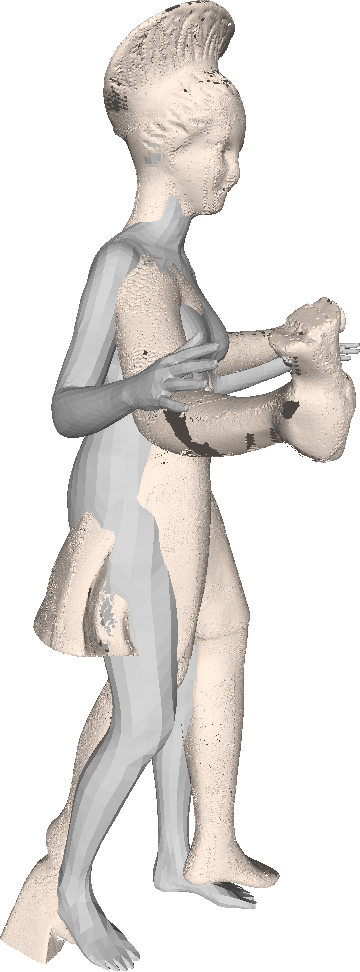} &
        \includegraphics[width=0.10\textwidth]{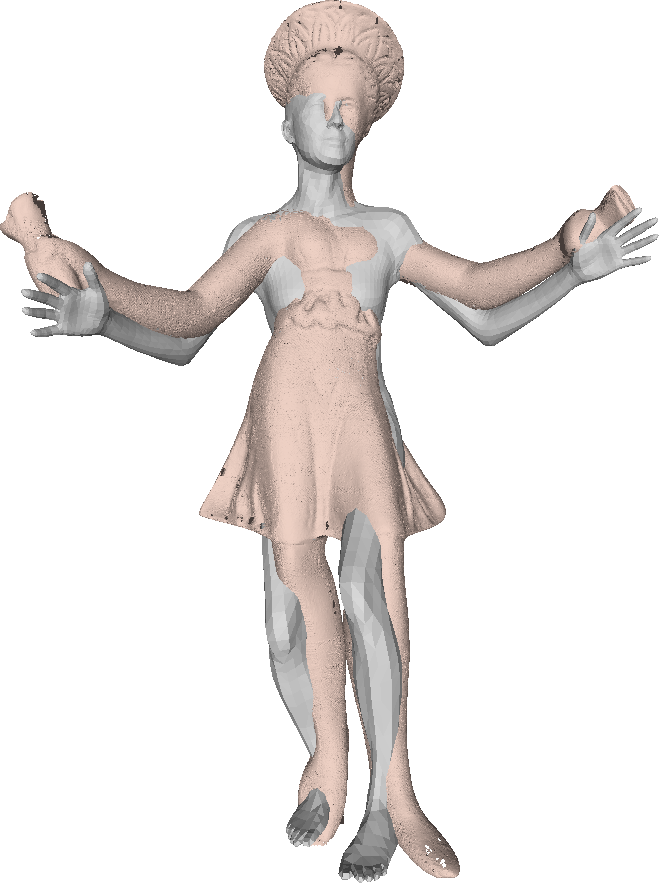}
        \includegraphics[width=0.06\textwidth]{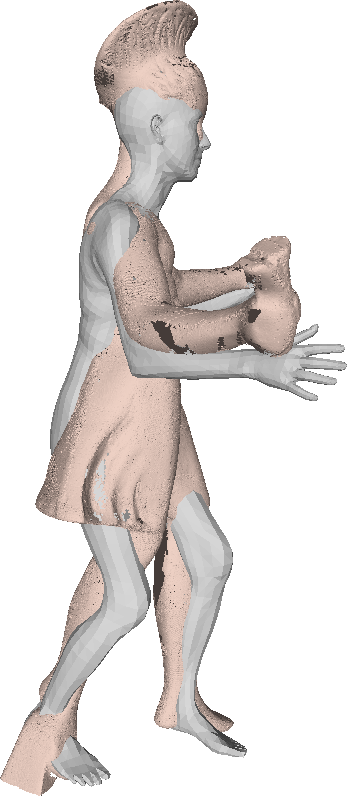} &
        \includegraphics[width=0.10\textwidth]{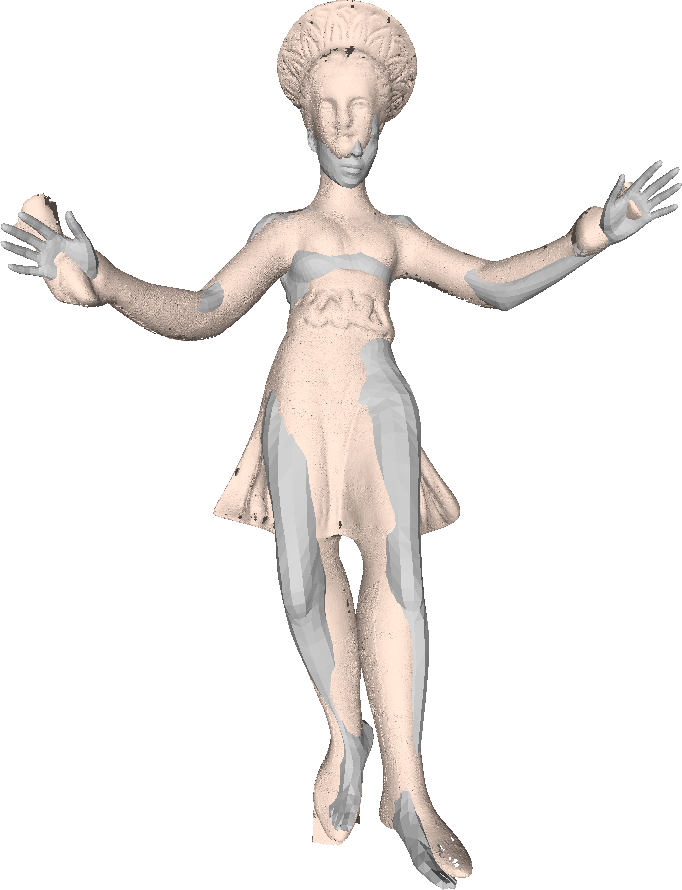}
        \includegraphics[width=0.055\textwidth]{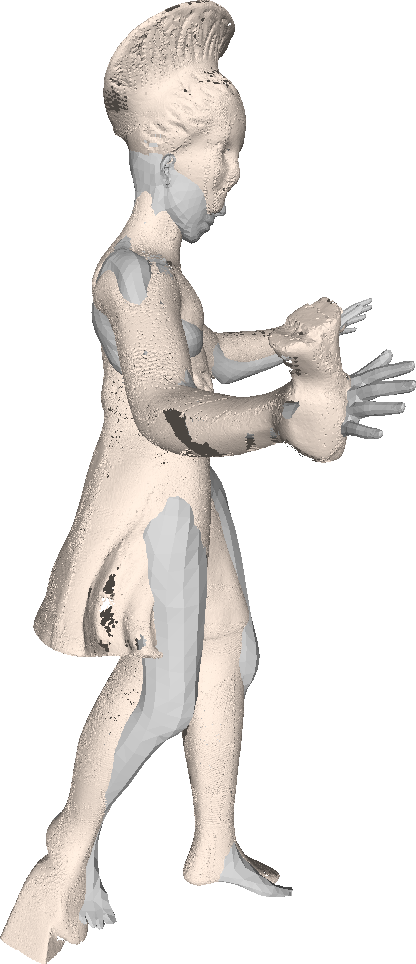} &
        \includegraphics[width=0.095\textwidth]{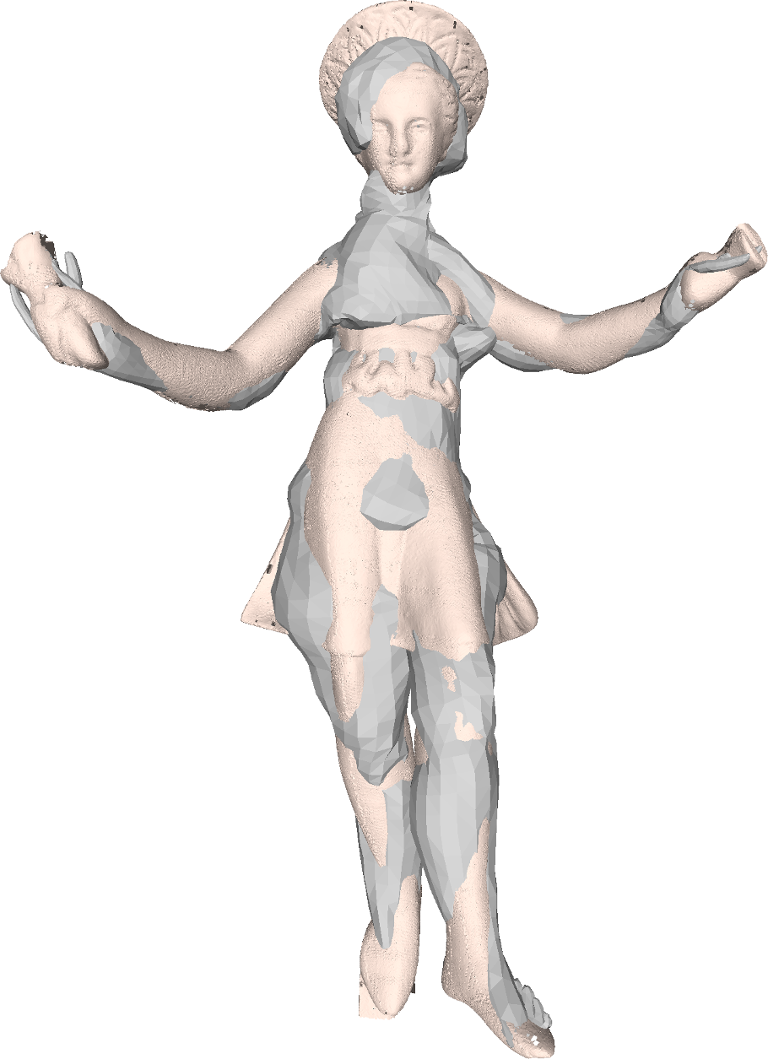}
        \includegraphics[width=0.048\textwidth]{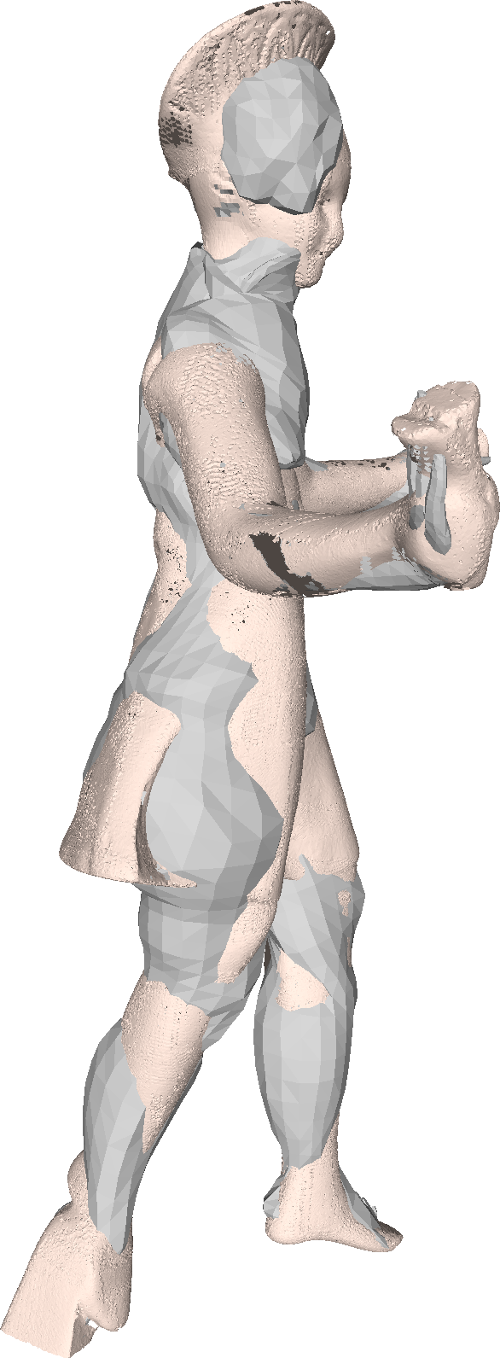}\\
        \includegraphics[width=0.10\textwidth]{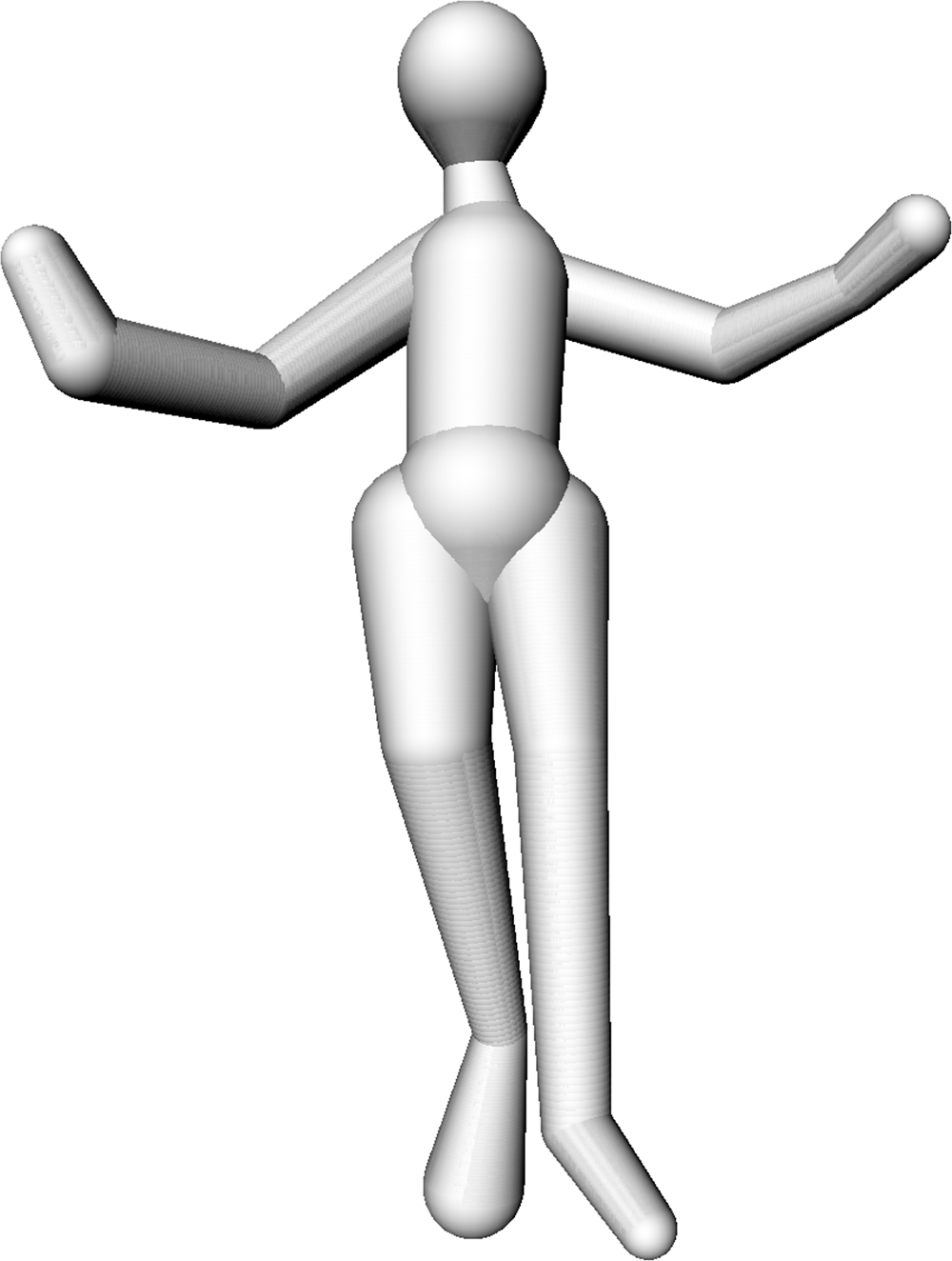}
        \includegraphics[width=0.05\textwidth]{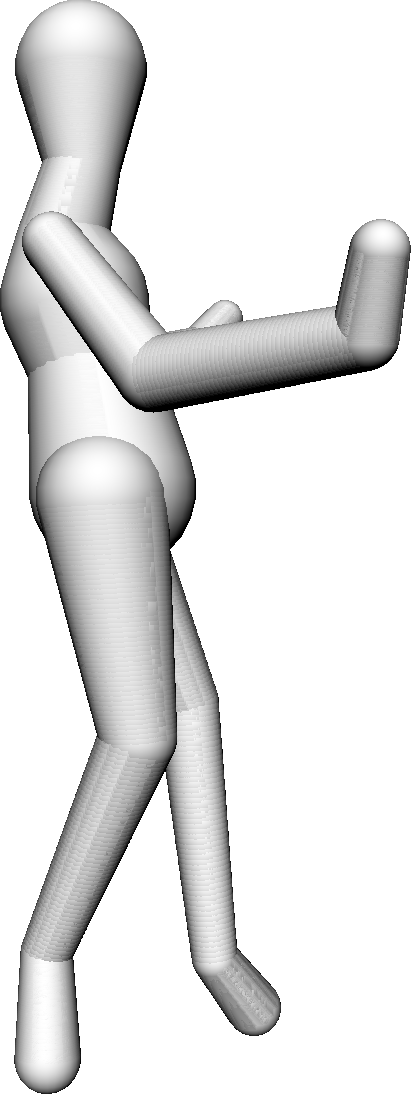} &
        \includegraphics[width=0.10\textwidth]{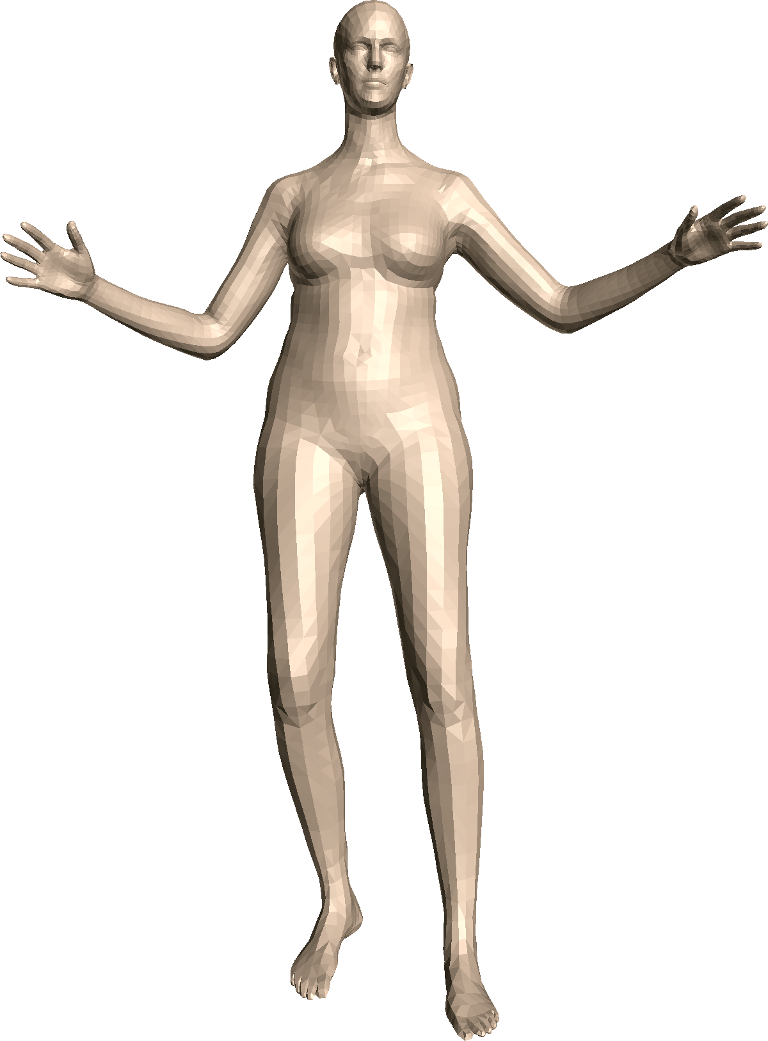}
        \includegraphics[width=0.045\textwidth]{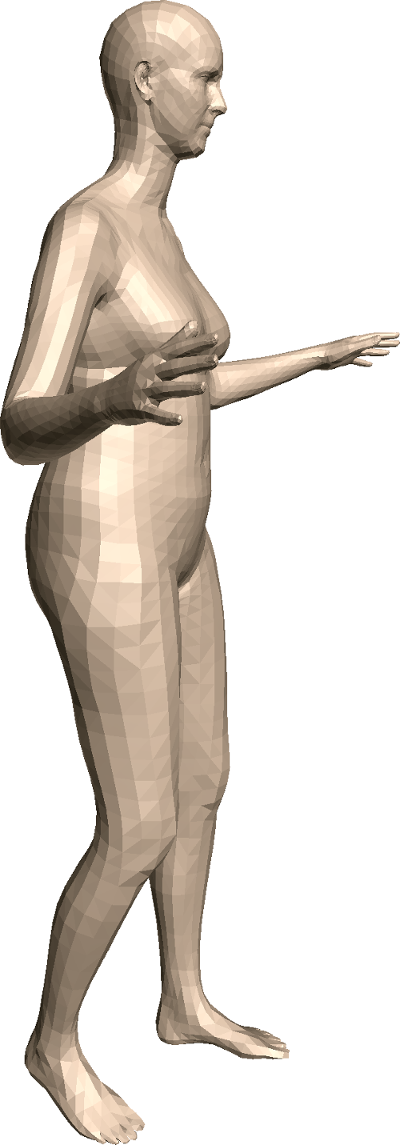} &
        \includegraphics[width=0.10\textwidth]{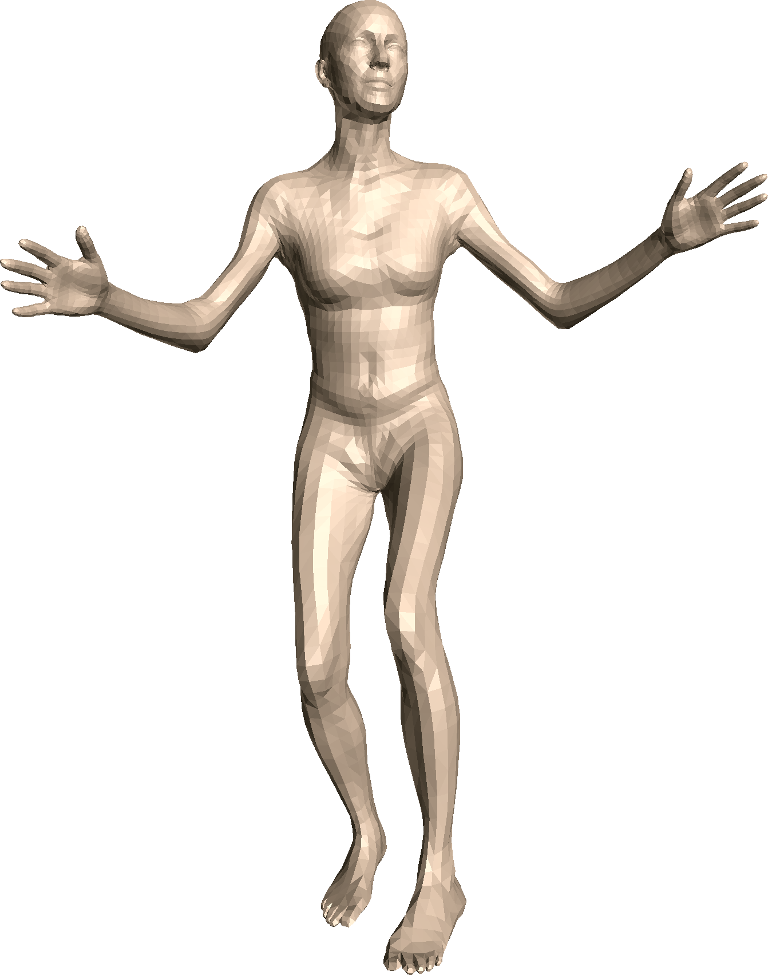}
        \includegraphics[width=0.055\textwidth]{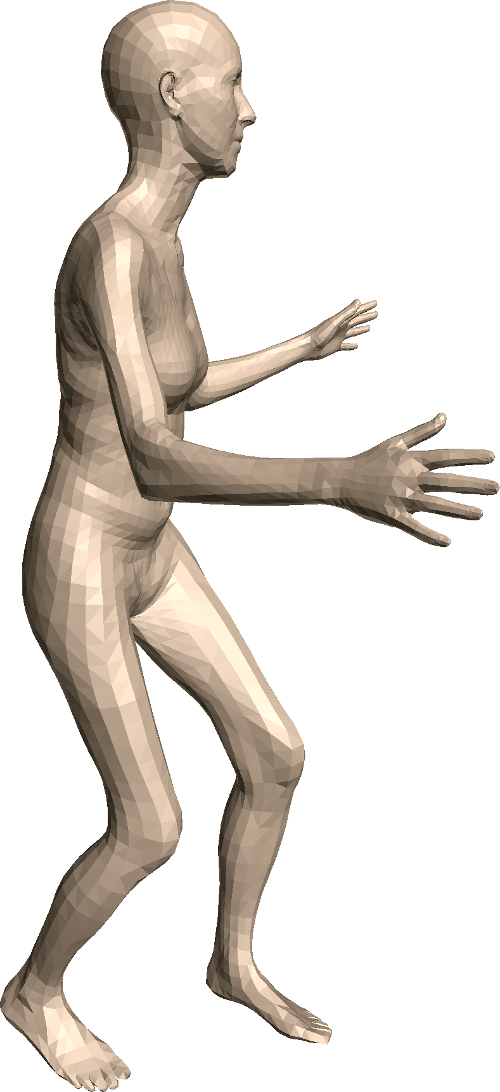} &
        \includegraphics[width=0.10\textwidth]{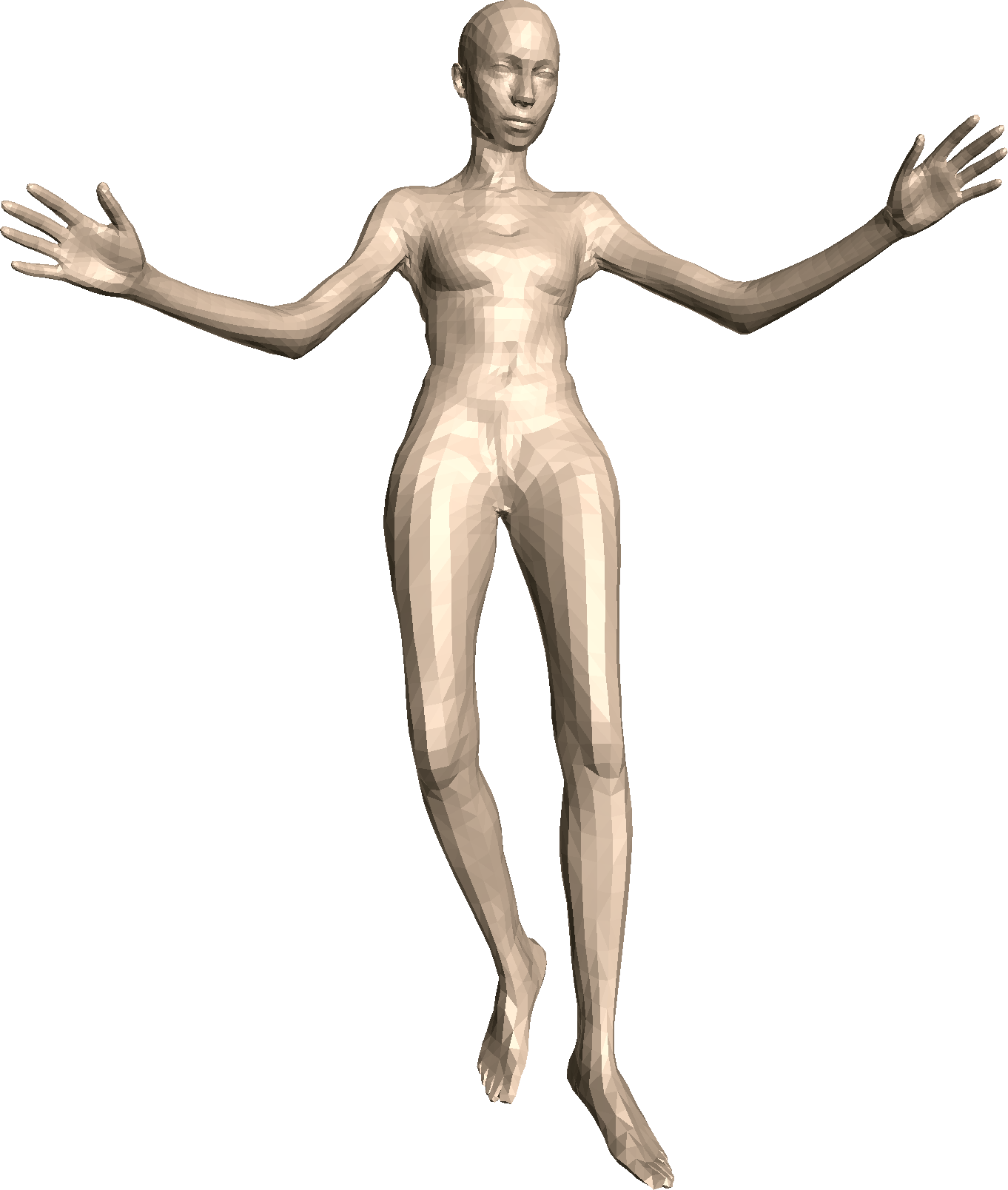}
        \includegraphics[width=0.05\textwidth]{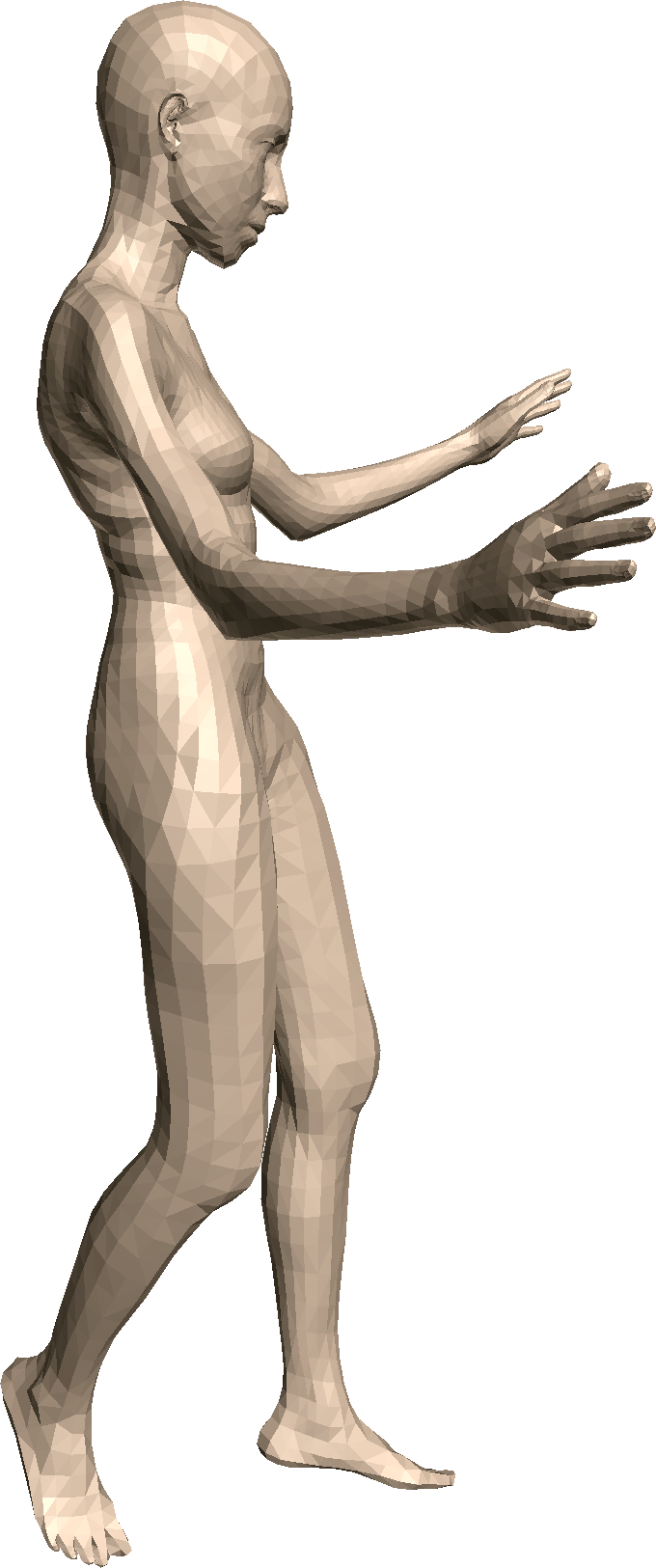} &
        \includegraphics[width=0.092\textwidth]{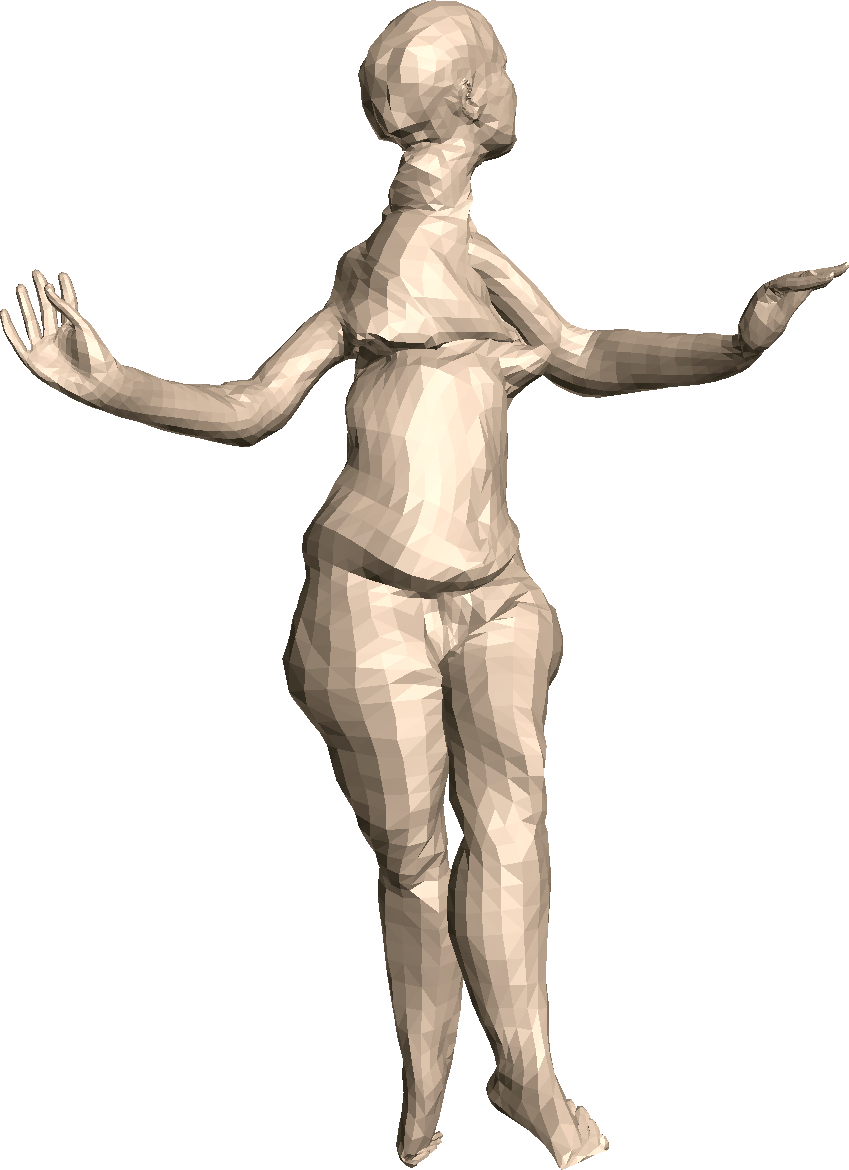}
        \includegraphics[width=0.042\textwidth]{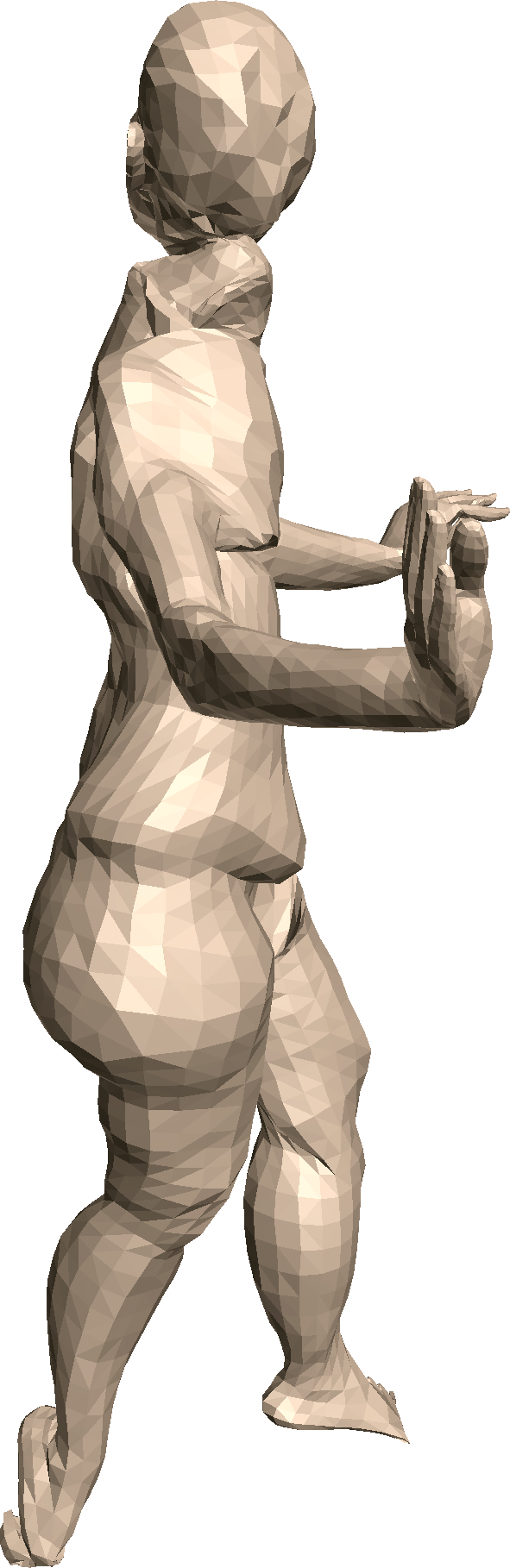} \\
        \multicolumn{1}{c}{\textbf{(a)} \textit{FAKIR}} & \multicolumn{1}{c}{\textbf{(b)} \textit{SMPLify + DeepCut 2D}} &  \multicolumn{1}{c}{\textbf{(c)} \textit{SMPLify + DeepCut 3D}} &
         \multicolumn{1}{c}{\textbf{(d)} \textit{SMPLify + FAKIR}} & \multicolumn{1}{c}{\textbf{(e)} \textit{FARM}}
   \end{tabular}

    \caption{Comparison with SMPLify on the Danseuse with Crotales. (a) FAKIR registration; (b) SMPLify using DeepCut predicted 3D joint positions on a single rendering; (c)  SMPLify using DeepCut predicted 3D joint positions on two rendered views; (d) SMPLify using 3D joint positions estimated by FAKIR; (e) FARM registration. (First row: overlayed registered model; Second row: registered model)}
    \label{fig:smpl_comp}

 \end{figure*}

While skeleton extraction is a much explored topic in geometry processing, we emphasize that extracting a computational geometry skeleton is very different than extracting an anatomical skeleton \cite{Huang2013,Tagliasacchi16}. Figure \ref{fig:compare_huang} show the $\ell^1$-medial skeleton extracted~\cite{Huang2013} on the Aphrodite and Danseuse with Crotales point sets -- to be compared with our results on Figures \ref{fig:pinocchio} and \ref{fig:smpl_comp}. This experiment shows that the geometrical skeleton definition is not enough for our purpose.

We compare FAKIR with Pinocchio~\cite{Baran07} in Figure \ref{fig:pinocchio}. The FAKIR algorithm yields a better skeleton registration, in particular for the shoulders and neck bones. As far as computation times are concerned, the Pinocchio method takes about $35s$ for a mesh with $138048$ vertices, which is roughly the same time as the $10$ iterations of FAKIR optimizing not only for the joint positions but also for the bone radii (38s). Furthermore, a single iteration of FAKIR takes 9s and already provides a better result with a much more plausible shoulders location. However it is important to note that the Pinocchio method does not require an initial skeleton position, while our method requires one of the joint to be not far from its optimal position (in this experiment we chose the pelvis joint).
 
We also compare compare FAKIR with the SMPLify method \cite{Bogo:ICCV:2015} (Figure \ref{fig:smpl_comp}). First we compute a  front-view rendering of the shape and run DeepCut \cite{Deepcut} to estimate the joint position used for SMPLify. The registration is clearly less accurate than ours (Figure \ref{fig:smpl_comp}b). We then extend SMPLify method to multi-view images using epipolar constraints to estimate a 3D joint positions from 2D joint positions obtained by DeepCut before applying SMPLify, which slightly improves the registration (Figure \ref{fig:smpl_comp}c). We also show that our articulated model regression can serve to initialize an SMPL model, leading to a better registration than with DeepCut (Figure \ref{fig:smpl_comp}d). However, the shape estimation of SMPL still cannot fit a statue with non-realistic body proportions.

Finally we compare FAKIR with the FARM method~\cite{FARM} a fully automatic method for registering an underlying model, such as SMPL, to a point set or a mesh. While this method exhibits excellent results on humanoid shapes, in the case of artistic statues, with irrealistic body proportions (Figure \ref{fig:pinocchio}) or with moderate garments (Figure \ref{fig:smpl_comp}), it cannot reconstruct a plausible model. Furthermore, scanned statues often exhibit inconsistent topologies, such as an arm glued to the body (Figure \ref{fig:pinocchio}) which is not handled by the FARM method.

\paragraph*{Computation Time}
The computational bottleneck of FAKIR lies in the assignment of each point several times during the optimization process. 
This assignment is updated after each bone parameter change.
However, the number of updates is related to the geometry of the surface and not to the number of sample points. Therefore, the overall complexity is linear with respect to the number of points. From an experimental point of view, FAKIR is a reasonably light algorithm: for a point cloud of $10 000$ points and the 22-bone human model, the first forward pass of FAKIR takes $2.5s$ and the computation time for one pass decreases to $1s$ in average afterwards.

\paragraph*{Limitations}
Despite its good results, FAKIR has some limitations. First, if two consecutive bones are aligned, their length estimation is not reliable, since the position of the middle joint is undetermined if no proportion constraint is set on the model. This limitation appears on the Aphrodite left leg (Figure \ref{fig:result}). We could also improve registration results by adding constraints of symmetry to the energy functions.  However these constraints should be quite loose, because of the unrealistic proportions of artistic statues. Furthermore, failure cases include some misalignments due to a local minimum (one of the arms of the mermaid in Figure \ref{fig:mermaid_res}, the feet of the shape on the top-right of Figure~\ref{fig:exp_tosca} and the back leg of the centaur in Figure \ref{fig:tosca_animals}.) Further local extrinsic refinement would improve the result. In addition, a process of local modification of the skeleton structure could be carried on, followed by an update of its registration.
Last, to avoid any manual intervention, we have tested an automatic approach by initializing the pelvis in the center of the bounding box of the points and by orienting it upwards. This initialization is effective for most standing statues (results provided in the supplementary material). Using principal component analysis to initialize the orientation of the body would also improve the registration for classical human poses (e.g. walking, sitting, lying down). However, the handling of the relaxation process at bifurcations could also be improved. Currently, the position of the pelvis is not updated by exploiting the whole information arising from the registration of all the incident branches. 

\section{Conclusion and perspectives}

We introduced a sphere-mesh anatomical model and a combined calibration and registration algorithm to estimate the anatomy and the pose of digitized archaeological statues. 
Our algorithm is useful when it is not possible to extract a shape template from a statistical analysis of examples representative of the diversity of poses and morphologies. 
Given the simplicity of our anatomical model, it is very easy to adapt it to many shapes, such as animals or imaginary creatures.  While our method already gives good results, a further improvement would be to handle the case of a clothed statue.

\paragraph*{Acknowledgments} 
This work was funded by the e-Roma project from the French Agence Nationale de la Recherche (ANR-16-CE38-0009).
The \emph{Dancer with crotales} model is a point set of the Farman Dataset \cite{ipol.2011.dalmm_ps}. The other 8 statues data are sampled on meshes from the Sketchfab website: the \emph{Venus de Milo} model is courtesy of Sketchfab user "tux", the \emph{Dancing Faun} model is courtesy of Moshe Caine and the other 5 statues models (\emph{Aphrodite}, \emph{Old Man Walking}, \emph{Esquiline Venus}, \emph{Old Fisherman} and \emph{Mermaid}) are courtesy of Geoffrey Marchal.
We thank Riccardo Marin for sharing his FARM registration results.
\bibliographystyle{eg-alpha}  
\bibliography{biblio}

\end{document}


\maketitle
\begin{abstract}
This supplementary material provides mathematical details for computing the distance and projection of a point onto a bone, and the optimization details for bone and joint parameters as they are performed in the FAKIR algorithm. It also provides additional FAKIR registration results and comparisons between registration obtained through our sequential optimization and a direct simultaneous optimization.
\end{abstract}  


\begin{appendices}
 
\section{One-bone distance computation \label{sec:1bonedist}}
\begin{figure}[ht]
     \centering
     \includegraphics[trim = 0mm 5mm 0mm 5mm, clip, width=0.8\linewidth]{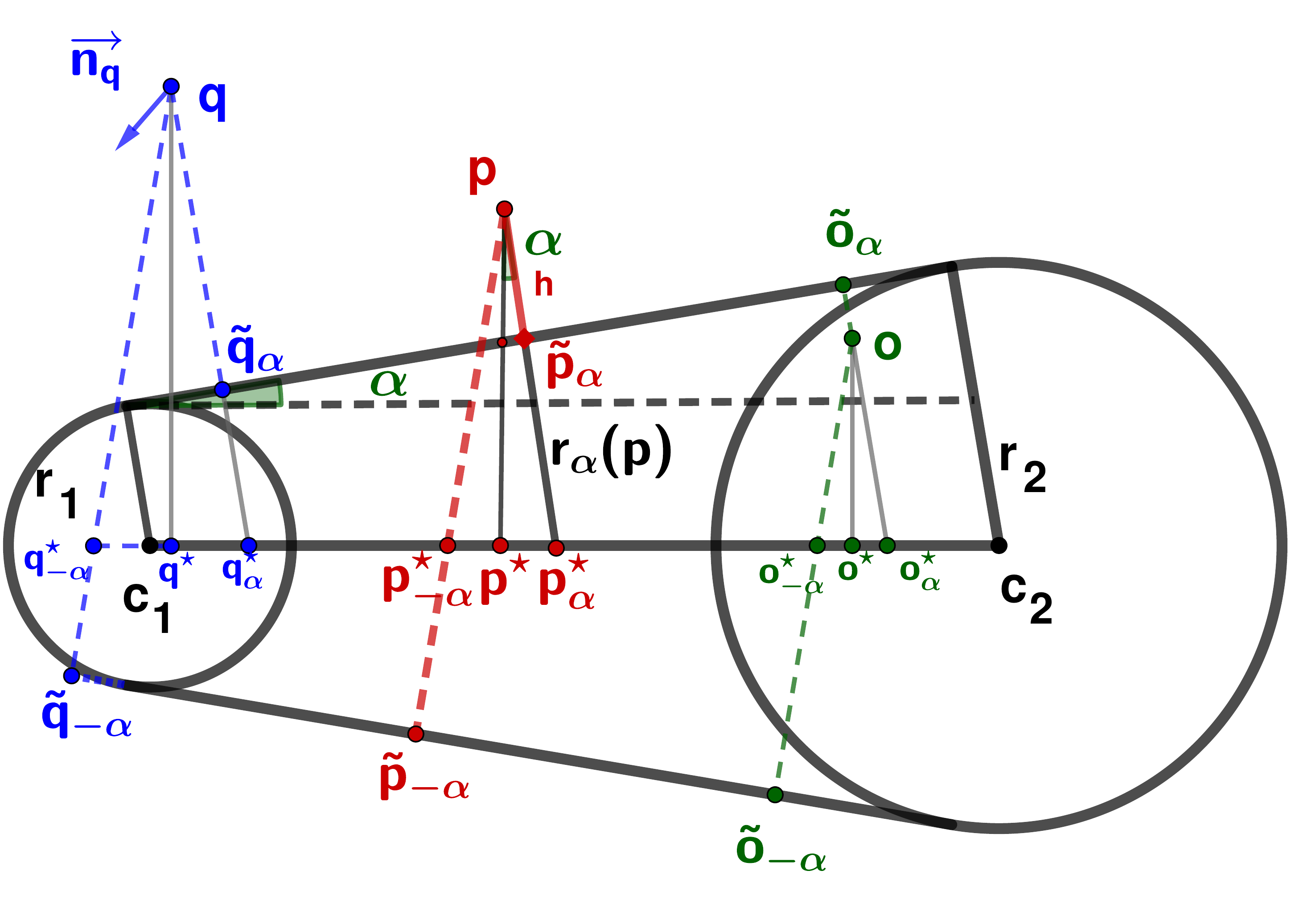}
     \caption{Various projection cases. $p$ has two possible projections $\tilde p_\alpha$ and $\tilde p_{-\alpha}$ depending on the orientation of the normal at $p$. Point $p^\star$ is the projection of $p$ on line $c_1c_2$. If the normal at $p$ is oriented upward $\tilde{p}=\tilde{p}_\alpha$. Otherwise, $\tilde{p}=\tilde{p}_{-\alpha}$. The same strategy is used to project points $q$ and $o$.}
     \label{fig:distance}
\end{figure}

In this appendix, we detail the projection of a point $p$ on a bone $B$ given an approximation of the oriented normal of point $p$. Instead of using the usual orthogonal projection on the bone, we constrain the projection $\tilde p$ to have a normal coherent with the one of $p$. This constraint is helpful when the bone lies far away from its corresponding point set: the point can then be projected on the ``right side'' of the bone. In the following, without loss of generality, let us assume $r_1 \leq r_2$. All the following computations depend on an angle $\alpha$ defined in Fig. \ref{fig:distance} and which can be expressed as $\alpha = \arctan\frac{|r_2-r_1|}{\sqrt{\|c_1c_2\|^2-(r_2-r_1)^2 }}$.
Let us first compute $p^{\star}$ the projection of $p$ on the oriented line $c_1c_2$, and two translations of these points along this line: $p^{\star}_\alpha$ at the distance $\|pp^{\star}\|\tan\alpha$ of $p^{\star}$ 
and $p^{\star}_{-\alpha}$ at the distance $-\|pp^{\star}\|\tan\alpha$,
as illustrated on Figure \ref{fig:distance}. Let $\tau_\alpha = \frac{p^{\star}_\alpha- c_1}{c_2 - c_1}$, so that 
$p^\star_\alpha$ can be expressed as $\tau_\alpha c_1 + (1-\tau_\alpha)c_2$. Different cases can occur:

\begin{itemize}
    \item $0<\tau_\alpha<1$: the point projects on the cone part of the bone. Let $\tilde{p}_\alpha$ be the intersection of segment $[p^{\star}_\alpha p]$ with the cone. $\tilde{p}_\alpha$ is the orthogonal projection of $p$ on the bone.
    If the normal to $\tilde p_\alpha$ has a positive scalar product with the normal of $p$, $\tilde p = \tilde p_\alpha$. Otherwise, normals are deemed inconsistent and $\tilde p = \tilde p_{-\alpha}$, \emph{i.e.} the farthest intersection of $pp^{\star}_{-\alpha}$ with the non-truncated cone. This situation occurs when the point $p$ is on the wrong side of the bone (i.e. its normal is inconsistent with the normal of its closest point on the bone).
    \item $\tau_\alpha<0$ (resp. $\tau_\alpha>1$): $\tilde{p}$ is the projection of $p$ on the sphere centered at $c_1$ (resp. $c_2$) with consistent normal direction, except
if this normal-constrained projection falls within the bone and not on the envelop. In that case, $p$ is on the wrong side of the bone, and we set $\tilde p = \tilde p_{-\alpha}$ on the other side of the non-truncated cone.
\end{itemize}

In any case, the distance between $p$ and its normal-constrained projection $\tilde{p}$ vanishes when $p$ is located near the surface of one bone, with a normal oriented consistently.
It may happen that the returned projection does not provide a point belonging to the surface of the bone: on Figure \ref{fig:distance}, $\tilde q_{-\alpha}$ is the normal-constrained projection of point $q$, but it is not on the surface of the bone. It corresponds to a case where the point is very far from the part of the bone which is coherent with its normal. During the registration process, $\tilde q_{-\alpha}$ will attract $q$ on the other side of the bone, so that the projection point will gradually be replaced by a more consistent one.

For completeness, let us express unsigned distance $d(p)=\|p - \tilde p\|$ in the various cases since they will be required in the following Levenberg-Marquardt optimization formulations. If $\tau_\alpha<0$ (resp. $\tau_\alpha>1$), $d(p)=\|c_1p\|-r_1$ (resp. $d(p)=\|c_2p\|-r_2$). If  $0\leq\tau\leq1$:

\begin{equation}
d(p) =
    \begin{cases}
    \|pp_\alpha^\star\|-r_{\alpha}(p)       & \text{ if } n_{\tilde{p}}\cdot n_p>0\\
    \|pp_{-\alpha}^\star\|+r_{-\alpha}(p)  & \text{ if }n_{\tilde{p}}\cdot n_p\leq0
    \end{cases}
\end{equation}

Since the radius of the cone varies linearly along line $c_1c_2$:

\begin{equation}
\begin{split}
  r_{\alpha}(p)=& \|\tilde{p_\alpha}p_\alpha^\star\| = (1-\tau_\alpha(p))r_1  +  \tau_\alpha(p) r_2 \\
  r_{-\alpha}(p) =& \|\tilde{p_{-\alpha}}p_{-\alpha}^\star\| = (1-\tau_{-\alpha}(p))r_1 + \tau_{-\alpha}(p) r_2 
 \end{split}
\end{equation}
with: $\tau_\alpha(p) = \frac{c_1p^\star_\alpha\cdot c_1c_2}{\|c_1c_2\|^2}$ and $\tau_{-\alpha}(p) = \frac{c_1p^\star_{-\alpha}\cdot c_1c_2}{\|c_1c_2\|^2}$. Furthermore $\|pp_\alpha^\star\| =\|pp_{-\alpha}^\star\| = \|pp^\star\|/\cos\alpha$.
Hence, for each bone, we first compute the $\alpha$ angle, then, for each point $p$, we compute its projection $p^\star$ on $c_1c_2$ and the corresponding $\tau_\alpha(p)$ yielding $r_{\pm\alpha}(p)$ and $\tilde p_{\pm\alpha}$.

\section{Optimization for one bone \label{annex:1bone_optim}}
Let us assume that $c_1$ (Fig. \ref{fig:distance}) is fixed and let us optimize for the pose and intrinsic parameters of bone $B$. In a local reference frame centered at $c_1$ with x-axis aligned with $c_1c_2$, $c_1$ has coordinates $(0,0,0)$ and $c_2$ has initial coordinates $(l,0,0)$. The rotation of the bone can be parameterized by a rotation of angle $\theta_1$ around the y-axis followed by a rotation of angle $\theta_2$ around the z-axis. The one-bone energy is invariant by rotation around the x-axis. After the double rotation, $c_2$ has coordinates $(l\cos\theta_2 \cos\theta_1,l\sin\theta_2,l\cos\theta_2\sin\theta_1)$.
Let us call $(x,y,z)$ the coordinates of point $p$ in this local coordinate system and express $d(p)$ with respect to parameters $\bm{\theta}=(\theta_1,\theta_2)$,$l$ and $\bm{r}=(r_1,r_2)$. We have:

$$\tan\alpha = \frac{r_2-r_1}{\sqrt{l^2-(r_2-r_1)^2}}, \cos\alpha = \frac{\sqrt{l^2-(r_2-r_1)^2}}{l}$$
$$\|c_1p\|^2=x^2+y^2+z^2$$ $$\|c_1p^\star\|=x\cos\theta_2\cos\theta_1+y\sin\theta_2+z\cos\theta_2\sin\theta_1$$
$$\|p^\star p\|^2=x^2+y^2+z^2-(x\cos\theta_2\cos\theta_1+y\sin\theta_2+z\cos\theta_2\sin\theta_1)^2$$
$$\|p^\star p_\alpha^\star\|=\|p^\star p\|\tan\alpha$$
$$\|p_\alpha^\star p\|=\frac{\|p^\star p\|}{\cos\alpha}$$
$$\tau_{\pm\alpha}(p) = \frac{\|c_1p^\star\|\pm\|p^\star p^\star_\alpha\|}{l}$$
$$\|c_2p\|^2=(x-l\cos\theta_2\cos\theta_1)^2+(y-l\sin\theta_2)^2+(z-l\cos\theta_2\sin\theta_1)^2$$

The one-bone energy function is (dropping the $k$ subscript for simplicity):
\begin{equation}
E(P,B(l,\bf{r}),\boldsymbol{\theta}) = \sum_{p\in P} d(p)^2
\end{equation}
The optimization is performed on three set of parameters in turn: angles $\bm{\theta}$, bone length $l$ and bone radii $\bm r$. 

the optimization for bone $B$ with respect to $\boldsymbol{\theta}$ writes:
\begin{equation}
\hat{\boldsymbol{\theta}}\equiv\argmin_{\boldsymbol{\theta}} E(P,B(l,\mathbf{r}),\boldsymbol{\theta})=\argmin_{\boldsymbol{\theta}}\sum_{p\in P}d(p,\boldsymbol{\theta})^2
\end{equation}

Following the Levenberg-Marquardt algorithm, at each iteration, parameter $\boldsymbol{\theta}$ is replaced by a new estimate $\boldsymbol{\theta}+\delta\boldsymbol{\theta}$, computed as:
\begin{equation}
\argmin_{\boldsymbol{\theta}}E(P,B(l,\mathbf{r}),\boldsymbol{\theta})
\approx \argmin_{\delta\boldsymbol{\theta}}E(P,B(l,\mathbf{r}),\boldsymbol{\theta}+\delta\boldsymbol{\theta})
\end{equation}

which is computed by taking:
$$\frac{\partial
E(P,B(l,\mathbf{r}),\boldsymbol{\theta}+\delta\boldsymbol{\theta})}
{\partial \delta\boldsymbol{\theta}}=\bold{0}$$
We finally get $\delta\boldsymbol{\theta}$: 
$$\delta\boldsymbol{\theta}=-[J^TJ+\lambda diag(J^TJ)]^{-1}J^T \bold{g}(\boldsymbol{\theta})$$

where $J=[J_1,J_2]$, $J_{i1}=\frac{\partial d(p_i)}{\partial \theta_1}$ and $J_{i2}=\frac{\partial d(p_i))}{\partial \theta_2}$ and $\bold g(\theta)$ is a column vector whose entries are $d(p,\boldsymbol\theta)$ for each point $p$. $\lambda$ is a damping factor set to $0.01$ initially and adapting it throughout iterations. 
    
In the following, we assume $0<\tau_\alpha(p)<1$ and $n_{\tilde{p}}\cdot n_p>0$. In this case, $p$ projects on $\tilde p_\alpha$ and $d(p) =\|pp_\alpha^\star\| - r_{\alpha}(p)$ with $r_\alpha(p) = (1 - \tau_\alpha(p)) r_1 + \tau_\alpha(p)r_2$, and $\tau_\alpha(p) = \frac{\|c_1p_\alpha^\star\|}{l}$. Hence:

\begin{equation}
\frac{\partial d(p)}{\partial \theta_1} = \frac{1}{\cos\alpha}\frac{\partial\|p^\star p\|}{\partial\theta_1} + (r_2-r_1)\frac{1}{l}(\frac{\partial\|c_1p^\star\|}{\partial\theta_1}+\tan\alpha\frac{\partial \|p^\star p\|}{\partial \theta_1})
\end{equation}

\begin{equation}
\frac{\partial d(p)}{\partial \theta_2} = \frac{1}{\cos\alpha}\frac{\partial\|p^\star p\|}{\partial\theta_2} + (r_2-r_1)\frac{1}{l}(\frac{\partial\|c_1p^\star\|}{\partial\theta_2}+\tan\alpha\frac{\partial \|p^\star p\|}{\partial \theta_2})
\end{equation}

The full expression for the derivatives can be easily derived given the expressions for $\|p_\alpha^\star p\|$, $\|c_1p^\star\|$, $\|p^\star p\|$ above.
The cases $\tau_\alpha(p)<0$, $\tau_\alpha(p)>1$ or $n_{\tilde{p}}\cdot n_p<0$ can be computed similarly.

\section{Optimization for a joint between two consecutive bones \label{annex:2bones_optim}} 

\begin{figure}
        \begin{subfigure}[ht]{0.45\linewidth}
            \includegraphics[trim = 0mm 130mm 0mm 0mm, clip, width=\linewidth]{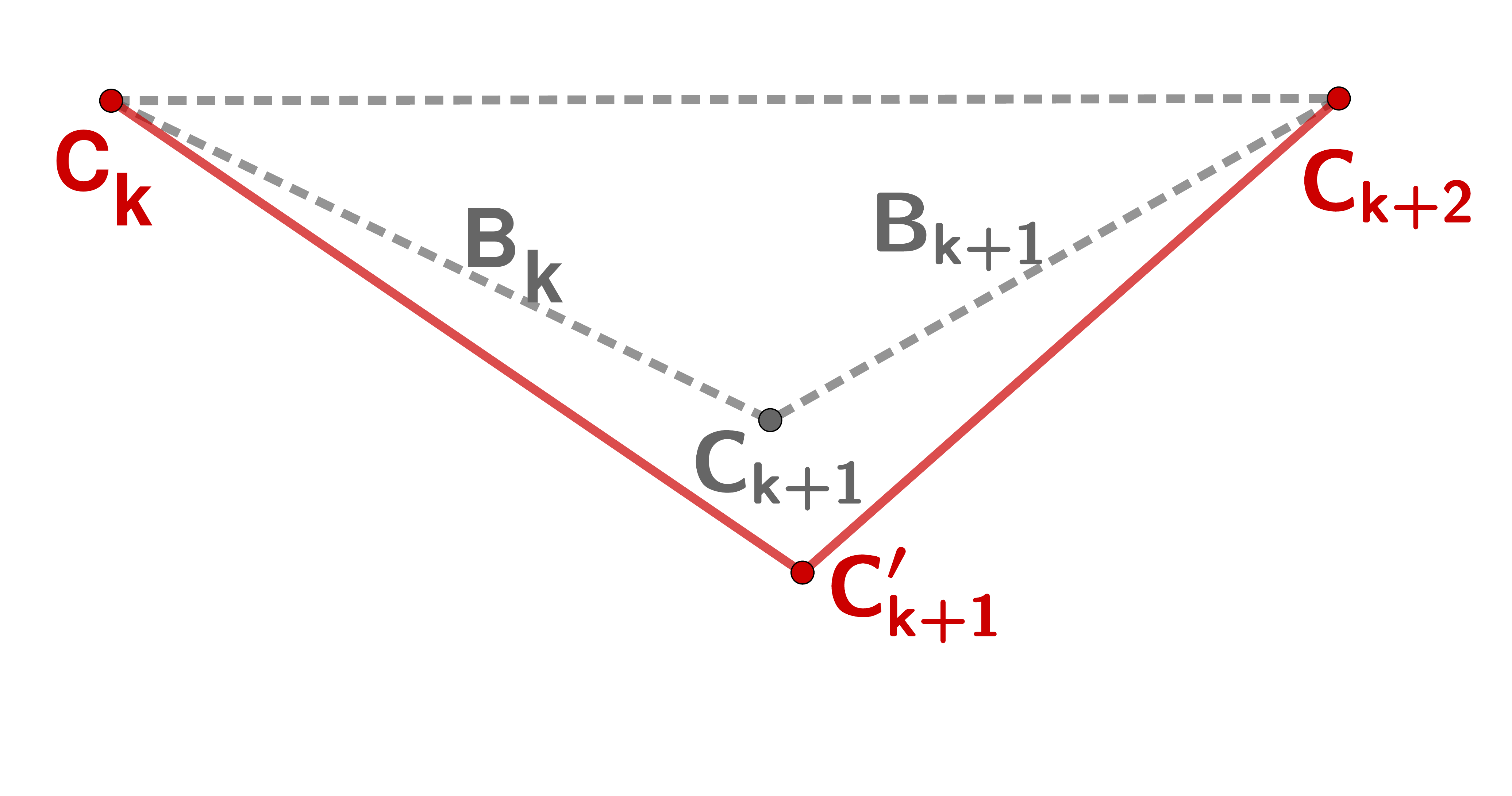}
            \caption{Rotate $B_k$ $\cup$  $B_{k+1}$ }
            \label{fig:2b_1}
        \end{subfigure}
        \begin{subfigure}[ht]{0.45\linewidth}
            \includegraphics[trim = 0mm 130mm 0mm 0mm, clip,width=\linewidth]{2bones_2}
            \caption{Refine length $l_k$ of $B_k$}
            \label{fig:2b_2}
        \end{subfigure}
    
        \begin{subfigure}[ht]{0.45\linewidth}
            \includegraphics[trim = 0mm 55mm 0mm 0mm, clip,width=\linewidth]{2bones_3}
            \caption{Refine length $l_{k+1}$ of $B_{k+1}$}
            \label{fig:2b_3}
        \end{subfigure}
         \begin{subfigure}[ht]{0.45\linewidth}
            \includegraphics[trim = 0mm 55mm 0mm 0mm, clip,width=\linewidth]{2bones_4}
            \caption{Refine radius $r_{k+1}$}
            \label{fig:2b_4}
        \end{subfigure}
    \caption{Pairwise Optimization. With fixed extremities $c_{k}$ and $c_{k+2}$, the pair of bones $B_k$ and $B_{k+1}$ is first rotated around axis $c_{k}c_{k+2}$ in order to minimize the two-bones energy. Then the lengths of the bones $B_k$ and $B_{k+1}$ and their common radius $r_{k+1}$ are optimized successively. After these updates, the point-to-bone assignment is recomputed. As the process is repeated the distances are more accurate since the point-to-bone assignment becomes more meaningful.}
    \label{fig:2bones}
\end{figure}

Let us consider the geometric optimization of the center of the joint between two bones by optimizing the two-bones energy with respect to the lengths $l_k$ and $l_{k+1}$. Each length is optimized in turn, with a side-effect on the value of the other length. The two-bones energy can be expressed as a function of $l_k$:  
\begin{equation}
E_{(k,k+1)}(l_k) = \sum_{p\in P_k} \|\tilde{p}_k -p\|^2 + \sum_{p\in P_{k+1}} \| \tilde{p}_{k+1} -p\|^2
\end{equation}

Following the Levenberg-Marquardt algorithm, at each iteration, each parameter $l_k$ is replaced by a new estimate $l_k+\delta l$:
\begin{equation}
    \argmin_{l_k} E_{(k,k+1)}(l_k)\approx \argmin_{\delta l}~  E_{(k,k+1)}(l_k+\delta l)
\end{equation}

By setting $\frac{\partial E_{(k,k+1)}(l_k+\delta l)}{\partial \delta l}=0$, we get:

\begin{equation}
\delta l = -\frac{\sum_{p\in P_k}d_k\frac{\partial d_k}{\partial l_k}+\sum_{p\in P_{k+1}}d_{k+1}\frac{\partial d_{k+1}}{\partial l_k}}{\sum_{p\in P_k}(\frac{\partial d_k}{\partial l_k})^2+\sum_{p\in P_{k+1}}(\frac{\partial d_{k+1}}{\partial l_k})^2}
\label{eq:deltal}
\end{equation}
where $d_k = \|p - \tilde p_k\|$ and $d_{k+1} = \|p - \tilde p_{k+1}\|$ are expressed as functions of $l_k$.

Let us detail the expression of $d_k$ with respect to $l_k$: during the pairwise optimization $c_k$ and $c_{k+2}$ remain fixed (Figure \ref{fig:2bones}).
Let $c_k$ be the origin of a local reference frame  with the x-axis aligned with $c_kc_{k+1}$.
In this frame, the coordinates write $c_k(0,0,0)$, $c_{k+1}(l_k,0,0)$ and $c_{k+2}(x_2,y_2,z_2)$ while a point $P$ has coordinates $(x,y,z)$. Then $c_{k+1}c_{k+2} = (x_2-l_k,y_2,z_2)$, $c_{k+1}p = (x-l_k,y,z)$.

Let us assume that $p$ projects on $\tilde p_\alpha$ (the case $\tilde p_{-\alpha}$ can be deduced with minor changes).
Using the same notation as in Figure \ref{fig:distance} and appendix \ref{annex:1bone_optim}, recall that $d_k =  \|p - \tilde p_k\| = \|pp^\star_{\alpha}\| - r_\alpha(p)$. Since when optimizing $l_k$ the orthogonal projection on $c_kc_{k+1}$ does not change, $\|pp^\star\|$ remains the same. However both $\alpha$ and $r_\alpha(p)$ change. Since $r_\alpha(p)= (1-\tau_\alpha(p)) r_k + \tau_\alpha(p)r_{k+1}$ with $\tau_\alpha(p)=\frac{\|c_kp_\alpha^\star\|}{l_k}$, we get:

\begin{equation}
\frac{\partial d_k}{\partial l_k} = -\frac{\|pp^\star\|}{\cos^2\alpha} \frac{\partial \cos\alpha}{\partial l_k} - (r_{k+1}-r_{k})\frac{\partial \tau_\alpha(p)}{\partial l_k}
\end{equation}

Simple geometric considerations give $\cos\alpha = \sqrt{1-\frac{(r_{k+1}-r_k)^2}{l_k^2}}$, $\tau_\alpha(p)=\frac{\|c_kp^\star\| + \|pp^\star\|\tan\alpha}{l_k}$ and $\tan\alpha = \frac{r_{k+1}-r_k}{\sqrt{l_k^2 - (r_{k+1} - r_k)^2}}$, whose differentiation with respect to $l_k$ is easy. 

One must also express distances $d_{k+1}$ as functions of $l_k$. In that case, the projection on bone $B_{k+1}$ is slightly different, since the position of point $c_{k+1}$ changes with $l_k$. The formulas are only slightly modified by it, but this time $\|pp_\star\|$ also depends on $l_k$. We get:
\begin{equation}
\frac{\partial l_{k+1}}{\partial l_k} = \frac{1}{\cos\alpha}\frac{\partial \|pp^\star\|}{\partial l_k}-\frac{\|pp^\star\|}{\cos^2\alpha} \frac{\partial \cos\alpha}{\partial l_k} - (r_{k+2}-r_{k+1})\frac{\partial \tau_\alpha}{\partial l_k}
\end{equation}

The full expression for the derivatives can be easily computed using the following formulas:

$$\cos\alpha = \sqrt{1 - \frac{(r_{k+2} - r_{k+1})^2}{(x_2-l_k)^2+y_2^2+z_2^2}}$$

$$\tau_\alpha(p) = \sqrt{\frac{(x-l_k)^2+y^2+z^2}{(x_2-l_k)^2+y_2^2+z_2^2}}$$

$$\|c_{k+1}p^\star\| = \frac{c_{k+1}p\cdot c_{k+1}c_{k+2}}{\|c_{k+1}c_{k+2}\|} = \frac{(x-l_k)(x_2-l_k)+yy_2+zz_2}{(x_2-l_k)^2+y_2^2+z_2^2} $$

Plugging all the derivatives in Equation \ref{eq:deltal} yields $\delta l$, and $l_k$ can be updated as $\hat{l}_k = l_k+\delta l$. This impacts the position of $c_{k+1}$, whose new position is computed as $\hat c_{k+1} = c_k + \hat l_k \frac{c_kc_{k+1}}{c_kc_{k+1}}$, and $l_{k+1}$ is recomputed as : $l_{k+1} = \|\hat c_{k+1}c_{k+2}\|$.

The two-bones energy $E_{k,k+1}$ is then optimized with respect to $l_{k+1}$. This optimization is symmetric to the $l_k$ case above and can be easily adapted. Finally, the optimization of the radius of the common joint and rotation angle around axis $c_kc_{k+2}$ are done in a similar manner.

\section{Importance of the optimization order for registering a chain of bones. }
During our optimization process our approach takes advantage of the articulated property of our model by processing bones in a specific order. Here, we run an experiment to illustrate that the order in which the optimizations are made is crucial.

To do so, we replace our iterations of sequential optimizations followed by point to bone reassignment by iterations of simultaneous parameter optimizations followed by point reassignment. At each iteration, the positions of all the joints are simultaneously optimized by minimizing their two-bone energies as if the adjacent joints remained fixed. The free joints at the extremities are also optimized at the same time by minimizing their one-bone energy as if their non free joint remaining fixed. However the parameter and pose change is not applied right away after each optimization but simultaneously once all updates have been computed.

\begin{figure*}[ht]
    \begin{center}
        \begin{subfigure}[ht]{0.15\linewidth}
            \includegraphics[width=\linewidth]{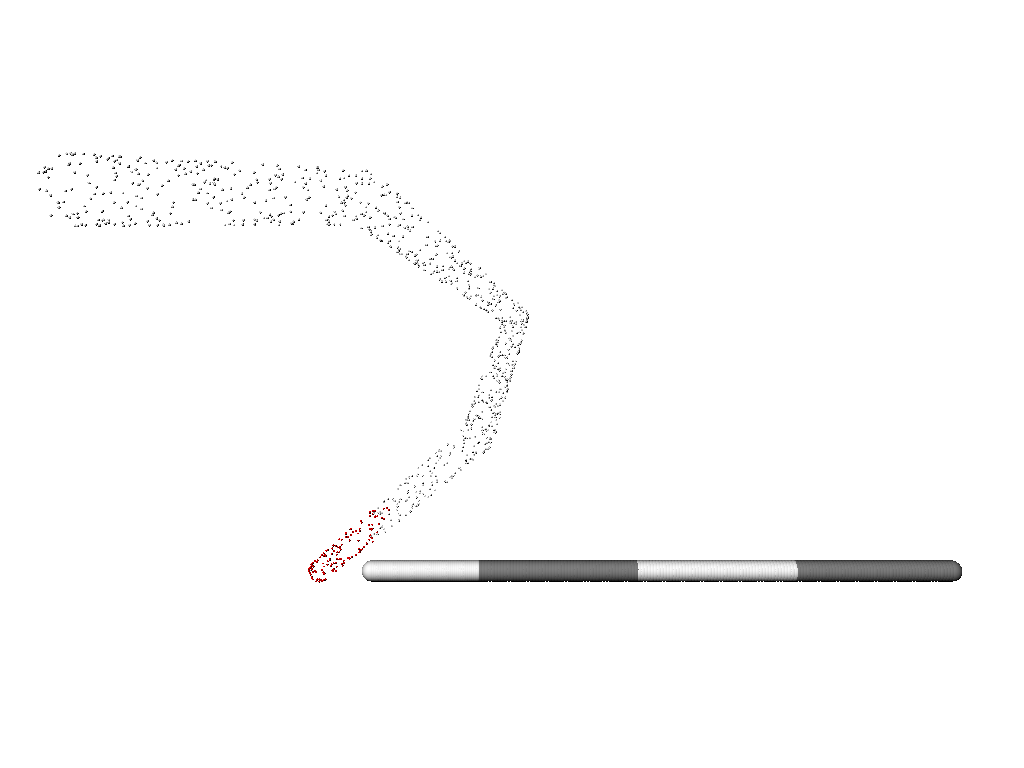}
        \end{subfigure}
        ~
        \begin{subfigure}[ht]{0.15\linewidth}
            \includegraphics[width=\linewidth]{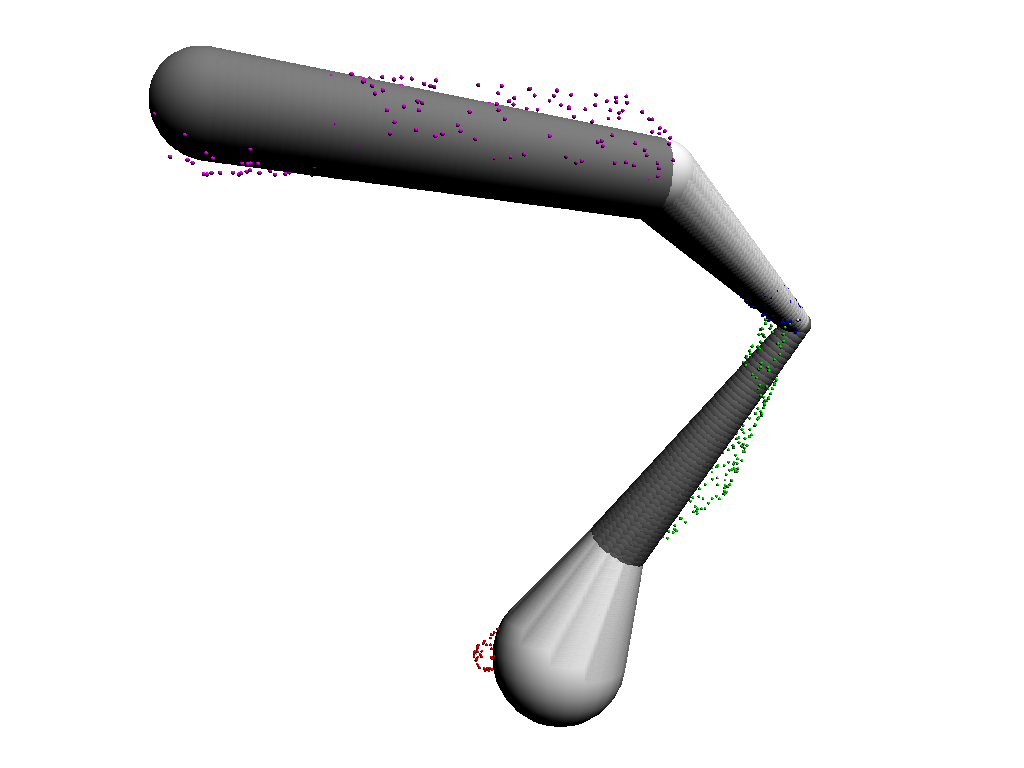}
        \end{subfigure}
        ~
        \begin{subfigure}[ht]{0.15\linewidth}
            \includegraphics[width=\linewidth]{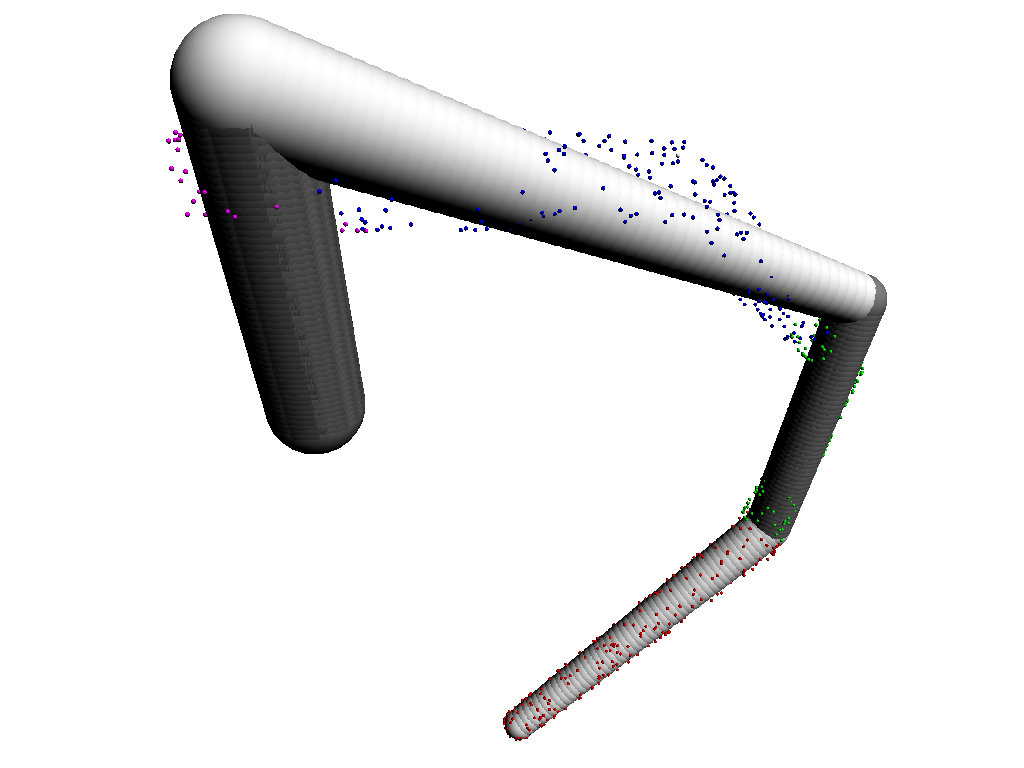}
        \end{subfigure}
        ~
         \begin{subfigure}[ht]{0.15\linewidth}
            \includegraphics[width=\linewidth]{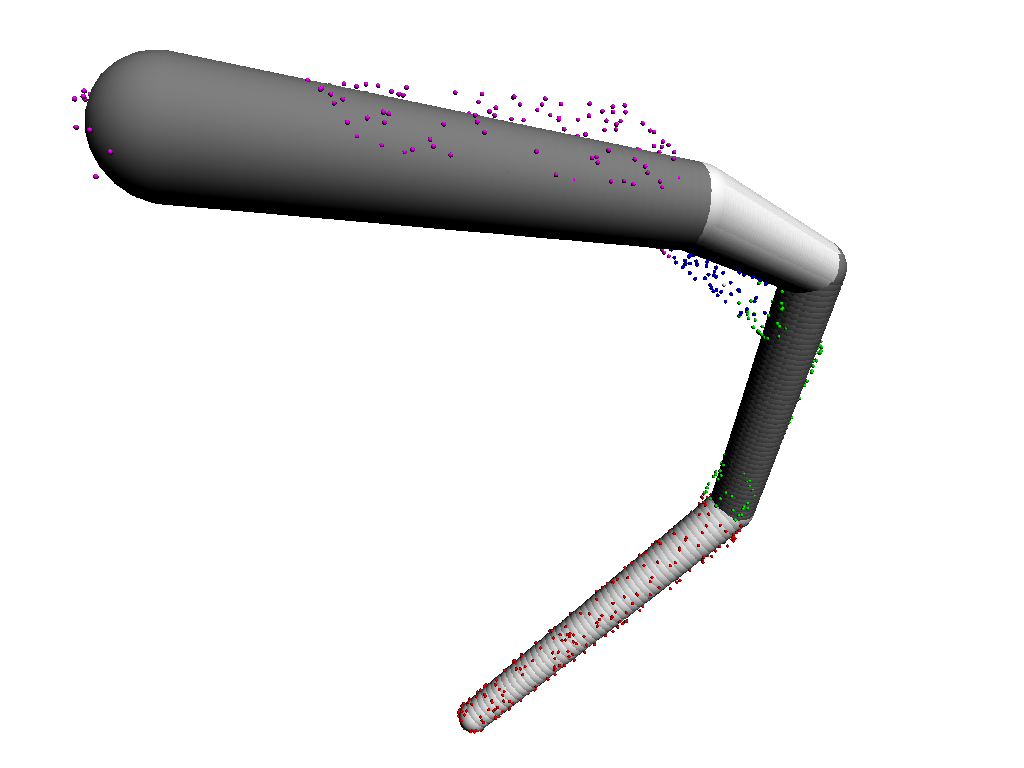}
        \end{subfigure}
        ~
        \begin{subfigure}[ht]{0.15\linewidth}
            \includegraphics[width=\linewidth]{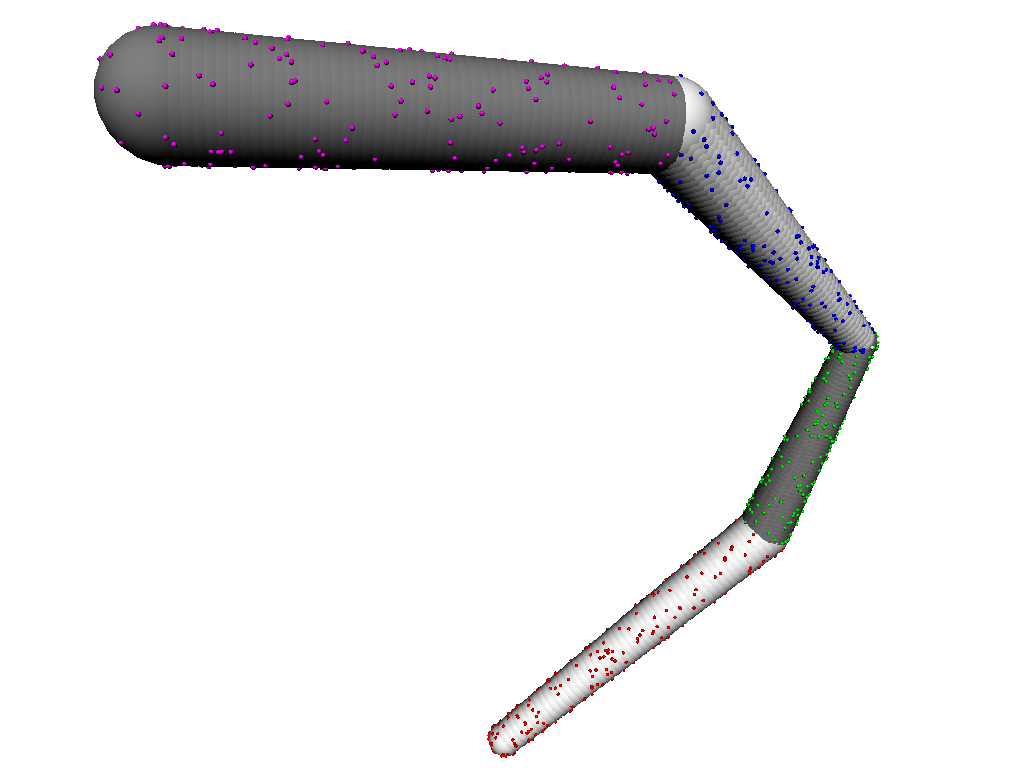}
        \end{subfigure}
    \end{center}
    \caption{Simultaneous Optimization for a chain of bones applied after the first Fakir forward pass. From left to right: initial position (after our forward step), position after 10 iterations, position after 35 iterations, position after convergence (43 iterations).}
    \label{fig:opt_simu}
\end{figure*}

\begin{figure}[ht]
    \begin{center}
    \begin{subfigure}[t]{0.4\linewidth}
        \includegraphics[width=\linewidth]{ite_0}
        \caption{Initial position for a simultaneous optimization }
    \end{subfigure}
    ~
    \begin{subfigure}[t]{0.4\linewidth}
        \includegraphics[width=\linewidth]{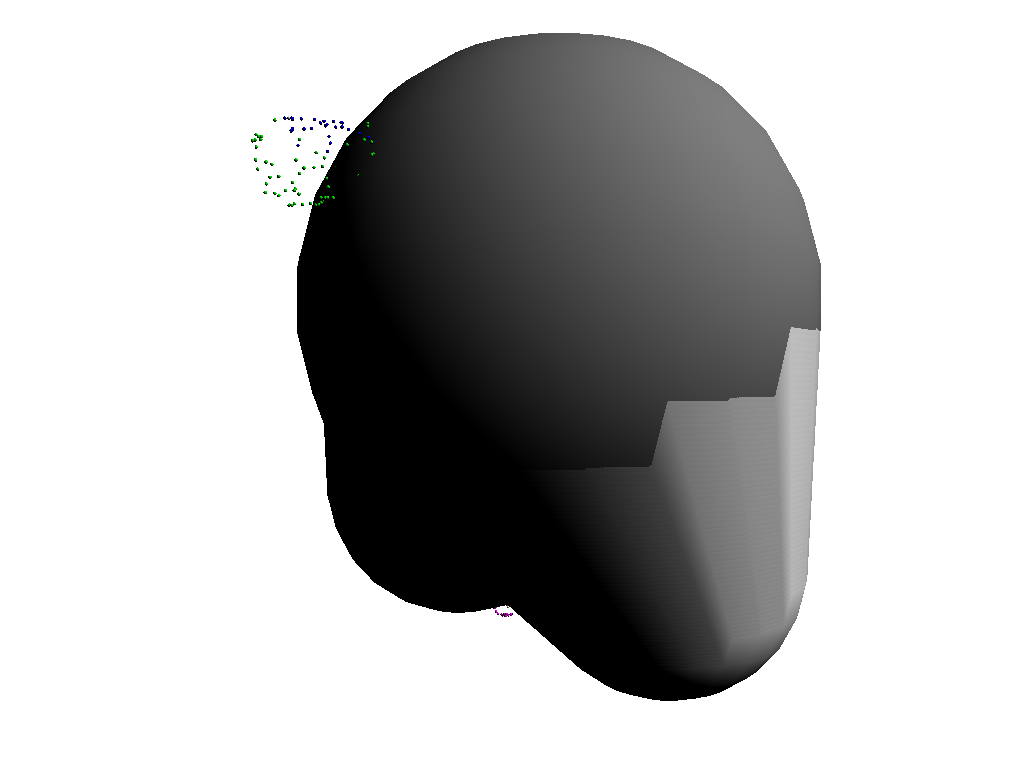}
        \caption{Position after 50 iterations }
    \end{subfigure}
    \end{center}
    \caption{Simultaneous Optimization for a chain of bones applied directly from the initial position.}
    \label{fig:opt_debut}
\end{figure}

If this optimization is run after our forward step (which is useful to bring each bone close to relevant data), it takes 43 iterations to converge, against 9 iterations only with our approach (See Figure \ref{fig:opt_simu} for an illustration of the stages). 
On the contrary, if the simultaneous optimization is run directly from the initial position, the method fails to converge. Figure \ref{fig:opt_debut} shows the result after 50 simultaneous optimizations steps.

\section{Importance of the normal-constrained projection}
\begin{figure}
  \begin{center}
         \begin{subfigure}[ht]{0.20\linewidth}
        \includegraphics[width=\linewidth]{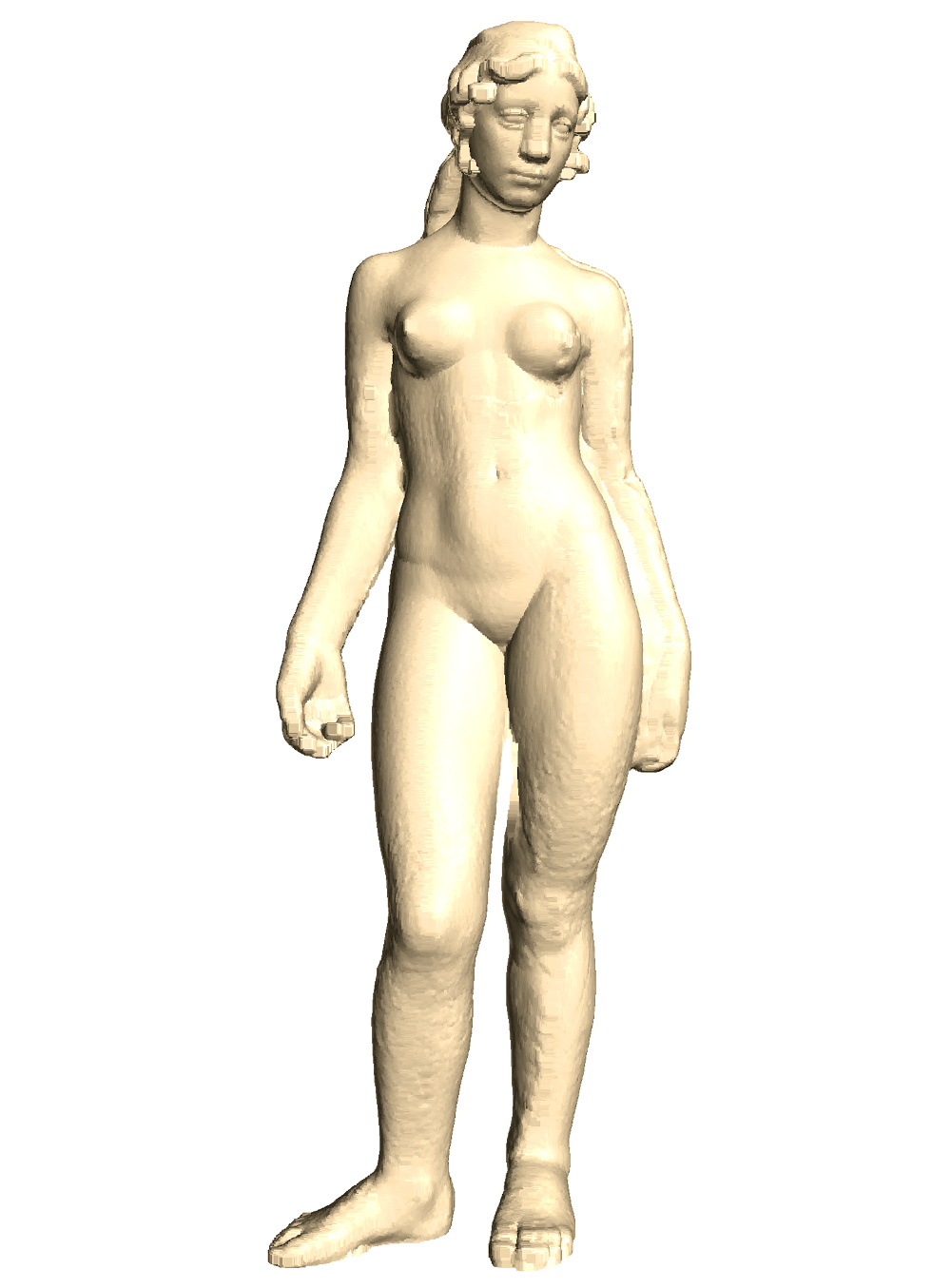}
        \end{subfigure}
        ~
        \begin{subfigure}[ht]{0.20\linewidth}
        \includegraphics[width=\linewidth]{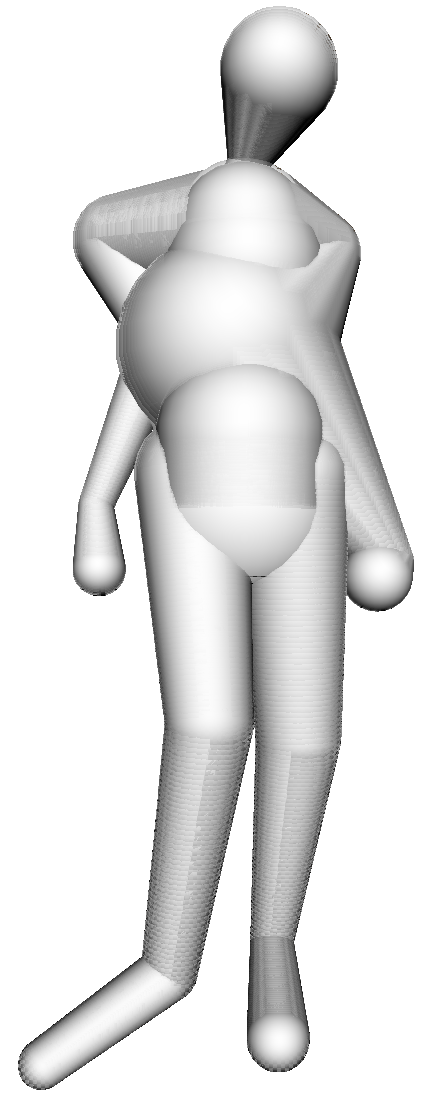}
        \end{subfigure}
        ~
        \begin{subfigure}[ht]{0.21\linewidth}
            \includegraphics[width=\linewidth]{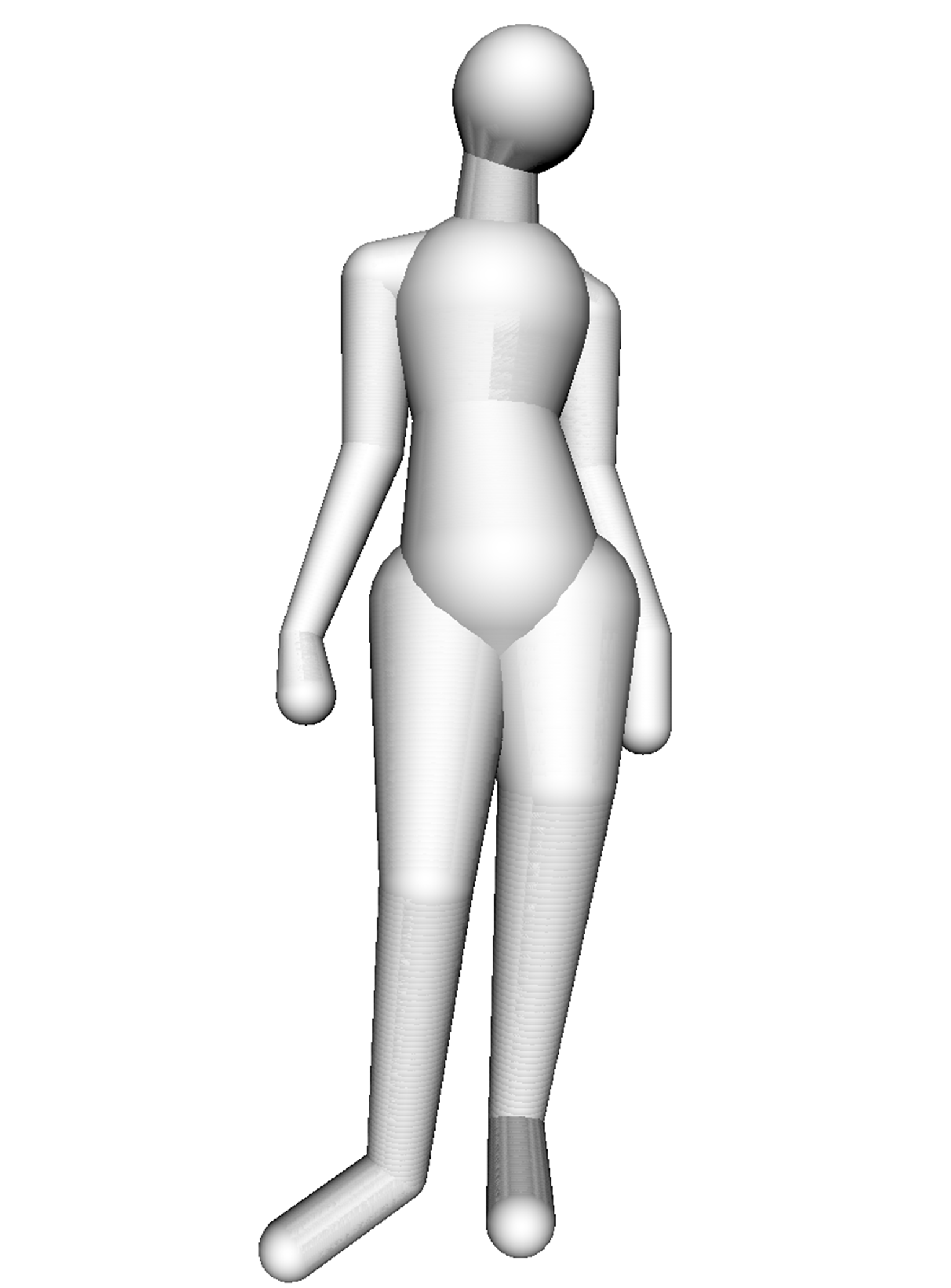}
        \end{subfigure}
        \end{center}
    \vspace{-5mm}
    \caption{From left to right: Aphrodite statue, registration result without the normal-constrained projection, registration result with our normal-constrained projection}
    \label{fig:constrained_projection}
\end{figure}

In this section we demonstrate that the normal-constrained projection both improves the result of the registration and the computation time. It is especially true in the case of the Aphrodite statue (Figure \ref{fig:constrained_projection}). Indeed, for this statue, the arms cling to the body which leads to wrong assignment of points when no normal information is used, yielding an unrealistic statue pose. The normal-constrained projection, on the contrary, permits to recover a good pose of the arms.
Furthermore, it takes $16.9s$ and 2 iterations for the algorithm using normal-constrained projection, against $32.8s$ and  8 iterations -- to converge to a wrong registration -- otherwise (number of points: $38954$). 
On simpler cases, like the Dancing Faun (see the main paper for a rendering of the shape), both methods manage to converge to the correct registration, but the computation time and number of iterations are still lower for the normal-constrained projection (8.7s, 6 iterations) than for the simple orthogonal projection (10.4s, 10 iterations).

\section{Automatic initialization test}
\begin{figure}
  \begin{center}
        \includegraphics[width=0.3\linewidth]{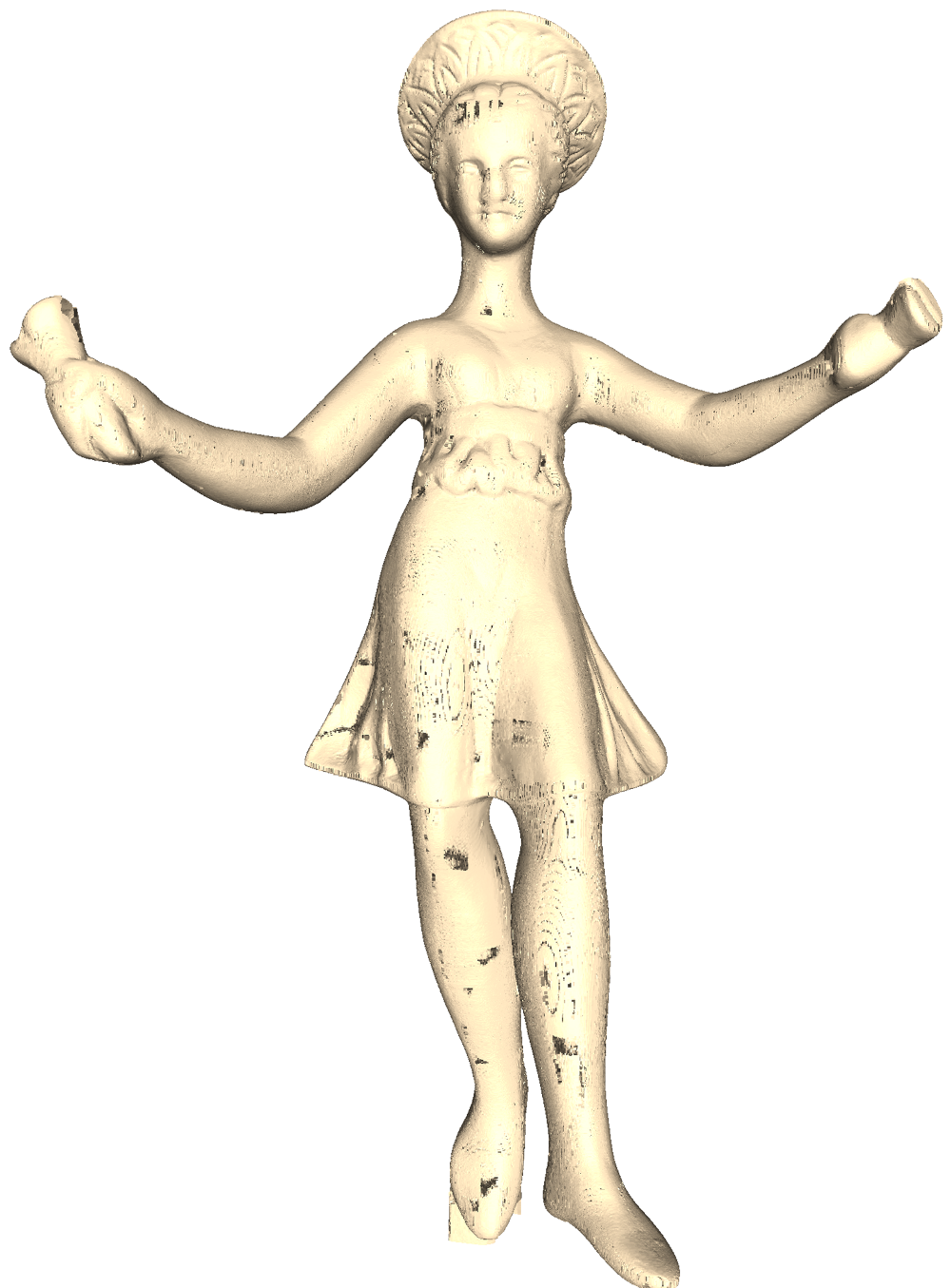}
        \includegraphics[width=0.3\linewidth]{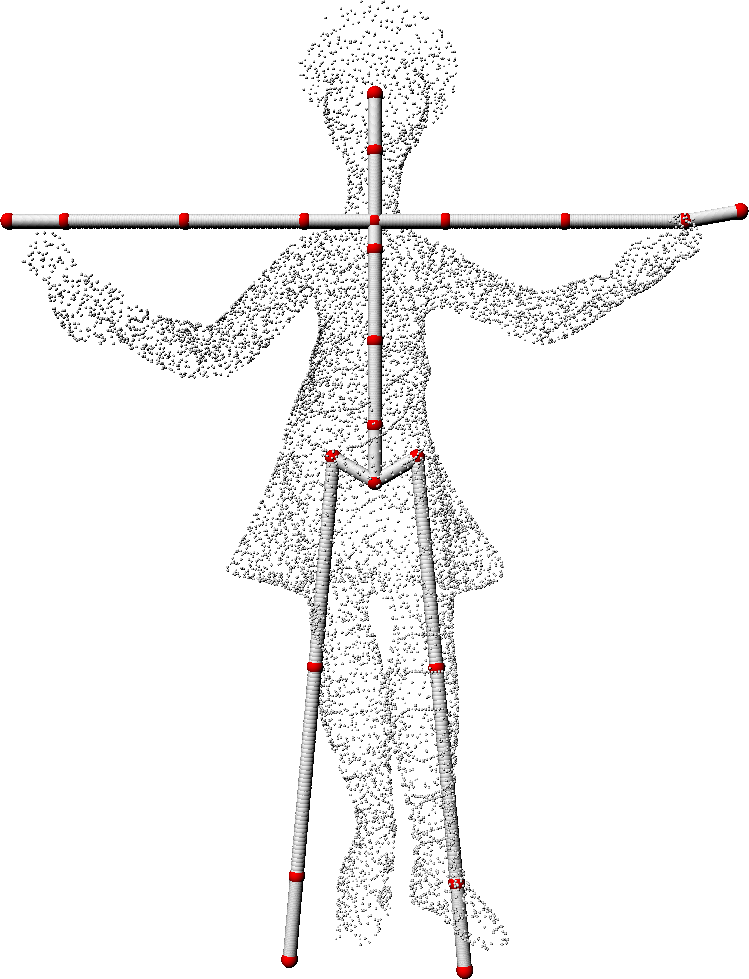}
        \includegraphics[width=0.3\linewidth]{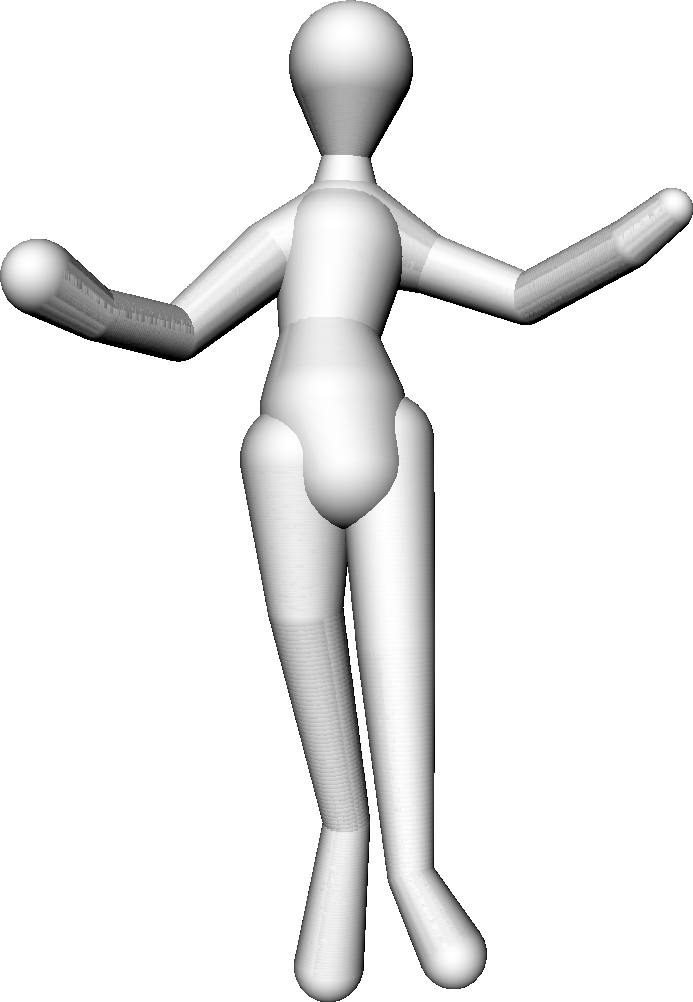}
  \end{center}
  \caption{From left to right: the Dancer with Crotales statue, initialization position at bounding box center of point cloud, automatic registration result.}
  \label{fig:result_auto_1}
\end{figure}

\begin{figure}
  \begin{center}
        \includegraphics[width=0.3\linewidth]{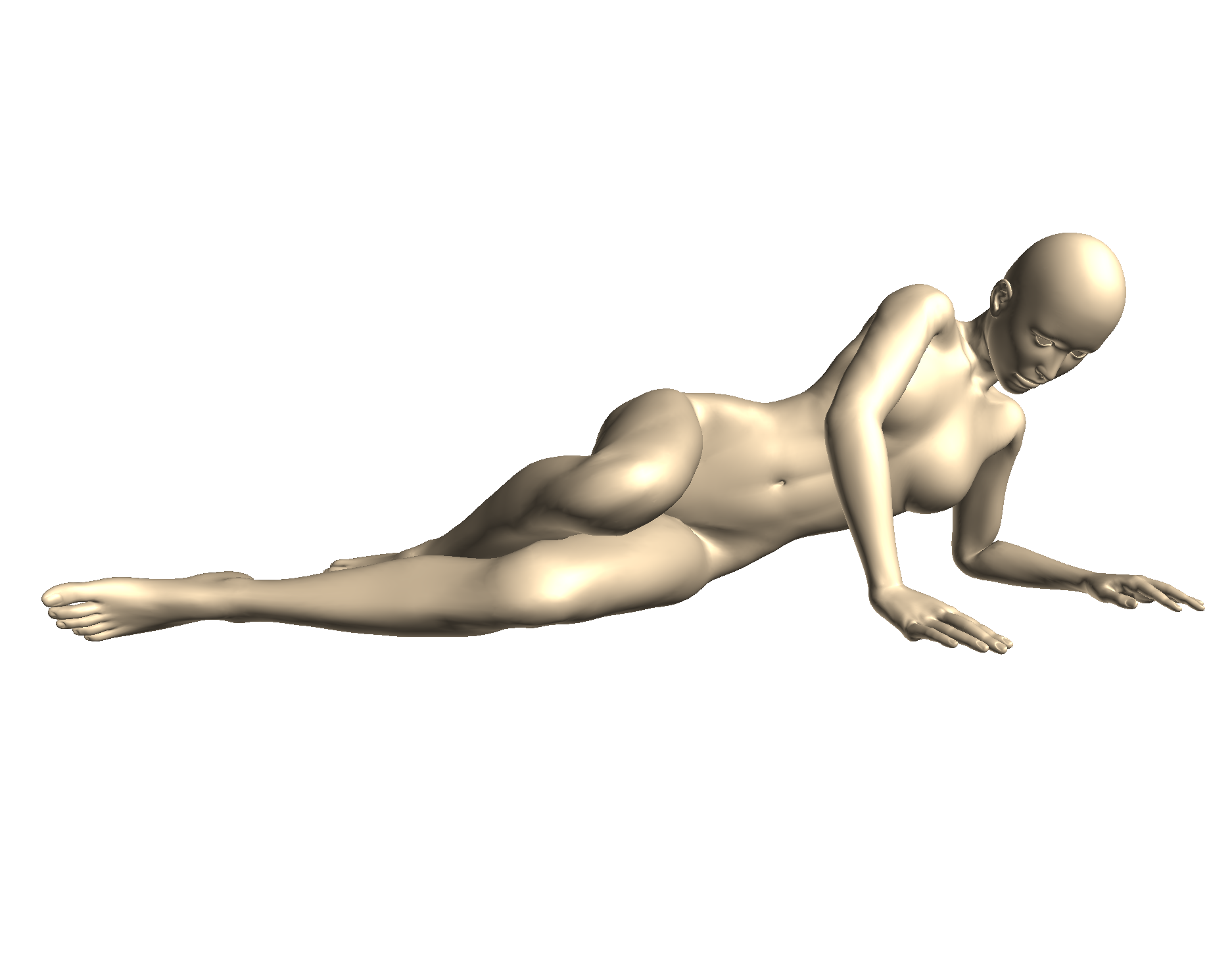}
        \includegraphics[width=0.3\linewidth]{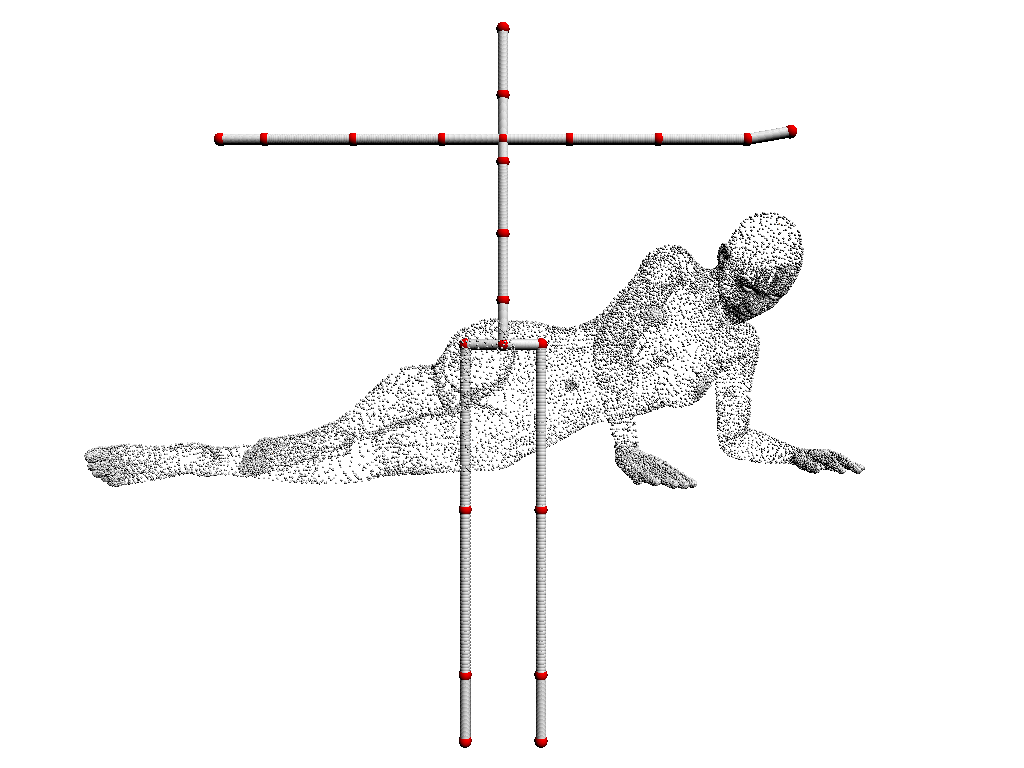}
        \includegraphics[width=0.3\linewidth]{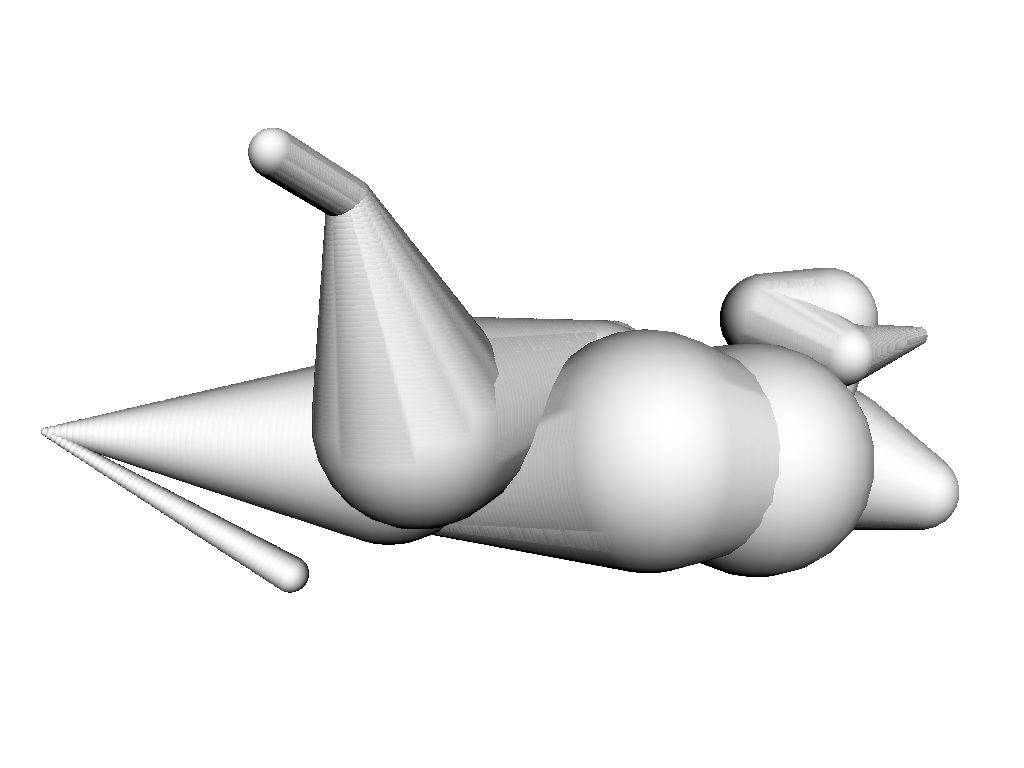}
  \end{center}
  \caption{A failure case for automatic initialization. From left to right: the Victoria from TOSCA data set, initialization position at bounding box center of point cloud, automatic registration result.}
  \label{fig:result_auto_2}
\end{figure}

Our attempt to automatically initialize FAKIR by placing the pelvis in the center of the bounding box is effective for many statues. The Figure \ref{fig:result_auto_1} illustrates that this works with a vertically oriented statue, while the Figure \ref{fig:result_auto_2} shows a failure result, due to the fact that the center of the bounding box does not give any information about the orientation of the body. Thus the chains of bones are not aligned with the appropriate points, which leads to a local minimum. An improved version using principal component analysis of the points to initialize the orientation of the pelvis and adding loose constraints on the length of the bones could improve the registration.

\section{Additional registration results}

  \begin{figure*}
  \begin{center}
        \includegraphics[width=0.12\linewidth]{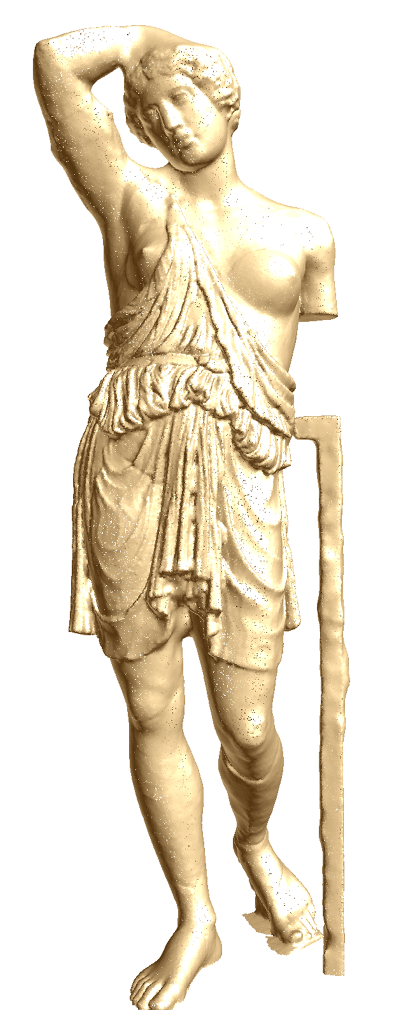}
        \includegraphics[width=0.12\linewidth]{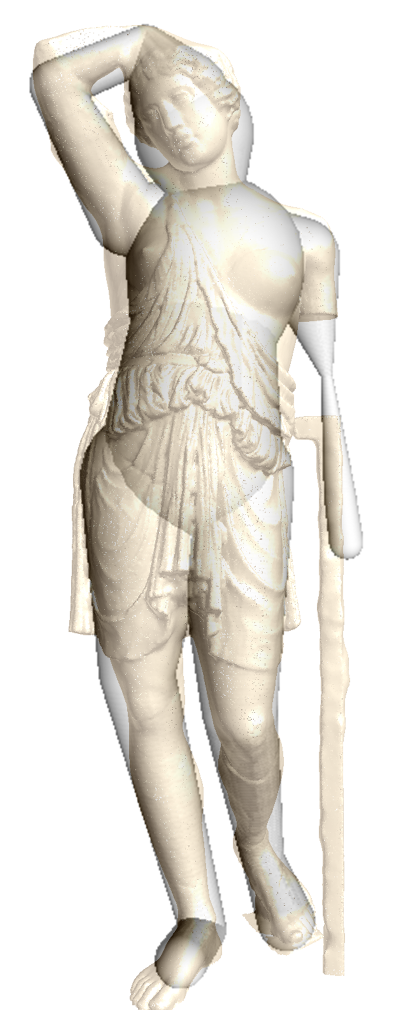}
        \includegraphics[width=0.12\linewidth]{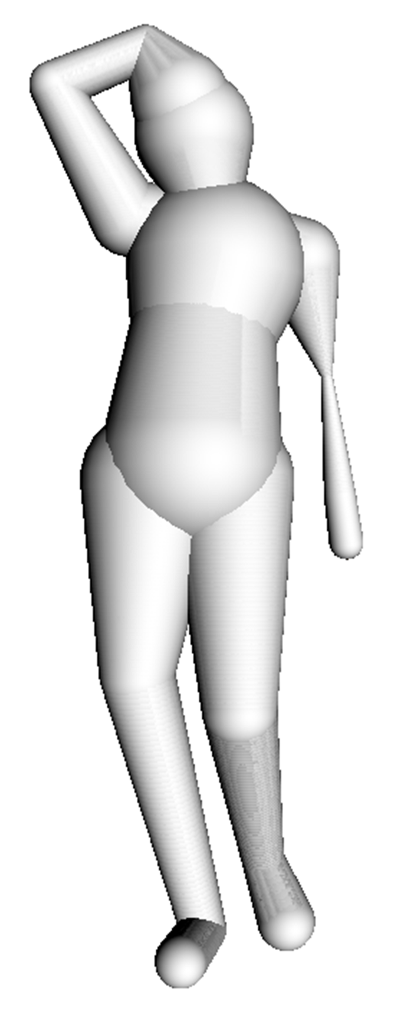}
        \includegraphics[width=0.13\linewidth]{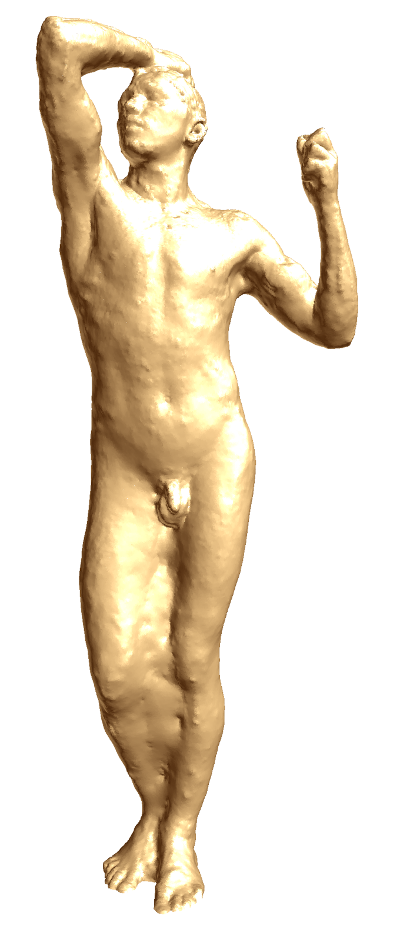}
        \includegraphics[width=0.13\linewidth]{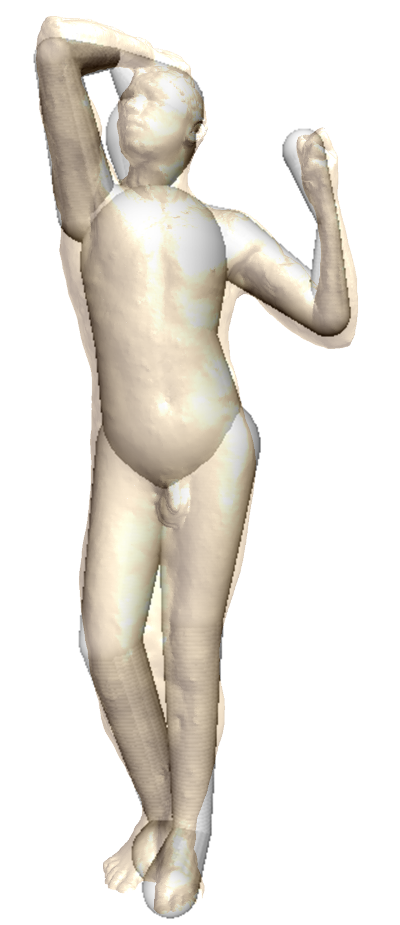}
        \includegraphics[width=0.13\linewidth]{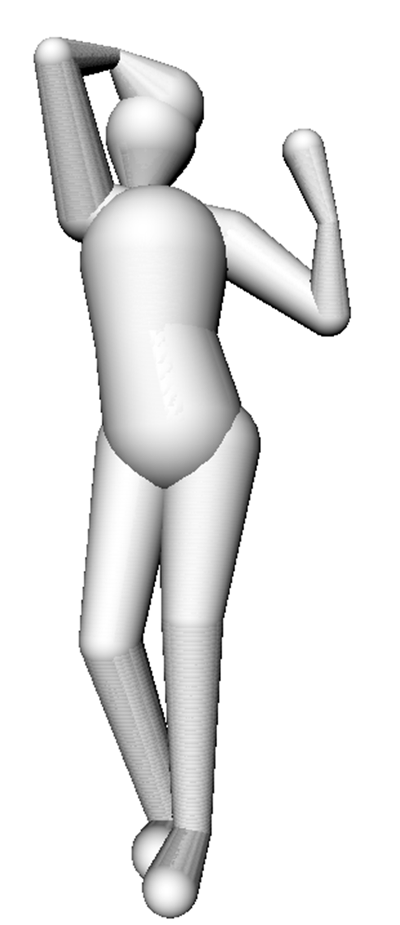}
        
        \includegraphics[width=0.11\linewidth]{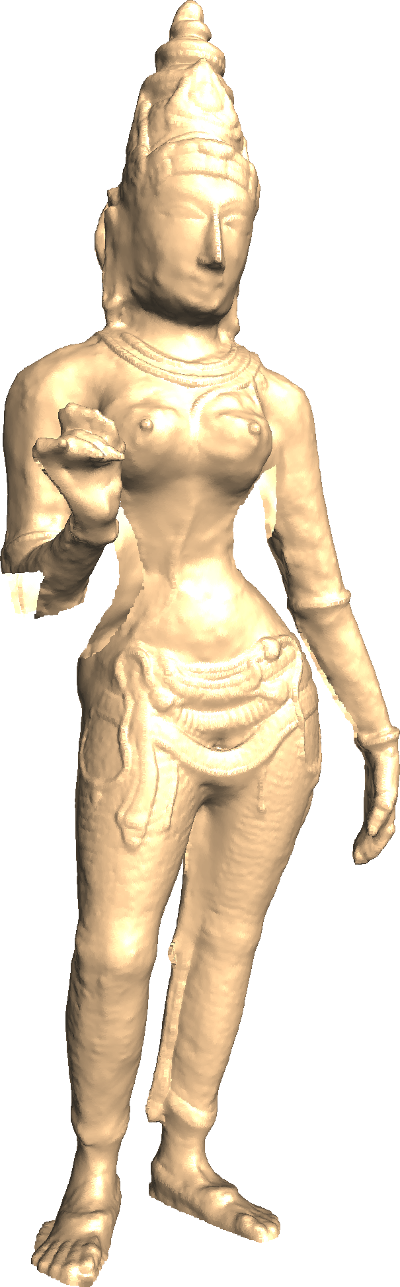}
        \includegraphics[width=0.11\linewidth]{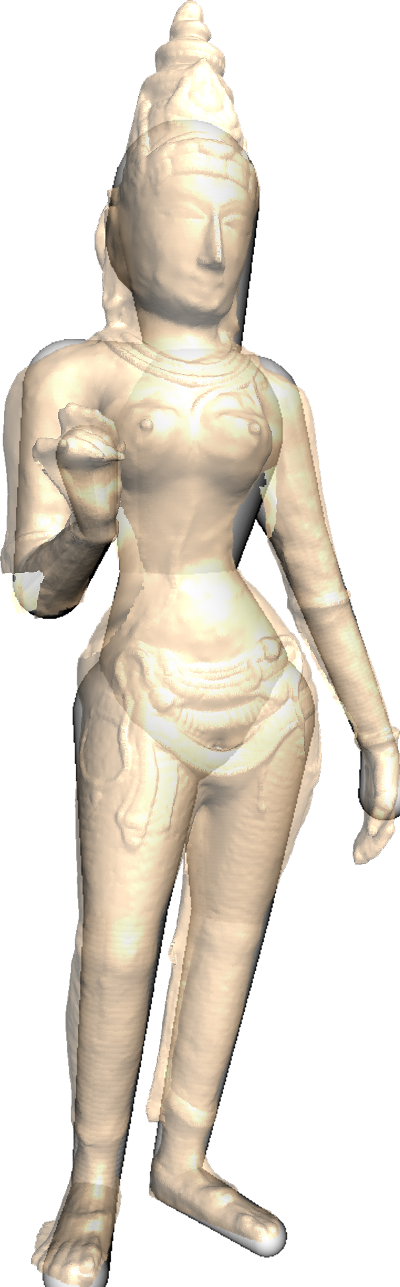}
        \includegraphics[width=0.11\linewidth]{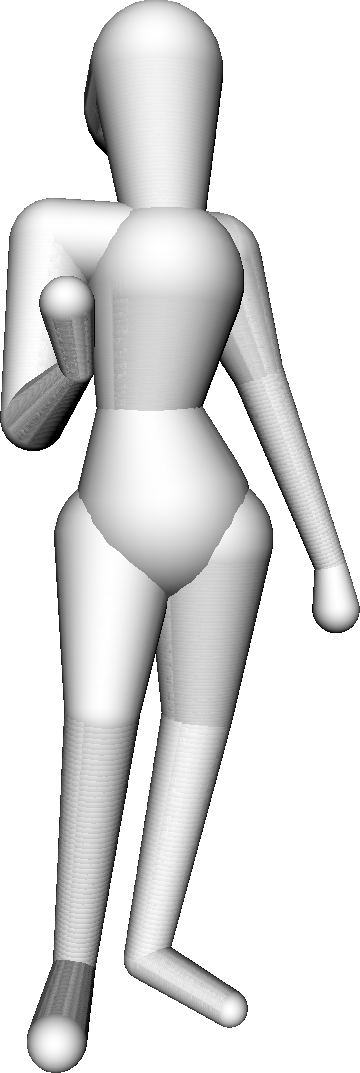}
        \end{center}
    \caption{Additional statue registration results. Top-left: \emph{Wounded Amazon} (Rome, 150 A.D., Nye Carlsberg Glyptotek, Copenhagen, Denmark); Top-right: \emph{Age of Bronze} (Auguste Rodin), top-right; Bottom: \emph{The Goddess Parvati} (South India, Circa 1200 A.C.). For each result, we show the initial point set, overlay of the data and the registered model, and the registered model alone.}
    \label{fig:result}
\end{figure*}

\begin{figure*}
  \begin{center}
        \includegraphics[width=0.3\linewidth]{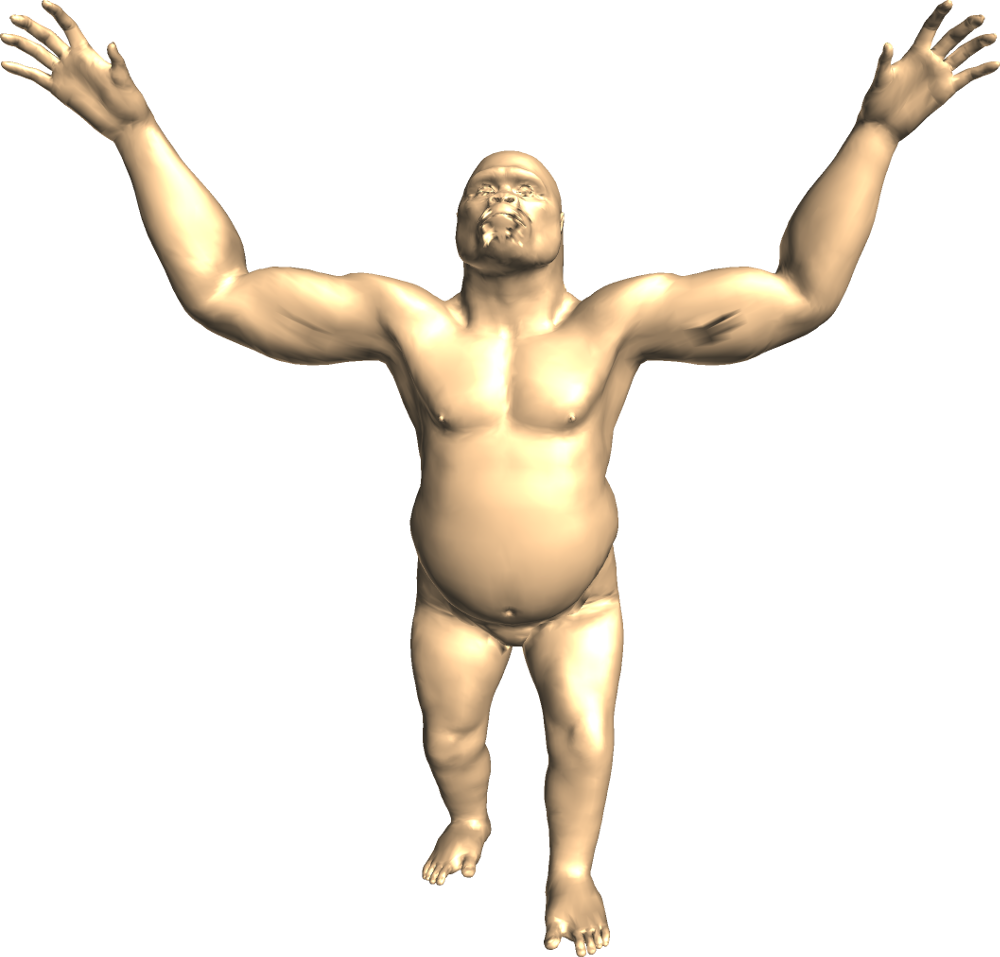}
        \includegraphics[width=0.3\linewidth]{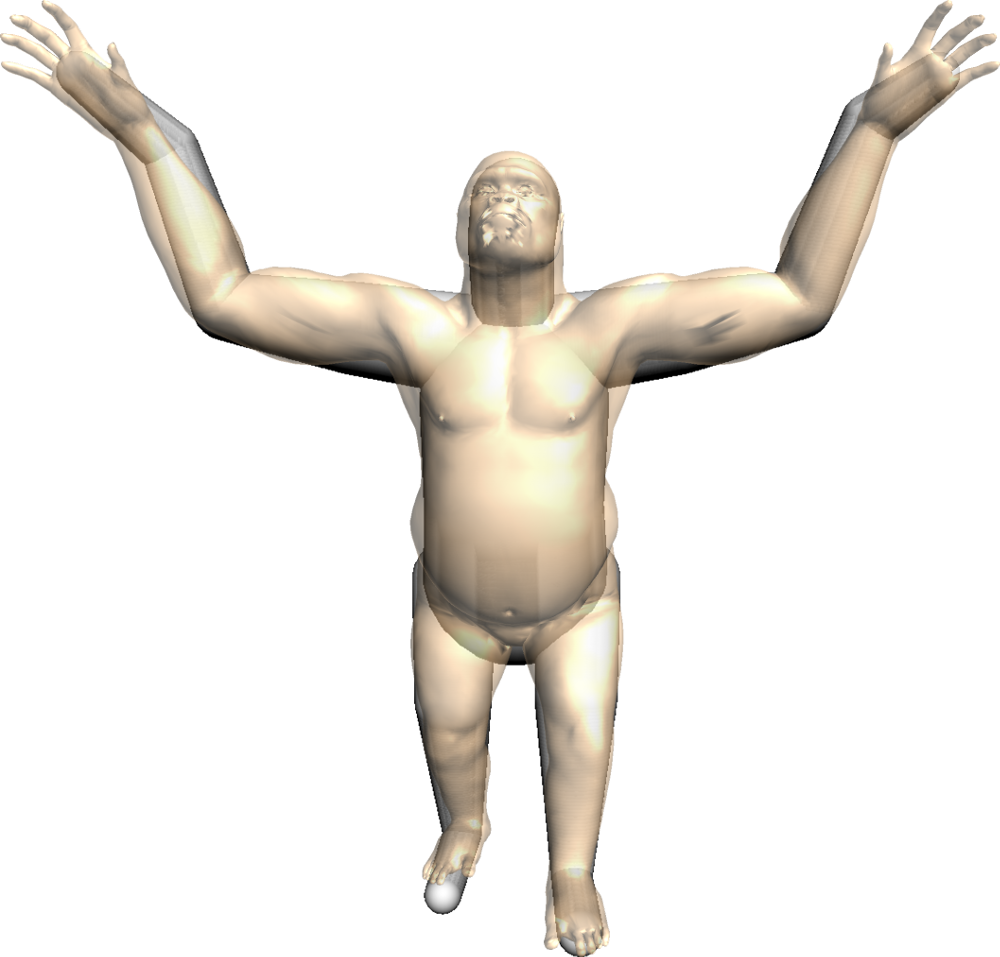}
        \includegraphics[width=0.28\linewidth]{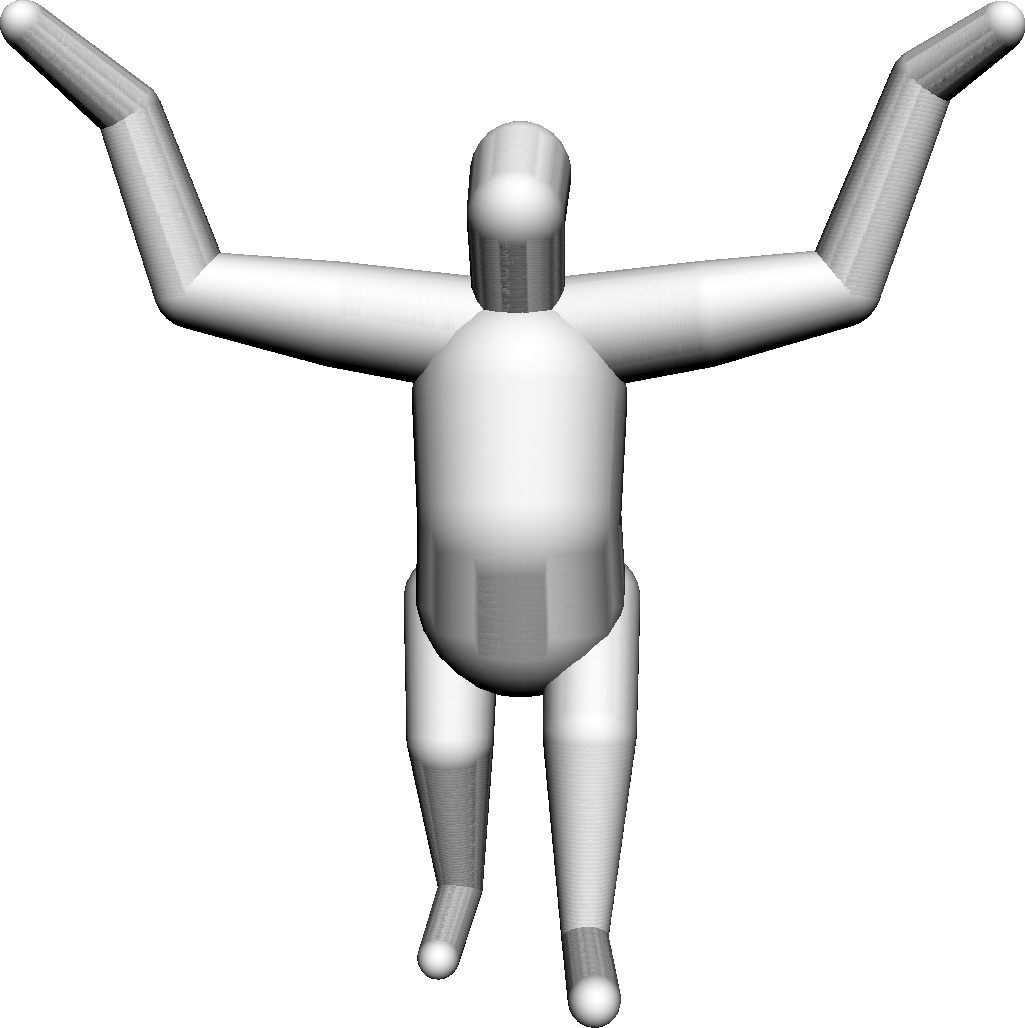}
  \end{center}
  \caption{Additional result on the Gorilla from the TOSCA dataset (initial point set; overlay of the data and the registered model; registered model alone).}
  \label{fig:result2}
\end{figure*}

We show on Figures \ref{fig:result} and \ref{fig:result2} four additional results, including one with a missing part (the arm) and some garments which do not hinder the registration. The Goddess Parvati and the Gorilla show the good performance of FAKIR on data which do not have realistic human proportions, while still using the standard human skeleton presented in the paper.

\end{appendices}